\documentclass{aa}
\usepackage{graphics,graphicx}
\usepackage[varg]{txfonts}
\usepackage{rotating}
\usepackage{lscape}
\usepackage{rotating}
\usepackage{natbib}
\usepackage{longtable}
\newcommand{\asec}{$^{\prime\prime}$}

\def\H{N$_{2}$H$^{+}$}
\def\D{N$_{2}$D$^{+}$}

\def\AMM{NH$_3$}
\def\DAMM{NH$_2$D}
\def\oDAMM{{\it ortho-}NH$_2$D}
\def\pDAMM{{\it para-}NH$_2$D}
\def\FORM{H$_2$CO}

\def\CII{\mbox{C$^{18}$O}}

\def\METH{CH$_3$OH}
\def\METHI{$^{13}$CH$_3$OH}
\def\DMETH{CH$_2$DOH}
\def\METHD{CH$_3$OD}

\def\HII{H{\sc ii}}

\def\kms{\mbox{km~s$^{-1}$}}
\def\cmc{cm$^{-3}$}
\def\cmq{cm$^{-2}$}

\def\Vlsr{$V_{\rm LSR}$}
\def\Dfrac{$D_{\rm frac}$}
\def\Tex{\mbox{$T_{\rm ex}$}}

\def\Trot{\mbox{$T_{\rm rot}$}}

\def\deltav{\mbox{$\Delta V$}}

\begin{document}

\title{Deuteration and evolution in the massive star formation process: the role of surface chemistry
}
%\subtitle{...}
\author{F. Fontani \inst{1} 
             \and G. Busquet \inst{2,3} 
              \and Aina Palau \inst{4} 
             \and  P. Caselli \inst{5}
         % \and Q. Zhang \inst{3}   
         % \and J. Brand \inst{4}  
         % \and G. Busquet \inst{4} 
          \and \'A. S\'anchez-Monge \inst{6} 
          \and J.C. Tan \inst{7,8}
          \and M. Audard \inst{9}
                 }
\offprints{F. Fontani, \email{fontani@arcetri.astro.it}}
\institute{INAF - Osservatorio Astrofisico di Arcetri, L.go E. Fermi 5, I-50125, Firenze, Italy
%	   \and
%             Institut de Radio-Astronomie Millim\'etrique, 300 rue de la Piscine, Domaine Universitaire, 38406 Saint Martin  d'H\`eres, France
         \and
          Instituto de Astrof\'isica de Andaluc\'ia, CSIC, Glorieta de la Astronom\'ia, E-18008, Granada, Spain
          \and
          INAF - Istituto di Astrofisica e Planetologia Spaziali, via Fosso del Cavaliere 100, I-00133, Roma, Italy
          \and
%           Institut de Ci\`encies de l'Espai (CSIC-IEEC), Campus UAB-Facultat de Ci\`encies, Torre C5-parell 2, 08193, Bellaterra, Spain
           Centro de Radioastronom\'ia y Astrof\'isica, Universidad Nacional Aut\'onoma de M\'exico, P.O. Box 3-72, 58090 Morelia, Michoac\'an, M\'exico
	  \and
	 Max-Planck-Institut f\"ur extraterrestrische Physik (MPE), Giessenbachstr., D-85741 Garching, Germany 
	   \and 
	 I. Physikalisches Institut der Universit\"at zu K\"oln, Z\"ulpicher Strasse 77, 50937 K\"oln, Germany  
	 % \and
	 % INAF-Istituto di Astrofisica e Planetologia Spaziali, via Fosso del Cavaliare 100, 00133 Roma, Italy
	  \and 
           Department of Astronomy, University of Florida, Gainesville, FL 32611, USA
           \and 
            Department of Physics, University of Florida, Gainesville, FL 32611, USA
%           ISDC Data Center for Astrophysics, University of Geneva, Ch. d'Ecogia 16, 1290 Versoix, Switzerland 
           \and
           Department of Astronomy, University of Geneva, Ch. d'Ecogia 16, 1290 Versoix 
%           Geneva Observatory, University of Geneva, ch. des Maillettes 51, 1290 Versoix, Switzerland
	   } 
\date{Received date; accepted date}

%\markboth{Fontani et al.: Mol160}{}
\titlerunning{Deuteration in massive star formation}
\authorrunning{Fontani et al.}

 \abstract{
 %The column density
 %ratio of a molecule containing D to its counterpart containing H, called deuterated
 %fraction (\Dfrac ), can be used as an 
 %evolutionary tracer in the low-mass star formation process.
 An ever growing number of observational and theoretical evidence suggests that
 the deuterated fraction (column density ratio between a species containing 
 D and its hydrogenated counterpart, \Dfrac ) is an evolutionary indicator
 both in the low- and the high-mass star formation process. However,
 the role of surface chemistry in these studies has not been quantified
 from an observational point of view.}
%We have started a survey of deuterated molecules towards massive 
%dense cores associated with different stages of the massive star 
%formation process (from massive starless cores to ultracompact
%\HII\ regions) to investigate the relation between the percentage
%of the deuterated species with respect to the main isotopologue,
%or deuterated fraction (\Dfrac ), and the core evolutionary stage.}
%The first results indicate that \Dfrac(\H ) drops 
%of an order of magnitude in evolved cores (Fontani et al.~\citeyear{fontani11}), 
%and is thus an excellent tracer of the earliest (starless) phases.}
{Because many abundant species, like \AMM ,
H$_2$CO and \METH , are actively produced on ice mantles
of dust grains during the early cold phases, their \Dfrac\ 
is expected to evolve differently from that of species
formed only (or predominantly) in the gas, like \H , HNC, HCN
and their deuterated isotopologues. The differences are expected to be relevant 
especially after the protostellar birth, in which the temperature rises up 
causing the evaporation of ice mantles.}
{In order to compare how the deuterated fractions of species formed only in the gas and 
partially or uniquely on grain surfaces evolve with time, 
we observed rotational transitions of \METH , \METHI , \DMETH , \METHD\
at 3 and 1.3~mm, and of \DAMM\ at 3~mm with the IRAM-30m telescope, and 
 the inversion transitions (1,1) and (2,2) of \AMM\ with the GBT, towards most
 of the cores already observed by Fontani et al.~(\citeyear{fontani11}, \citeyear{fontani14}) 
 in \H , \D , HNC, DNC.}
 {NH$_2$D is detected in all but two cores, regardless of the evolutionary
 stage. \Dfrac (\AMM ) is on average above 0.1, and does not change significantly 
 from the earliest to the most evolved phases, although the highest average
 value is found in the protostellar phase ($\sim 0.3$).
 Few lines of \DMETH\ and \METHD\ are clearly detected, and only 
 towards protostellar cores or externally heated starless cores. In quiescent
 starless cores, we have found only one doubtful detection of \DMETH .}
{This work clearly confirms an expected different evolutionary trend of the species
formed exclusively in the gas (\D\ and \H ) and those formed 
partially (\DAMM\ and \AMM ) or totally (\DMETH\ and \METH )
on grain mantles. The study also reinforces the idea that \Dfrac (\H )
is the best tracer of massive starless cores, while 
high values of \Dfrac (\METH ) seem rather good
tracers of the early protostellar phases, at which the evaporation/sputtering 
of the grain mantles is most efficient.}
  

\maketitle
%
%________________________________________________________________

\section{Introduction}
\label{Introduction}

Theory and observations suggest that the abundance of
deuterated molecules in dense star-forming cores
is related to the core evolution. The formation of deuterated molecules
is favoured by the combination of low temperatures
($T\leq 20$ K) and high-densities ($n\geq 10^4$ \cmc ),
which on one hand boosts the depletion of CO and other neutrals, and on
the other hand makes the relative abundance between a species containing
D and its hydrogenated counterpart (the so-called deuterated fraction, \Dfrac )
higher by 3-4 orders of magnitude with respect to the [D/H] interstellar
abundance ($\sim 10^{-5}$, e.g.~Oliveira et al.~\citeyear{oliveira03}),
due to the endothermicity of their backward reactions
(see e.g.~Millar et al.~\citeyear{millar89}, Gerlich et al.~\citeyear{gerlich02}).
After protostellar birth, the young stellar object formed at the core 
centre heats up its surrounding material, and the temperature enhancement
favours the progressive destruction of deuterated species and,
consequently, makes \Dfrac\ decrease (see e.g. Caselli et al. 2002).
Observations of {\it low-mass star-forming cores} have confirmed this
theoretical scenario:  
both the column density ratio \Dfrac (\H ) and the column density
of {\it ortho-}H$_2$D$^+$, the parent species of most of the deuterated
molecules formed in the gas (e.g.~DCO$^+$, \D , DNC), increase 
in starless cores close to the onset of gravitational collapse, and
then, after the formation of the protostar, decrease as the core evolves
(Crapsi et al.~2005, Emprechtinger et al.~2009, Caselli et al.~2008).
Growing observational evidence suggests that high values of \Dfrac\ are 
typical also in high-mass star-forming cores (e.g.~Fontani
et al.~\citeyear{fontani06}, Pillai et al.~\citeyear{pillai07}, 
\citeyear{pillai11}, Miettinen et al.~\citeyear{miettinen11}), and that \Dfrac\ 
of some species could be an evolutionary indicator also in the intermediate-
and high-mass regime 
(e.g. Busquet et al.~\citeyear{busquet10}, Fontani et al.~\citeyear{fontani11}, Sakai et al.~\citeyear{sakai12}).

To investigate the relation between \Dfrac\ and core evolution in
the high-mass regime in a systematic way, our team started a survey of deuterated 
molecules in about 30 dense cores carefully selected and almost 
equally divided among the three evolutionary phases in which we can roughly
divide observationally the high-mass star formation process
(see e.g.~Beuther et al.~\citeyear{beuther07} and Tan et al.~\citeyear{tan14}): 
high-mass starless cores (HMSCs), high-mass protostellar objects (HMPOs) and 
ultracompact \HII\ regions (UC \HII s).
%, with the aim of investigating if some 
%of these species are evolutionary indicators in the high-mass regime. 
In brief, the targets were selected as follows: the HMSCs had to be dense 
molecular cores not associated with indicators of star formation; the 
HMPOs had to show outflows, and/or infrared sources, and/or faint 
($S_{\rm 3.6 cm}< 1$mJy) radio continuum emission; the UCHIIs had to 
be associated with stronger ($S_{\rm 3.6 cm}\geq 1$mJy) radio continuum.
In selecting the sources, we rejected cores blended with nearby cores
to avoid confusion and make the emission of the targeted core dominant. 

In the first study, we (Fontani et al.~\citeyear{fontani11}, hereafter paperI) 
presented the results obtained from spectroscopic observations of 
millimeter rotational transitions of \H\ and \D\ obtained with the IRAM-30m telescope, 
where we showed that \Dfrac (\H ) is $\sim 0.26$ in HMSCs,
and drops by about an order of magnitude in the HMPO and 
UC \HII\ stages. These results are consistent with the fact that
deuteration of \H\ starts from the reaction H$_2$D$^+$ + N$_2$ $\rightarrow$ \D\ + H$_2$, 
efficient only at temperatures
$\leq 20$~K (Gerlich et al.~\citeyear{gerlich02}). In a following study, 
focused on DNC/HNC, Fontani et al.~(\citeyear{fontani14}, paperII)
showed that \Dfrac (HNC) also decreases from the pre- to the 
proto--stellar phase, but much more moderately, indicating that the ratio
\D -to-\H\ is more appropriate to identify massive starless cores. 
This is consistent with the prediction that DNC can also
easily form when the gas gets warmer, because the route reaction for the
deuteration of HNC is linked to CH$_2$D$^+$, which can stay abundant
up to temperatures of 70~K (e.g.~Leurini et al.~\citeyear{leurini06}).
However, \H , HNC and their deuterated isotopologues can form mainly 
(HNC, DNC) or solely (\H , \D ) in the gas phase.
Other important molecules, like \AMM , \FORM , \METH\ and their deuterated forms, 
can be produced on dust grain surfaces (e.g. Aikawa et al. 2005), and theoretical
models show that this can make relevant differences in their \Dfrac , especially 
during the protostellar phase in which grain mantles evaporate 
(Aikawa et al.~\citeyear{aikawa12}).

%during the cold starless phase, 
%one expects an increase of \Dfrac (\METH ) from HMSCs to young HMPOs, 
%where grain mantles have just evaporated after the protostellar birth
%as seen recently in Orion BN/KL (Peng et al.~2012), 
%contrary to the sharp decrease of \Dfrac (\H );
%then, in more evolved cores (old HMPOs and UCHIIs)

In this work we investigate the role of surface chemistry by means of 
measurements of \Dfrac (\AMM ) and \Dfrac(\METH ). 
Because methanol and its deuterated forms 
can be produced {\it only on grain surfaces} (see e.g. Parise et al.~\citeyear{parise02},
Garrod et al.~\citeyear{garrod07}), \Dfrac(\H ) and \Dfrac(\METH ) represent the two 
"extreme conditions" under which deuteration can occur: on grain surfaces only 
(\METH ) and in gas only (\H ). Therefore, the results obtained in this work and
in paperI can be used as reference for the deuteration process of any
other species formed potentially both in the gas and on dust grains.  
In Sect.~\ref{obs} we present the source sample and give an overview of the 
technical details of the observations. The main results are presented and 
discussed in Sects.~\ref{res} and \ref{discu}, respectively. 
A summary with the main conclusions
of the work is given in Sect.~\ref{summary}.

\section{Source list, observations and data reduction}
\label{obs}

\subsection{Source list}

We targeted the same sources studied in paperI, to avoid 
possible biases due to the source selection when comparing the
deuterated fractions.
Table~\ref{tab_sources} contains the list of the observed sources selected
as explained briefly in Sect.~\ref{Introduction}. In particular, three HMSCs have been classified
as "warm" cores because they show evidence of heating from external
sources (see paperI for details). More information extracted from the literature about 
the star forming regions in which the sources lie are given in Table A.1 of paperI. 
To the list of HMSCs reported in paperI, we have added the source G028--C3, 
selected applying the same selection criteria as for the other HMSCs. 

\subsection{IRAM-30m observations}

{\it \bf Run-1}: towards all sources in Table~\ref{tab_sources}, observations of 
the {\it ortho-} and {\it para-}NH$_2$D($1_{1,1}-1_{0,1}$) line were obtained 
simultaneously to the \D\ and \H\ observations described in
paperI. Table~\ref{tab_mol} lists the main observational parameters.
We refer to Sect.~2 of paperI for any other technical detail related to these 
observations.

\noindent
{\it \bf Run-2}: we performed \METH\ and \DMETH\ observations towards all sources 
observed in paperI from the 6th to the 9th of February, 2013. 
We observed simultaneously two bands at 3 and 1.3~mm, covering 
some important rotational transitions of \METH , \METHI\ and \DMETH .
Table~\ref{tab_mol} presents the observed spectral windows and some main
technical observational parameters. 
The atmospheric conditions were very stable during the whole observing
period, with precipitable water vapour usually below $\sim 2$ mm.
The observations were made in wobbler--switching mode. Pointing 
was checked almost every hour on nearby quasars or bright
\HII\ regions. The data were calibrated with the chopper wheel 
technique (see Kutner \& Ulich~\citeyear{kutner}), with a calibration
uncertainty of $\sim 20 \%$. The spectra were obtained in
antenna temperature units, $T_{\rm A}^{*}$, and then converted to
main beam brightness temperature, $T_{\rm MB}$, via the relation
$T_{\rm A}^{*}=T_{\rm MB} (B_{\rm eff}/F_{\rm eff}$). 
The spectra were obtained with the Fast Fourier Transform Spectrometers (FTS),
providing a broad band of $\sim 8$ GHz simultaneously
at 3 and 1.3~mm (see Table~\ref{tab_mol} for details).
All calibrated spectra were analyzed using the 
GILDAS\footnote{The GILDAS software is available at http://www.iram.fr/ IRAMFR/GILDAS} software developed 
at the IRAM and the Observatoire de Grenoble. The rest frequencies used for 
the line identification have been taken from the Cologne Molecular Database 
for Spectroscopy (CDMS, http://www.astro.uni-koeln.de/cdms; M\"uller et al.~\citeyear{miller01}, 
\citeyear{miller05})

\subsection{GBT observations}

The ammonia (1,1) and (2,2) inversion transitions (rest frequencies 
23.6944955 and 23.7226336 GHz, respectively) were observed 
with the 100 m Robert C. Byrd Green Bank Telescope\footnote{The National Radio Astronomy Observatory is a facility 
of the National Science Foundation operated under cooperative 
agreement by Associated Universities, Inc.} (GBT) during the 
13th and 21st March and the 4th and 21th April 2013. 
The GBT spectrometer backend was configured to simultaneously 
observe the two transitions in separate spectral windows, using 
bands of 50 MHz and spectral resolution of 12.2070 kHz, 
corresponding  to ~0.154 \kms\ for both lines. The main
observational parameters are listed in Table~\ref{tab_mol}. 
The data were taken using in-band frequency switching with a throw 
of 7.5~MHz. The beam FWHM was approximately 32\asec .
The pointing was checked at hourly intervals on a nearby quasar, 
with corrections approximately 2\asec --3\asec . Flux calibration 
was performed on 3C123 and NGC7027. The absolute flux accuracy is ~10--20\%.
Data reduction and calibrations were done using the 
GBTIDL\footnote{GBTIDL is an interactive package for reduction and 
analysis of spectral line data taken with the GBT. 
See http://gbtidl.nrao.edu/} package, and subsequently converted to CLASS format. 

\begin{table}
\begin{center}
\caption[] {List of the observed sources. Col.~4 shows the velocity at which
we centred the spectra, corresponding to the systemic velocity. More
information (e.g.~source distances, bolometric luminosities of the associated star 
forming regions, reference papers) are given in Table~1 of paperI.}
\label{tab_sources}
\normalsize
\begin{tabular}{llll}
\hline \hline
source& RA(J2000) & Dec(J2000) & \Vlsr\ \\ %& $d$ & $L_{\rm bol}$ & Ref. \\
    & h m s & $o$ $\prime$  $\prime\prime$ & \kms\  \\ % & kpc & \soll\ & \\
\cline{1-4}
\multicolumn{4}{c}{HMSC} \\
\cline{1-4}
I00117--MM2   & 00:14:26.3     & +64:28:28 & $-36.3$ \\ %& 1.8  & $10^{3.1}$ & (1) \\
AFGL5142--EC \tablefootmark{w}  & 05:30:48.7	&  +33:47:53 & $-3.9$ \\ %& 1.8 & $10^{3.6}$ & (2) \\
05358--mm3    \tablefootmark{w}  & 05:39:12.5 & +35:45:55 & $-17.6$ \\ %& 1.8 & $10^{3.8}$ & (3,11) \\ 
G034--G2(MM2) & 18:56:50.0 & +01:23:08 &  $+43.6$ \\ %& 2.9 & $10^{1.6}$\tablefootmark{r} & (4)  \\
G034--F2(MM7) & 18:53:19.1  & +01:26:53 & $+57.7$ \\ % & 3.7 & $10^{1.9}$\tablefootmark{r}   & (4)  \\
G034--F1(MM8) & 18:53:16.5  & +01:26:10 & $+57.7$ \\ % & 3.7 & -- & (4) \\
G028--C1(MM9) & 18:42:46.9  & $-$04:04:08 & $+78.3$ \\ %  & 5.0  & -- & (4) \\
G028--C3(MM11)\tablefootmark{a}   & 18:42:44  & $-$04:01:54 & $+78.3$ \\
I20293--WC    & 20:31:10.7  &   +40:03:28 & $+6.3$ \\ %& 2.0 & $10^{3.6}$ & (5,6) \\
I22134--G     \tablefootmark{w}  & 22:15:10.5  &	+58:48:59 & $-18.3$ \\ % & 2.6 & $10^{4.1}$ & (7) \\
I22134--B     & 22:15:05.8 &  +58:48:59 & $-18.3$ \\ %& 2.6  & $10^{4.1}$ & (7) \\
% 22134+5834-3 (HMSC) & 22:15:06.77   &  +58:48:49.30 & \\
\cline{1-4}
\multicolumn{4}{c}{HMPO}   \\
\cline{1-4}
I00117--MM1   & 00:14:26.1      & +64:28:44 & $-36.3$  \\ %& 1.8 & $10^{3.1}$ & (1) \\
I04579--VLA1  & 05:01:39.9 &  +47:07:21 & $-17.0$ \\ %  & 2.5 & $10^{3.6}$ &  (8) \\
AFGL5142--MM  & 05:30:48.0     & +33:47:54 &  $-3.9$ \\ %& 1.8  & $10^{3.6}$ & (2) \\
05358--mm1    & 05:39:13.1 & +35:45:51 & $-17.6$ \\ %& 1.8  & $10^{3.8}$ & (3) \\
18089--1732   & 18:11:51.4 & $-$17:31:28 & $+32.7$ \\ % & 3.6 & $10^{4.5}$ & (9) \\
18517+0437    & 18:54:14.2 & +04:41:41 & $+43.7$ \\ %  & 2.9 & $10^{4.1}$ & (10) \\
G75--core     & 20:21:44.0 &    +37:26:38 & $+0.2$  \\ %& 3.8 & $10^{4.8}$ & (11,12) \\
I20293--MM1   & 20:31:12.8 &	+40:03:23 & $+6.3$ \\ %& 2.0 & $10^{3.6}$ & (5) \\
I21307        & 21:32:30.6  &    +51:02:16  & $-46.7$ \\ %& 3.2 & $10^{3.6}$ & (13) \\ 
I23385        & 23:40:54.5 &      +61:10:28 & $-50.5$  \\ % & 4.9 & $10^{4.2}$ & (14) \\
\cline{1-4}
\multicolumn{4}{c}{UC \HII }   \\
\cline{1-4}
%G75.78+0.74 (HC HII) & 20:21:44.01 &	+37:26:37.6 & \\
%05137(UC)\tablefootmark{c}  &  05:17:13.3 &  +39:22:14 & $-25.4$  & \\
G5.89--0.39   & 18:00:30.5    &    $-$24:04:01 & $+9.0$ \\ %& 1.28 & $10^{5.1}$ & (15,16) \\
I19035--VLA1  & 19:06:01.5 &   +06:46:35 & $+32.4$ \\ % & 2.2 & $10^{3.9}$ & (11) \\
19410+2336    & 19:43:11.4 &    +23:44:06 & $+22.4$ \\ % & 2.1 & $10^{4.0}$ & (17) \\
ON1           & 20:10:09.1  &	 +31:31:36 & $+12.0$  \\ %& 2.5 & $10^{4.3}$ & (18,19) \\
I22134--VLA1  & 22:15:09.2 &    +58:49:08 & $-18.3$ \\ %& 2.6 & $10^{4.1}$ & (11) \\
23033+5951    & 23:05:24.6 & +60:08:09 & $-53.0$  \\ %& 3.5 & $10^{4.0}$ & (17) \\
NGC7538-IRS9  & 23:14:01.8   &   +61:27:20 & $-57.0$ \\ % & 2.8 & $10^{4.6}$ & (8) \\
\hline
\end{tabular}
\end{center}
\tablefoot{
\tablefoottext{a}{Source not included in paperI,
selected from Butler \& Tan~(\citeyear{bet}). See also Butler et al.~(\citeyear{butler14});}
%\tablefoottext{b}{Observed in \H\ (1--0), \H\ (3--2), and \D\ (2--1);}
%\tablefoottext{c}{Observed in \H\ (1--0) and \D\ (2--1);}
\tablefoottext{w}{"warm" ($T \geq 20$~K) HMSCs externally heated (see paperI).}
%\tablefoottext{r}{Luminosity of the core and not of the whole associated star-forming region (Rathborne et al.~\citeyear{rathborne});}
%%\tablefoottext{n}{non-perturbed;} 
%\tablefoottext{1}{Palau et al.~(\citeyear{palau10})}
%\tablefoottext{2}{Busquet et al.~(\citeyear{busquet11})}
%\tablefoottext{3}{Beuther et al.~(\citeyear{beuther07b})}
%\tablefoottext{4}{Butler \& Tan~(\citeyear{bet})}
%\tablefoottext{5}{Palau et al.~(\citeyear{palau07})}
%\tablefoottext{6}{Busquet et al.~(\citeyear{busquet})}
%\tablefoottext{7}{Busquet~(\citeyear{busquetphd})}
%\tablefoottext{8}{S\'anchez-Monge et al.~(\citeyear{sanchez})}
%\tablefoottext{9}{Beuther et al.~(\citeyear{beuther04})}
%\tablefoottext{10}{Schnee \& Carpenter~(\citeyear{schnee})}
%\tablefoottext{11}{S\'anchez-Monge~(\citeyear{sanchez11})}
%\tablefoottext{12}{Ando et al.~(\citeyear{ando})}
%\tablefoottext{13}{Fontani et al.~(\citeyear{fonta04a})}
%%\tablefoottext{14}{S\'anchez-Monge et al.~(\citeyear{sanchez10})}
%\tablefoottext{14}{Fontani et al.~(\citeyear{fonta04b})}
%\tablefoottext{15}{Hunter et al.~(\citeyear{hunter})}
%\tablefoottext{16}{Motogi et al.~(\citeyear{motogi})}
%\tablefoottext{17}{Beuther et al.~(\citeyear{beuther02})}
%\tablefoottext{18}{Su et al.~(\citeyear{su})}
%\tablefoottext{19}{Nagayama et al.~(\citeyear{nagayama})}
}
\end{table}

\normalsize
\begin{table*}
\begin{center}
\caption[] {Observed transitions and technical parameters}
\label{tab_mol}
\begin{tabular}{llcccc}
\hline \hline
molecular line & line rest frequency & HPBW  & $\Delta v$ & $T_{\rm sys}$ & $\eta_{\rm MB}$ \\
  & (GHz) & (\asec ) & (\kms ) & K &  \\
  \cline{1-6}
\multicolumn{6}{c}{IRAM-30m Telescope}   \\
\cline{1-6}
{\it ortho-}NH$_2$D($1_{1,1}-1_{0,1}$) & 85.9263 & $\sim 28$ & 0.136 & $\sim 85 - 120$ &  0.85 \\
{\it para-}NH$_2$D($1_{1,1}-1_{0,1}$)  & 110.1536 & $\sim 22$ & 0.106 & $\sim 95 -125$ & 0.83 \\
\METH (3mm-band)  & 89.11 -- 96.89\tablefootmark{a}  & 27\tablefootmark{b} & $\sim 0.62$\tablefootmark{c} & $\sim 100-120$ & 0.84 \\
\METH (1mm-band)  & 216.0 -- 223.78\tablefootmark{a}   & 11\tablefootmark{b} & $\sim 0.26$\tablefootmark{c} & $\sim 200-300$  &  0.66 \\
  \cline{1-6}
\multicolumn{6}{c}{Green Bank Telescope}   \\
\cline{1-6}
NH$_3$(1,1) & 23.6945  & $\sim 32$ & $\sim 0.15$ & $\sim 50-100$ & $\sim 0.81$ \\
NH$_3$(2,2) & 23.7226 & $\sim 32$ & $\sim 0.15$ & $\sim 50-100$ & $\sim 0.81$ \\
\hline
\end{tabular}
\tablefoot{
\tablefoottext{a}{Total spectral window covered by the FTS correlator. Please see Tables~\ref{tab_fit3mm}
and \ref{tab_fit1mm} to see the transitions detected in it.}
\tablefoottext{b}{Telescope HPBW at the central frequency of the spectral window.}
\tablefoottext{c}{Maximum spectral resolution obtained with FTS.}
}
\end{center}
\end{table*}

\section{Results and derivation of physical parameters}
\label{res}

\subsection{\AMM\ and \DAMM\ }
\label{res_amm}

\subsubsection{Detection summary and parameters derived directly from the fits}
\label{res_line_par}

{\it \bf \AMM}: the \AMM (1,1) and (2,2) inversion lines have been detected with excellent
signal-to-noise ratio in all sources observed. 
The spectra of all HMSCs, HMPOs and UC \HII s are shown in Figs.~\ref{spectra_nh3_HMSC}, 
\ref{spectra_nh3_HMPO} and \ref{spectra_nh3_UCHII}, respectively, of Appendix-A.
Both transitions consist of 18 hyperfine components, grouped in 5 lines: the main 
one at the center of the spectrum, and four satellites symmetrically placed in
frequency with respect to the main one (see Ho \& Townes~\citeyear{ho83} for details). 
The spectra have been fit considering this hyperfine structure when the 
satellites are detected. When they are not, we adopted a simplified approach in which
we fitted the main line with a Gaussian curve. This simplified method was used for
8 of the (2,2) spectra observed, in which the satellites have not been detected.
The fit procedure has given good results (with
very low residuals, see Figs.~\ref{spectra_nh3_HMSC}, 
\ref{spectra_nh3_HMPO} and \ref{spectra_nh3_UCHII}) using both methods.
The simplified approach in principle tends to overestimate the 
intrinsic line width, as the main line is in reality a blending of several 
hyperfine components. To quantify this, we have taken a (2,2) spectrum 
with hyperfine structure nicely fit (one spectrum per evolutionary group),
applied the simplified method and compared the derived line width with 
that obtained from the accurate method (the hyperfine fit method). From this
comparison, we quantify an overestimate of at most the 10\%
of the true intrinsic line width.
Nevertheless, because the column density in this approach is computed
from the integral of the line (see Sects.~\ref{sect_amm} and \ref{sect_damm}), 
this overestimation does not influence the calculation of neither the column density 
nor the deuterated fraction.

The line parameters derived from these fit procedures are listed in 
Tables~\ref{line_par_amm11} and \ref{line_par_amm22}. The accurate method has given a 
well-constrained value of the optical depth of the main component of the (1,1) line 
($\tau_{\rm m}(1,1)$/$\Delta \tau_{\rm m} (1,1)\geq 3$) 
for all objects except for I04579--VLA1, for which the line is optically thin. 
The average $\tau_{\rm m}(1,1)$ is $\sim 1$, with no significant differences
between the three evolutionary groups, while $\tau_{\rm m}(2,2)$ is usually
smaller than 1. The average line widths of the (1,1) lines are $\sim 1.7$, $\sim 2.3$ and 
$\sim 2.6$ \kms\ for the HMSC, HMPO and UC \HII\ groups (standard deviation
0.5, 0.6 and 0.7 \kms , respectively), and tend to increase with evolution, as expected
(S\'anchez-Monge et al.~\citeyear{sanchez13}).

%\begin{sidewaystable*}
\begin{table*}
\begin{center}
\caption[]{Derived line parameters of \AMM\ (1,1). All lines have been fit taking 
into account the hyperfine structure as explained in Sect.~\ref{res_line_par}. 
Cols.~2--5 report the output parameters of the fitting procedure 
($A \times \tau_{m}$ = $f[J_{\nu}(T_{\rm ex})-J_{\nu}(T_{\rm BG})]$, where
$f$ is the filling factor, assumed to be unity, $J_{\nu}(T_{\rm ex})$ and $J_{\nu}(T_{\rm BG})$
are the equivalent Rayleigh-Jeans excitation and background temperatures, respectively,
and $\tau_{m}$ is the opacity of the main group of hyperfine components;
$V_{\rm peak}$ = peak velocity; $\Delta v$ = full width at half maximum
corrected for hyperfine splitting; $\tau_{m}$ = opacity of the main group of hyperfine components)
for the (1,1) line, and Col.~6 lists the excitation temperature of the transition
derived as explained in Sect.~\ref{sect_amm}. 
%For both sources marked with 'G' or 'T', \Tex$_{2,2}$ cannot be estimated.
The uncertainties obtained from either the fitting procedure (parameters
in Cols.~2 -- 5) or from the propagation of errors (Col.~6) are in 
parentheses.}
\normalsize
\label{line_par_amm11}
\begin{tabular}{cccccc}
\hline \hline
source & $A \times \tau_{m}(1,1)$ & $V_{\rm peak}(1,1)$     & $\Delta v(1,1)$ & $\tau_{m}(1,1)$ & \Tex$_{1,1}$   \\
  &    & (\kms ) &  (\kms )                        & & (K)   \\
\hline 
\cline{1-6}
\multicolumn{6}{c}{HMSCs}   \\
\cline{1-6}
I00117--MM2 & 1.83(0.06) &  --36.16(0.01)                 & 1.71(0.03)  & 0.71(0.09) & 15(4) \\
AFGL5142-EC & 3.69(0.03) &  --2.936(0.004)    & 2.44(0.01)   & 0.77(0.02)         & 7.4(0.2) \\
05358--mm3  & 5.32(0.01) &  --16.258(0.004)   & 1.989(0.005) & 0.85(0.01)       & 8.86(0.07) \\
G034--G2    & 3.68(0.04) &  41.854(0.007)      & 2.25(0.01)   & 1.52(0.04)        & 5.0(0.1) \\
G028--C1    & 2.69(0.01) &  79.810(0.007)      & 2.30(0.01)   & 2.50(0.07)      & 3.67(0.05) \\
G028--C3       &           2.89(0.08) &  80.858(0.007) & 1.15(0.02) & 1.9(0.1)       & 4.1(0.2) \\
I20293--WC  & 5.46(0.02) &  6.419(0.004)    & 2.080(0.006)   & 1.31(0.01)       & 6.76(0.07) \\
I22134--G   & 2.41(0.07) &  --18.643(0.006)& 1.33(0.02) & 0.40(0.07) & 9(2)               \\   
I22134--B   & 1.72(0.08) &  --18.800(0.01) & 1.15(0.03)   & 0.6(0.1)    & 5.4(0.9)        \\   
\cline{1-6}
\multicolumn{6}{c}{HMPOs}   \\
\cline{1-6}
I00117--MM1  &  1.59(0.03)  & --36.32(0.01)   & 1.59(0.04)    &  0.13(0.03) & 5.2(0.6)        \\      
I04579--VLA1 &  0.272(0.01)  & --16.73(0.03)  & 1.73(0.07)  &  0.1\tablefootmark{e}  & --\tablefootmark{f}      \\      
AFGL5142--MM & 3.524(0.001) & --3.072(0.002)  & 2.644(0.007) &  0.75(0.01)   & 7.28(0.01)     \\      
05358--mm1   & 4.636(0.003) & --16.318(0.003)  & 2.064(0.001) &  0.80(0.01)     & 8.39(0.02)   \\     
18089--1732\tablefootmark{c}  & 8.301(0.006) & 33.02(0.01)   & 3.241(0.004) & 2.53(0.01) & 5.9(0.1) \\
18517+0437   & 1.76(0.03)  &  43.908(0.009)    & 2.52(0.03)  &  0.43(0.04)               & 6.7(0.7) \\
G75--core    &  2.99(0.04)  &  0.067(0.009)    & 3.42(0.02)  &  0.50(0.03)               & 8.6(0.6) \\
I20293--MM1  &  8.40(0.04)  &  6.058(0.003)           & 1.739(0.004) &  1.15(0.02)       & 9.9(0.2) \\
I21307 & 0.61(0.05)  &  --46.57(0.04)        & 1.9(0.1)     &  0.8(0.2)                 & 3.4(0.5)  \\
I23385\tablefootmark{d}  & 0.84(0.03) & --50.21(0.03) &  2.09(0.08)  &  0.15(0.05) & 8(5)      \\      
\cline{1-6}
\multicolumn{6}{c}{UC \HII s}   \\
\cline{1-6}
G5.89--0.39    & 5.63(0.02)  &  8.70(0.01)                 & 3.745(0.002) &  0.65(0.01) & 11.2(0.1) \\
I19035--VLA1   & 1.90(0.03)  &  32.56(0.01)                & 3.64(0.03)  &   1.08(0.05) & 4.4(0.2)  \\
19410+2336    &  12.054(0.005) &  22.458(0.001)         & 1.389(0.001) &  1.05(0.01)  & 14.12(0.02) \\
ON1           &  13.25(0.02)   &  10.985(0.001)    & 2.886(0.005) &  1.58(0.01)      & 10.98(0.02)  \\
23033+5951 & 4.96(0.07)  & --53.444(0.006)              & 1.95(0.02)  &  0.98(0.04)   & 7.7(0.4)    \\
NGC7538--IRS9 & 3.94(0.04)  & --57.31(0.01)                & 2.17(0.03)  &  1.00(0.01) & 6.54(0.08) \\
\hline 
\end{tabular}
\end{center}
\tablefoot{
\tablefoottext{a}{Integrated area of the main group of hyperfine components, in K \kms ;}
\tablefoottext{b}{Peak intensity of the main group of hyperfine components, in K;}
%\tablefoottext{bt}{After the $-$ sign, we give the
%peak intensity (in K) of the main group of hyperfine components;}
\tablefoottext{c}{The spectrum shows two velocity components (Fig.~\ref{spectra_nh3_HMPO}). 
Only the fit to the stronger component is shown;}
\tablefoottext{d}{The spectrum shows two velocity components (Fig.~\ref{spectra_nh3_HMPO}).
Fontani et al.~(\citeyear{fonta04b}) also found these two components
in \CII\ and attributed the one centred at $\sim -50$\kms\ to the HMPO. 
Only the fit to this component is shown.}
\tablefoottext{e}{derived from the hyperfine fit procedure;} 
\tablefoottext{f}{an average value of 6.5~K, computed from the HMPOs with well-constrained opacity, is assumed.}
}
\end{table*}
%\end{sidewaystable*}

\begin{table*}
\begin{center}
\caption[]{Same as Table~\ref{line_par_amm11} for the \AMM\ (2,2) transitions.
For the sources with 'HFS' in Col.~2, 
the line hyperfine structure has been fit and the 
same output parameters in Cols. 2--6 of Table ~\ref{line_par_amm11}
are given in Cols.~3, 5, 6, 7, 9. 
For the sources with 'G' in Col.~2, the satellites of the (2,2)
line are undetected, so that the main group of hyperfine components has been fit 
with a single Gaussian. For these objects, we give
integrated area (in K \kms , Col.~4) and peak intensity (in K, Col.~8) of this Gaussian, respectively.
For the sources marked with a 'T' in Col.~2, we clearly detect the satellites in
the (2,2) transition, and performed a good fit to the hyperfine structure,
but the line is optically thin. Hence, in Col.~8
we also give the peak temperature of the main group of hyperfine
components, which is the parameter used to derive the \AMM\ total column density
in this case (see Sect.~\ref{sect_amm}).
The uncertainties of all parameters are in 
parentheses.}
\normalsize
\label{line_par_amm22}
\begin{tabular}{ccccccccc}
\hline \hline
  & & $A \times \tau_{m}(2,2)$ & $\int T_{{\rm MB}}{\rm d}v$\tablefootmark{a} & $V_{\rm peak}(2,2)$     & $\Delta v(2,2)$ & $\tau_{m}(2,2)$ & $T_{{\rm peak}2,2}$\tablefootmark{b} & \Tex$_{2,2}$  \\
      &  & & (K \kms )   & (\kms ) &  (\kms )                        & & (K) & (K) \\
\hline 
\cline{1-9}
\multicolumn{9}{c}{HMSCs}   \\
\cline{1-9}
I00117--MM2 & G  & & 0.91(0.04) &  --36.36(0.04) &  1.82(0.09) & & 0.47(0.04) & --\tablefootmark{e} \\
AFGL5142-EC & HFS  & 1.73(0.06) & &  --2.972(0.007) & 2.73(0.03) & 0.21(0.07) & & 11(4) \\		  
05358--mm3  & T & 2.31(0.01) & &  --16.283(0.005) & 2.23(0.01) & 0.1 & 2.15(0.05) & --\tablefootmark{e}  \\
G034--G2    & T & 0.67(0.02) & & 41.56(0.02) & 2.16(0.06) &   0.1 & 0.71(0.03) & --\tablefootmark{e}  \\  
G028--C1    & HFS & 0.57(0.07) & & 79.72(0.03) & 2.22(0.09) & 1.3(0.3) & & 3.0(0.3) \\
G028--C3       & G & & 0.40(0.03) & 80.87(0.04) & 1.4(0.1) & & 0.27(0.02) & --\tablefootmark{e}  \\	   
I20293--WC  & HFS & 1.8(0.1) & & 6.34(0.02) & 2.29(0.05) & 0.6(0.2) & & 5(1) \\
I22134--G   & G & & 1.30(0.04) & --18.81(0.02) &  1.50(0.05) &  & 0.81(0.03) & --\tablefootmark{e}  \\    
I22134--B   & G & & 0.55(0.04) & --18.93(0.05) &  1.5(0.1)  & & 0.35(0.02) & --\tablefootmark{e}  \\	  
\cline{1-9}
\multicolumn{9}{c}{HMPOs}   \\
\cline{1-9}
I00117--MM1   & G & & 0.97(0.04) & --36.48(0.03) & 1.67(0.08) & & 0.54(0.04) & --\tablefootmark{e}  \\
I04579--VLA1  & G & & 0.25(0.02) & --16.85(0.08) & 1.9(0.2) & & 0.12(0.05) & --\tablefootmark{e}  \\
AFGL5142--MM  & HFS & 1.80(0.06) & & --3.083(0.003) & 2.79(0.03) & 0.34(0.07) & & 8(1) \\
05358--mm1    & T & 2.078(0.008) & & --16.334(0.001) & 2.27(0.01) &0.1 & 1.96(0.05) & --\tablefootmark{e}  \\
18089--1732\tablefootmark{c}    & HFS & 5.58(0.03) & & 32.85(0.01) & 3.14(0.02) & 2.48(0.02) & & 4.9(0.1) \\
18517+0437    & HFS & 0.94(0.07) & & 43.77(0.02) & 2.58(0.07) &  0.24(0.08) & & 7(3) \\
G75--core     & T & 1.91(0.02) & & --0.12(0.01) & 3.76(0.03) &  0.1 & 1.88(0.05) & --\tablefootmark{e}  \\
I20293--MM1  & HFS & 2.6(0.1) & & 5.901(0.008) &  2.01(0.04) &  0.2(0.07) & & 20(5) \\
I21307     & G & & 0.57(0.04) & --46.71(0.08) &  2.35(0.21) & & 0.23(0.02) & --\tablefootmark{e}  \\
I23385\tablefootmark{d}  & G & & 0.82(0.09) & --50.5(0.1) & 1.9(0.2) & & 0.42(0.03) & --\tablefootmark{e}  \\
\cline{1-9}
\multicolumn{9}{c}{UC \HII s}   \\
\cline{1-9}
G5.89--0.39  & T & 3.62(0.01) & & 8.772(0.004) & 4.519(0.003) & 0.1& 3.65(0.05) & --\tablefootmark{e}  \\
I19035--VLA1 & HFS & 0.79(0.07) & & 32.53(0.03) & 3.9(0.1) &  0.7(0.2) & & 3.7(0.7) \\
19410+2336   & T & 4.22(0.02) & & 22.306(0.002) & 1.65(0.01) & 0.1& 3.82(0.05) & --\tablefootmark{e}  \\
ON1          & HFS & 6.1(0.1) & & 10.941(0.007) & 3.09(0.02) & 0.82(0.05) & & 10.2(0.8) \\
23033+5951  & T & 1.31(0.02) & & --53.69(0.02) & 2.36(0.03) & 0.1 & 1.22(0.02) & --\tablefootmark{e}  \\
NGC7538--IRS9 & T & 1.57(0.02) & & --57.47(0.02) & 2.53(0.04) & 0.1 & 1.62(0.03) & --\tablefootmark{e}  \\
\hline
\end{tabular}
\end{center}
\tablefoot{
\tablefoottext{a}{Integrated area of the main group of hyperfine components, in K \kms ;}
\tablefoottext{b}{Peak intensity of the main group of hyperfine components, in K;}
%\tablefoottext{bt}{After the $-$ sign, we give the
%peak intensity (in K) of the main group of hyperfine components;}
\tablefoottext{c}{The spectrum shows two velocity components (Fig.~\ref{spectra_nh3_HMPO}). 
Only the fit to the stronger component is shown;}
\tablefoottext{d}{The spectrum shows two velocity components (Fig.~\ref{spectra_nh3_HMPO}).
Fontani et al.~(\citeyear{fonta04b}) also found these two components
in \CII\ and attributed the one centred at $\sim -50$\kms\ to the HMPO. 
Only the fit to this component is shown.}
%\tablefoottext{e}{an average value of 6.5~K, computed from the HMPOs with well-constrained opacity, is assumed;} 
\tablefoottext{e}{\Tex$_{2,2}$ cannot be estimated. For these objects, in the calculations
described in Sect.~\ref{sect_amm} we have assumed \Tex$_{2,2}$ = \Tex$_{1,1}$.}
}
\end{table*}

\vspace{0.2cm}
\normalsize
\noindent
{\it \bf \DAMM }: the \oDAMM ($1_{1,1}-1_{0,1}$) line has been detected in all sources
observed except in two HMPOs (I04579--VLA1 and I21307). The detection
rate is thus $\sim 92\%$. The \pDAMM ($1_{1,1}-1_{0,1}$) line
has been detected in 13 out of 26 sources observed (detection rate of 50\%).
The spectra of both lines are shown in Figs.~\ref{spectra_damm} and \ref{spectra_damm_para}.
In this work we will use the \oDAMM ($1_{1,1}-1_{0,1}$) line to 
derive the physical parameters of our interest because of its higher detection rate
and signal-to-noise ratio. The line of the {\it para-} species will be used
to test whether the {\it ortho-}/{\it para-} ratio assumed to derive the total column
density in Sect.~\ref{sect_damm} is correct. 
Like ammonia, the \DAMM\ lines have been fit taking into account their 
hyperfine structure driven by the quadrupole moment of the Deuterium and Nitrogen nuclei
(see Olberg et al.~\citeyear{olberg} for details). 

In general, the procedure
has provided good fits to the spectra, except a few cases in which
deviations from the LTE (symmetric) pattern are seen
(e.g.~G028--C1, I20293--WC, I20293--MM1, 23033+5951, see Fig.~\ref{spectra_damm}).
To check if (and how) our simplified LTE approach gives results different
from those of a non-LTE analysis, we have run the non-LTE radiative transfer 
code RADEX\footnote{http://www.sron.rug.nl/~vdtak/radex/} (Van der Tak 
et al.~\citeyear{vandertak07})
in order to reproduce the
measured line ratios of the two lines ({\it ortho-}NH$_2$D and {\it para-}NH$_2$D).
The molecular data were taken from the LAMDA database (Sch\"oier et al.~\citeyear{schoier}) 
using the collisional rate coefficients with H$_2$ of Daniel et al.~(\citeyear{daniel14}).
We built grids of models with kinetic temperatures in the range 8 -- 25 K, H$_2$ volume densities
in the range $10^{3}-10^{8}$ \cmc , and total column densities of $10^{12}-10^{15}$ \cmq .
We assumed line widths of 1.5 \kms\ and an {\it ortho}-to-{\it para-} ratio of 3. 
The 'best estimate' of the column densities
that we find are roughly consistent with the values measured from the LTE approach,
but since we only have one line ratio, we cannot discriminate properly 
between the different non-LTE models. 
%A further step would be to reproduce the line shapes using LIME 
%(Brinch \& Hogerheijde~\citeyear{beh10}), a 3D line radiative transfer code, which is part of
%ARTIST\footnote{http://youngstars.nbi.dk/artist/Welcome.html/}.
%However, the sofware requires an input physical model and a detailed 
%knowledge of the physical structure of the sources (e.g. temperature and density profiles), 
%which are unknown and would introduce only additional sources of uncertainties. 
Therefore, with only one line ratio, all we can say is that the column densities of NH$_2$D 
derived assuming LTE conditions are consistent with the values expected from 
a non-LTE approach.

The average optical depth of the main hyperfine component 
derived from this fitting procedure is $\sim 1$ in 
all three evolutionary groups. For the sources for which the mentioned fitting 
procedure did not give good results (because of poor signal-to-noise ratio), 
we have fit the lines with Gaussians. As for the \AMM\ lines, this simplified 
method could overestimate the line widths by at most $\sim 10$\%,
and we find yet an increasing trend of the line widths going from the
HMSC phase to the HMPO and UC \HII\ phases, for which mean values
(and standard deviations) are: 1.4(0.6), 2.5(1.3) and 2.4(1) \kms , respectively.
All line parameters are listed in Table~\ref{line_par_damm}. 

\begin{table*}
\begin{center}
\caption[]{Derived line parameters of \oDAMM ($1_{1,1}-1_{0,1}$) for
all sources observed in this line. The spectra of the sources marked with a 'HFS'
in Col.~2 have been fit taking into account the hyperfine structure as described in
Sect.~\ref{res_line_par}. For these, cols.~3, 5, 6 and 7 give: $A \times \tau_{m}$, peak velocity, 
line width and $\tau_{m}$, respectively (see Table~\ref{line_par_amm11}). 
%For the others (optically thin lines or not
%well-constrained opacity), indicated with an {\bf a}, 
%Cols.~3 and 6 represent total integrated area (in K \kms ) and peak intensity
%(in K), respectively. 
Col.~9 lists the excitation temperatures computed as explained in Sect.~\ref{sect_damm}.
The sources marked with a G in Col.~2 have optically thin lines or not
well-constrained opacity. These have been fit with a Gaussian
function, so that Cols.~4 and 8 represent total integrated area 
($\int T_{\rm MB}{\rm d}v$, in K \kms ) and peak intensity ($T_{\rm peak}$,
in K) of this Gaussian, respectively. 
The uncertainties obtained from either the fitting procedure (parameters
in Cols.~3 -- 8) or from the propagation of errors (Col.~9) are in parentheses.
}
\label{line_par_damm}
\begin{tabular}{lllllllll}
\hline \hline
source &  & $A \times \tau_{m}$ & $\int T_{\rm MB}{\rm d}v$ & $V_{\rm peak}$     & $\Delta v$ & $\tau_{m}$ & $T_{\rm peak}$  & \Tex  \\
   &   &  & (K \kms )   &    (\kms )   &   (\kms )    &    &    (K)   & (K)   \\
\cline{1-9}
\multicolumn{9}{c}{HMSC}   \\
I00117--MM2                  &       HFS           &  0.15(0.02)    &     &  --35.54(0.08)  &   2.1(0.2)      & 0.44(0.16) & & 7.1(0.1) \\
AFGL5142--EC               &      HFS            &  0.51(0.03)   &      &  --2.54(0.03)     &   1.92(0.07)  & 0.8(0.1)  &  & 7.9(0.1)   \\
05358--mm3                   &      HFS             &  0.32(0.04)   &     &  --16.04(0.05)  &   1.29(0.09) & 1.2(0.4)    &  & 6.9(0.06) \\
G034--G2(MM2)             &      HFS             &  0.22(0.01)   &    &   41.74(0.03)      &   1.01(0.07) &  0.83(0.03) &  & 6.94(0.01) \\
G034--F2(MM7)              &     HFS              &  0.19(0.02)  &      &   58.12(0.03)    &   1.30(0.07) &  0.6(0.2)  &   & 7.1(0.1)   \\
G034--F1(MM8) & G &  & 0.36(0.04)        &  56.3(0.03)            &   0.6(0.1)      & & 0.08(0.01) &  --\tablefootmark{b}               \\
G028--C1(MM9) & G & &  0.75(0.04)      &  80.20(0.08)            &   2.8(0.1)       &  & 0.13(0.01) &  --\tablefootmark{b}              \\
G028--C3(MM11)   & G & & 0.114(0.02)  &  81.07(0.07)         &   0.9(0.2)	     &  & 0.04(0.01) & --\tablefootmark{b}                      \\
I20293--WC    &            HFS                       &  1.25(0.03)  &     &    7.15(0.01)  &   1.28(0.02) &  2.06(0.09) &  & 7.84(0.04) \\
I22134--G    & G &  & 0.20(0.03)      &   --18.5(0.1)               &   1.5(0.2)	     &  & 0.05(0.01)  & --\tablefootmark{b}     \\
I22134--B    & G &  & 0.25(0.03)    &  --18.95(0.04)          &   0.86(0.09)    &  & 0.15(0.02) &  --\tablefootmark{b}                  \\
\cline{1-9}
\multicolumn{9}{c}{HMPO}   \\
I00117--MM1  & G &  & 0.39(0.03)  &  --35.94(0.14)                &    2.6(0.6)    & &  0.07(0.01) &  --\tablefootmark{b}                            \\
I04579--VLA1  & &  & $\leq 0.10$\tablefootmark{a}  &  --                                  &     --    & -- & --                                      \\
AFGL5142--MM     &      HFS                  &  0.51(0.03)   & &  --2.867(0.002)              &   2.25(0.07) & 0.9(0.1) & &  7.77(0.08)              \\
05358--mm1       &      HFS                 &  0.14(0.03)   &  & --16.07(0.07)           &   1.6(0.2)     &    0.4(0.1)  &  & 7.2(0.2)               \\
18089--1732         &        HFS                &  0.89(0.06)   &  &  34.44(0.04)          &  1.60(0.07) &    2.0(0.2)  &  & 7.40(0.05)              \\
18517+0437          &      HFS               &  0.12(0.02)  &  &  43.9(0.1)             &   2.2(0.3)    &  1.1(0.4)  &    &  6.52(0.03)              \\
G75--core        & G &   &     0.31(0.03)	 & --1.3(0.5)                  &    5.7(1.7)    & & 0.04(0.01) & --\tablefootmark{b}                 \\
I20293--MM1      &        HFS                  &  0.708(0.006)  & &   5.69(0.02)             &  1.63(0.02) &  0.68(0.03) & & 8.97(0.08)              \\
I21307                   &          &    &  $\leq 0.08$\tablefootmark{a}	  & --                &    --         &       &  --       & --       \\
I23385\tablefootmark{c}   & G &  & 0.15(0.02) & --49.4(0.3)        &  2.7(0.7)      &  & 0.03(0.01) & --\tablefootmark{b}                            \\
\cline{1-9}
\multicolumn{9}{c}{UC \HII }   \\
G5.89--0.39                    &     HFS             &  0.15(0.03)  &   &  7.9(0.2)        &   2.2(0.2)      &  1.8(0.5) &    & 6.45(0.02)    \\
I19035--VLA1    & G    &  &  0.64(0.03)	 &  32.7(0.2)        &   4.1(0.5)  &    &  0.09(0.01) & --\tablefootmark{b}                           \\
19410+2336           &       HFS                   &  0.59(0.02) &  &  22.72(0.01)   &   1.51(0.02)  &  0.53(0.07) & & 9.2(0.2)               \\
ON1                          &        HFS            &  0.19(0.02)  & &  11.01(0.06)       &    3.2(0.2)     &  0.50(0.15) &  & 7.3(0.1)      \\
I22134--VLA1   & G   &  & 0.12(0.02)   & --18.86(0.07)    &   1.1(0.2)      & & 0.05(0.02) & --\tablefootmark{b}                              \\
23033+5951         &      HFS                      &  0.30(0.02) &  & --53.28(0.03)     &   1.36(0.06) & 0.6(0.2) &   & 7.5(0.1)              \\
NGC7538-IRS9\tablefootmark{c}  & G &  & 0.16(0.03) & --56.9(0.4)      &   3(1)	   & & 0.03(0.01) & --\tablefootmark{b}                       \\
\hline
\end{tabular}
\end{center}
\tablefoot{
\tablefoottext{a}{Upper limit to the integrated line intensity from the equation $\int T_{\rm MB}{\rm d}v = \frac{\Delta v}{2\sqrt{ln2/\pi }}T_{\rm MB}^{peak}$,
assuming the 3$\sigma$ rms level of the spectrum as $T_{\rm MB}^{peak}$, and an average value of $\Delta v$ from
the other sources;}
\tablefoottext{b}{\Tex\ cannot be derived from the fit results, therefore the average value of 
the sources with well-constrained opacity (7.5~K) is assumed;}
\tablefoottext{c}{marginal detection.}
}
\end{table*}
\normalsize

\subsubsection{\AMM\ rotation temperature and total column density}
\label{sect_amm}

From the \AMM (1,1) and (2,2) line parameters, we have obtained rotation 
temperatures, $T_{\rm rot}$, adopting two methods: for the nine sources 
having $\tau_{\rm (2,2)}$/$\Delta \tau_{\rm (2,2)}\geq 3$
and $\tau_{\rm (2,2)} > 0.1$, we have derived first the excitation 
temperature of the (1,1) and (2,2) lines (\Tex$_{1,1}$ and \Tex$_{2,2}$,
respectively) independently using Eq.~A.2 of 
Busquet et al.~(\citeyear{busquet09}), and, from these, the column 
densities of the two levels, $N_{(2,2)}$ and $N_{(1,1)}$, from the 
relations given in Anglada et al.~(\citeyear{anglada}). Note that,
although Eq. A.2 in Busquet et al.~(\citeyear{busquet09}) is derived
for the (1,1) line, it is valid also for the (2,2) line given the 
small difference in frequency between the two transitions. Then, 
the rotation temperature has been derived from the relation:
\begin{equation}
T_{\rm rot} = \frac{-41.5}{{\rm ln}[(3/5)(N_{(2,2)}/N_{(1,1)})]}\;.
\end{equation}
For sources with an optically thin (2,2) line, or with $\tau_{\rm (2,2)}$ 
not determined because the satellites are undetected, \Tex$_{2,2}$
is assumed to be equal to \Tex$_{1,1}$. This hypothesis is justified
by the good agreement between the two excitation temperatures in the
sources in which they can both be measured (see Sect.~\ref{sect_tex}). 
Under this assumption, we have applied eq.~A.4 
in Busquet et al.~(\citeyear{busquet09}), which utilises the peak intensity of
the main hyperfine component of the (2,2) line. 

In both methods, the total \AMM\ column density, $N$(\AMM ), has been
calculated from Eq.~A.6 in Busquet et al.~(\citeyear{busquet09}).
Both \Trot\ and $N$(\AMM ) are listed in Table~\ref{tab_coldens_amm}.
Rotation temperatures range from 11.7 to 29 K, and on average they are 
$\sim 17$, $\sim 22$ and $\sim 22$ K for HMSCs, HMPOs 
and UC \HII s, respectively (standard deviations are 2.6, 3.5 and 4~K, 
respectively). Separately, quiescent and "warm" HMSCs have
mean temperatures of 16 and 20~K (standard deviations of
2.6 and 1.2~K, respectively), which confirms the higher gas
temperature in the "warm" cores. Total \AMM\ column densities range 
from 5.6$\times 10^{13}$ to 3.6$\times 10^{15}$\cmq , and the
average values are 9.4$\times 10^{14}$,
9.3$\times 10^{14}$ and 1.6$\times 10^{15}$\cmq\ in the 
HMSC, HMPO and UC \HII\ groups, respectively.
We have assumed a unity filling factor because available
VLA interferometer ammonia maps of some of the targets 
show that the ammonia emission is extended and fills most 
of the GBT beam. Nevertheless, we stress that the emission
from the target cores is clearly dominant with respect to that of
nearby condensations (see Sanchez-Monge et al.~\citeyear{sanchez13}). 

\subsubsection{\DAMM\ total column density}
\label{sect_damm}

The \DAMM\ column densities have been computed from the
line parameters of the \oDAMM\ line following Eq. (1) in 
Busquet et al.~(\citeyear{busquet10}), which assumes the same \Tex\
for all the hyperfine components.  
\Tex\ was computed as described in Sect.~\ref{sect_amm}
for sources with opacity of the main component well-constrained.
For the others, we have assumed \Tex = 7.5~K, which is the average value 
derived from the sources with well-constrained opacity, and obtained the 
column density from Eq.~(A4) of Caselli et al.~(\citeyear{casellib}), 
valid for optically thin lines. 

Again, we have assumed a unity filling factor
because there are few high angular resolution observations 
of this line towards the targets from which the emitting region of \DAMM\
can be determined. This assumption is critical, as the 
\oDAMM\ line has a critical density of $\sim 10^6$\cmc , higher
than that of the inversion transitions of \AMM\ ($\sim 10^{3-4}$\cmc ).
However, while by neglecting the beam dilution the absolute values 
of the column densities can be certainly affected, the evolutionary trend
of the column density ratio should not be affected by this assumption
because the beam dilution is expected to be almost constant, and
thus it should introduce only a systematic correction (see also
paperI). Also,  observations at high angular resolution towards
massive star forming regions (Busquet et al.~\citeyear{busquet10},
Pillai et al.~\citeyear{pillai11}) indicate that the emission of \DAMM\ 
can be as extended as that of \AMM , despite the different
critical density. For example, the emitting region of \DAMM ($1_{1,1}-1_{0,1}$) 
and \AMM\ (1,1) in I20293--WC and I20293--MM1, both included
in our survey, is approximately the same (Busquet et al.~\citeyear{busquet10}).
$N$(\DAMM ) is listed in Table~\ref{tab_coldens_amm}.

\subsection{Methanol and deuterated methanol lines}
\label{res_meth}

In this work we focus the attention on the deuterated fraction of \METH , and on the 
physical quantities relevant to derive it (i.e.~temperature and total column density). 
Therefore, in what follows we will present the approach adopted to identify the 
lines from which \Dfrac (\METH ) will be derived (Sect.~\ref{sect_meth_ide}), 
the method to compute rotation temperature and total column density from the 
line parameters (Sect.~\ref{sect_meth_par}), and the deuterated fraction in the 
sources detected in \DMETH\ (Sect.~\ref{sect_dfrac_2}).
%We postpone a complete presentation and deeper analysis of these data
%to future works.

\subsubsection{Lines detected and fit procedure}
\label{sect_meth_ide}

Multiple \METH\ lines are detected in the observed spectral 
windows (Col.~1 of Table~\ref{tab_mol}) towards all sources, while 
\METHI\ lines are clearly detected in four HMSCs, seven HMPOs and
six UC \HII\ regions. \DMETH\ lines are detected only towards 6 objects:
three HMSCs and three HMPOs, and two out of the three HMSCs are
"warm" cores (see Sect.~\ref{obs}). Moreover, in two HMPOs
(AFGL5142--MM and 18089--1732), the \METHD\ ($5_{1,5}-4_{1,4}$A++) 
line at 1.3~mm has been detected, although in 18089--1732 this
could be blended with emission of (CH$_2$OH)$_2$. Tables~\ref{tab_fit3mm} and 
\ref{tab_fit1mm} give the line parameters obtained from Gaussian fits 
to the lines detected at the 3$\sigma$ level and not probably blended 
with other transitions. 

\begin{figure}
 \begin{center}
{\includegraphics[angle=0, width=9cm]{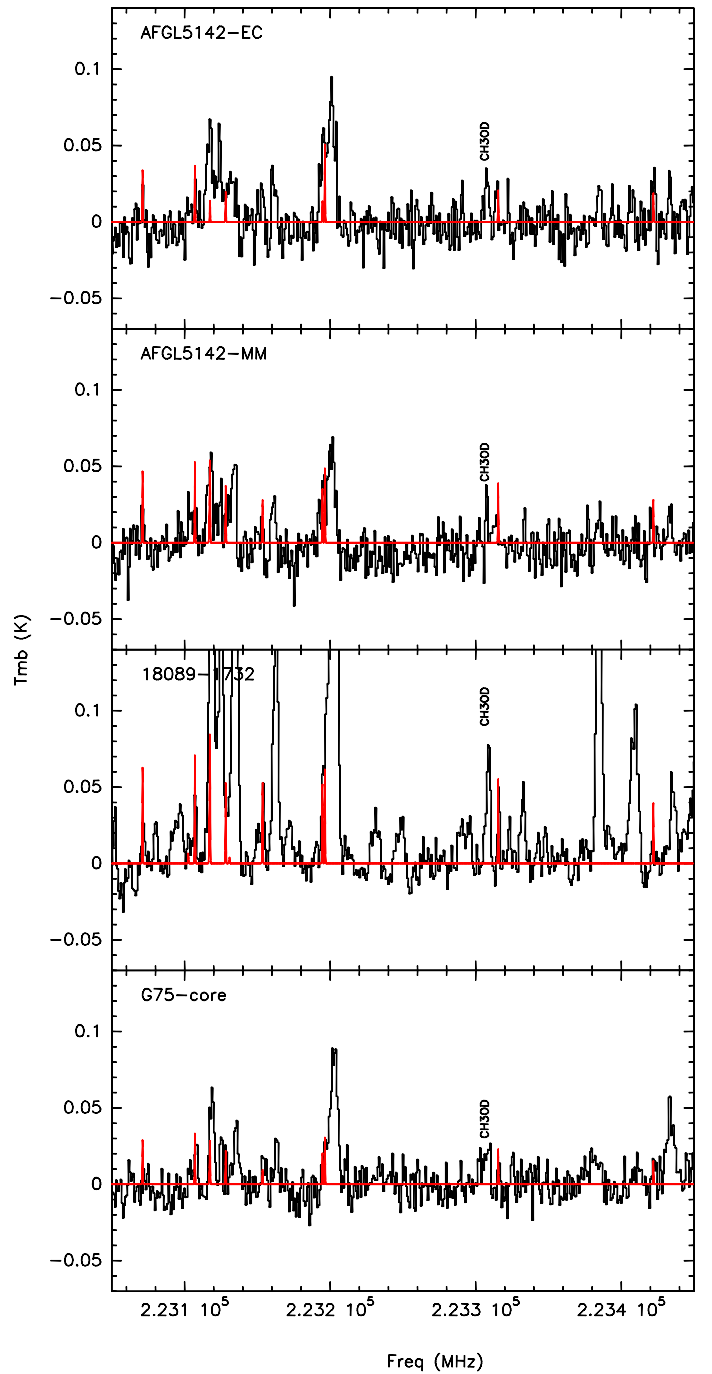}}
 \caption[] {\label{example_spec_ide} Example of spectra observed 
 at 1~mm with the \DMETH\ synthetic spectra (red line) used for the identification
 of the deuterated methanol lines superimposed on them.}
 \end{center}
\end{figure}

The detection of the deuterated lines has been double-checked by comparing 
observed and synthetic spectra. For this purpose, the observed spectra were 
smoothed to 1.0 \kms\ at 1 mm and to 2.5 \kms\ at 3 mm, to improve the 
signal-to-noise ratio. The \DMETH\ synthetic spectra were computed assuming 
LTE and optically thin emission as in Palau et al. (2011), and using the molecular 
data from the Jet Propulsion Laboratory (Pickett et al. 1998). To build the synthetic 
spectra, we adopted a line width of 1.5 \kms\ at 1~mm and 2.5 \kms\ at 3~mm, and 
used the rotational temperature listed in Table~\ref{tab_coldens_meth} derived from \METH . 
Examples of the synthetic spectra of CH2DOH overplotted on the observed spectra 
are shown in Fig.~\ref{example_spec_ide} (red line indicates the synthetic spectra of \DMETH ). 
The figure shows that several transitions are (marginally) detected at 1~mm 
in each of the four cases shown.

\subsubsection{Derivation of molecular column densities and rotation temperatures}
\label{sect_meth_par}

\begin{figure}
\begin{center}
\includegraphics[angle=0, width=7.2cm]{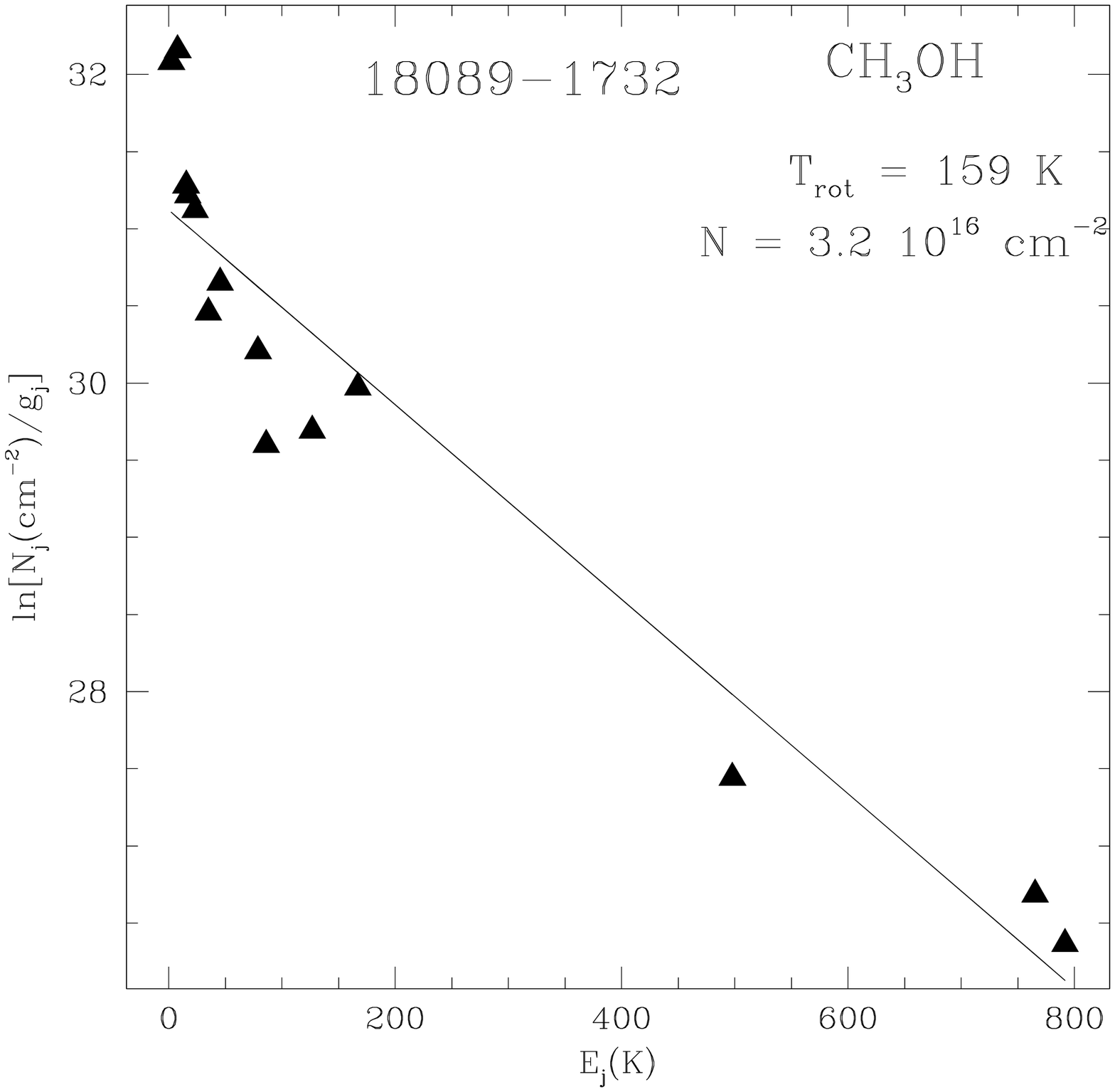}
\includegraphics[angle=0, width=7.2cm]{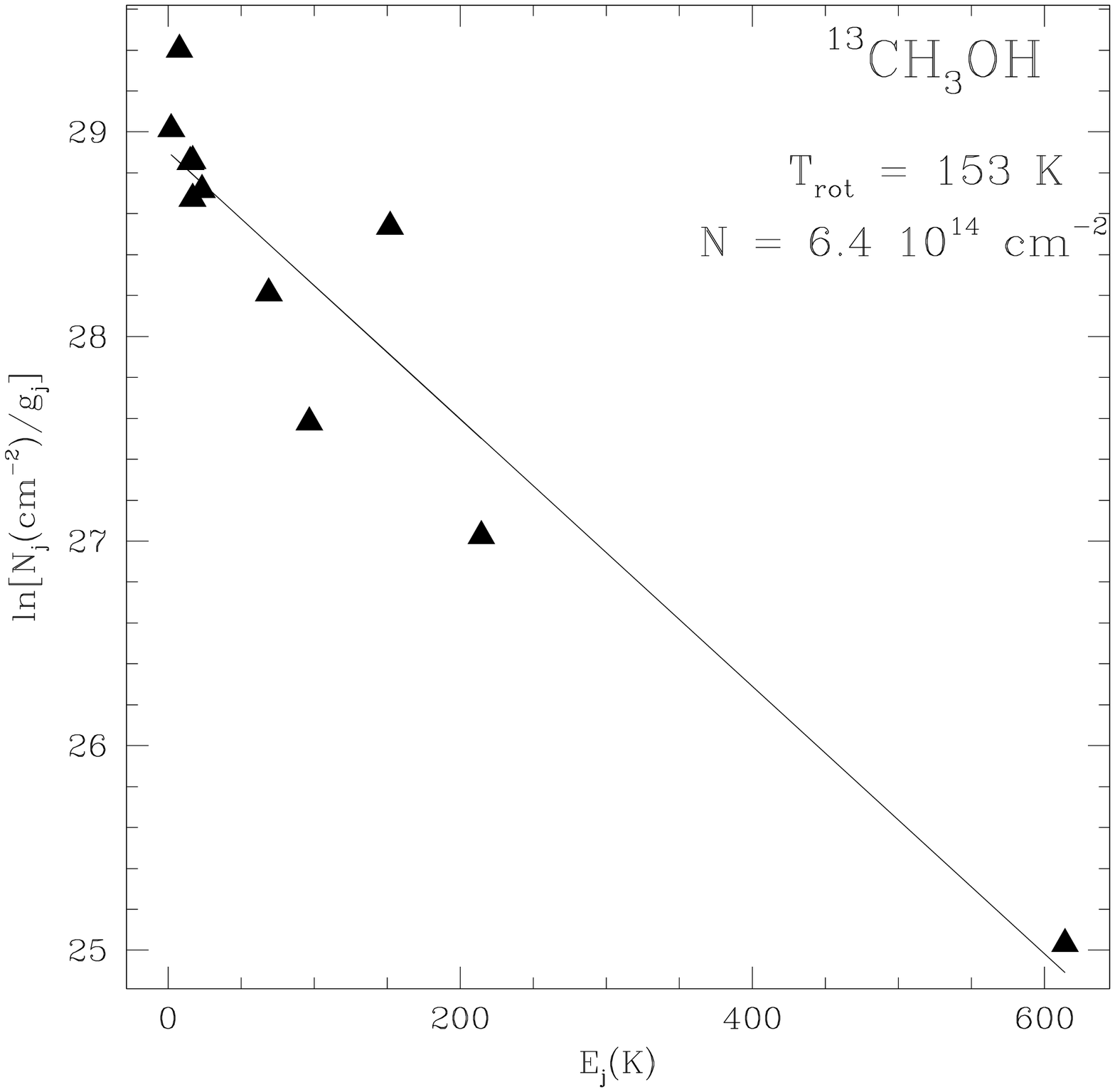}
\includegraphics[angle=0, width=7.2cm]{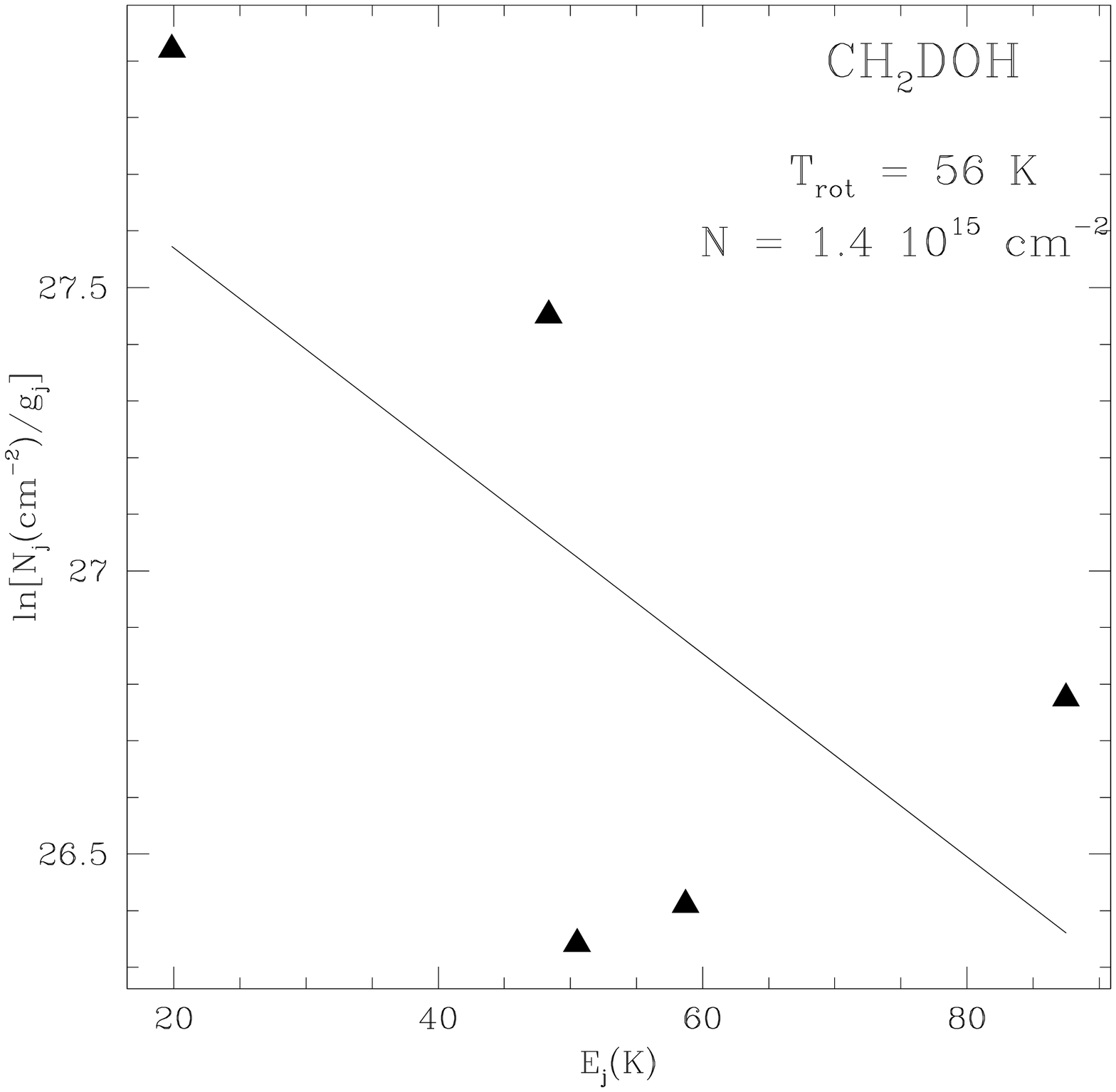}
 \caption[]
 {\label{rot_dia} Rotation diagrams obtained for 18089--1732
 from lines of \METH , \METHI\ and \DMETH\ (from top to bottom).
 Derived rotation temperatures and total column densities are shown
 in the top-right corner of each panel.
  }
\end{center}
\end{figure}

From the line parameters in Tables~\ref{tab_fit3mm} and \ref{tab_fit1mm},
we derived rotation temperature ($T_{\rm rot}$) and total column densities, 
$N$, of \METH , \METHI\ and \DMETH\ 
from the rotation diagram method. 
As an example, in Fig.~\ref{rot_dia} we show the rotation
diagrams obtained for 18089--1732. We will include all rotation
diagrams in an Appendix on-line.
The method has been applied when the number of
transitions detected was sufficient to built a "reliable" rotation diagram: 
for example, we rejected the results obtained from this method for sources 
in which rotation diagrams provide meaningless negative temperatures, or
for objects in which few lines associated with large uncertainties and/or 
similar energy of the upper levels have been detected.
Specifically, for \DMETH\ the rotation diagram method has given acceptable
results only for two sources, AFGL5142--MM and 18089--1732. However,
because in AFGL5142--MM we have only two lines, and in 18089--1732
the fit results are not very accurate (bottom panel in Fig.~\ref{rot_dia}),
the column densities have been derived also from Eq.~(A4) of Caselli et 
al.~(2002b), taking the strongest line detected and assuming the gas 
temperature equal to \Trot\ derived from \METH .

For \METHI , we have assumed that all transitions are optically thin; 
for \METH , we have first derived a rough estimate of the opacity by 
comparing two identical lines (specifically, we compared the 
$2_{(-1,2)}-1_{(-1,1)}$ and the 
$2_{(0,2)}-1_{(0,1)}$ transitions) of \METH\ and \METHI ,
and assumed an LTE abundance ratio of [$^{12}$C]/[$^{13}$C] = 77 
(Wilson \& Rood~\citeyear{wer}). From this check, we have derived 
low opacities (values smaller than 1) in all sources, so that
we have decided to compute $N$ and $T_{\rm rot}$ 
assuming optically thin conditions too. As for \AMM\ and \DAMM , 
the source sizes of \METH\ and \DMETH\ are unknown, but they are 
expected to be smaller than the beam size
and to have a comparable extent, based on observations 
at high angular resolution in Orion (Peng et al.~\citeyear{peng12}).
Therefore, to take into account the beam dilution, the column densities in the
rotation diagrams have been corrected by assuming the same source
sizes as in paperI, namely 6.5, 4.1 and 5.5\asec\ for HMSCs, HMPOs,
and UC \HII s, assuming that methanol and its deuterated forms
trace approximately the same material. This is a reasonable 
general assumption also from a theoretical point of view 
if deuterated methanol is formed from methanol through H--D 
substitution reactions on dust grains. In principle, 
\DMETH\ could be formed following other pathways, but the H--D 
substitution reaction on solid ices remains the
most efficient one (Nagaoka et al.~\citeyear{nagaoka}). Moreover, 
due to the lack of direct measurements in the cores, assuming a 
different source size for methanol and their deuterated forms 
would be an arbitrary choice not supported by observations. 

For the deuterated species for which
only one line is detected, and for sources in which $N$(\METHI )
cannot be derived from rotation diagrams, we derived $N$ 
using Eq.~(A4) of Caselli et al.~(\citeyear{casellib}) from one transition
only assuming optically thin conditions and adopting as excitation
temperature the rotation temperature derived from \METH , available in all sources.
The partition functions of all species have been calculated from the approximated
expressions valid for asymmetric rotors provided, e.g., in Ratajczak et 
al.~(\citeyear{ratajczak}, see also Parise~\citeyear{parise04}).  
The results of this analysis are presented in Table~\ref{tab_coldens_meth}.

\section{Discussion}
\label{discu}

\subsection{The {\it ortho-}/{\it para-} ratio of \DAMM }
\label{sect_tex}

The total column density of \DAMM\ has been derived from lines of 
{\it ortho-}\DAMM\ taking into account the statistical o/p ratio (3:1).
In the sources detected also in the {\it para-}\DAMM\ line at $\sim 110$~GHz
(see Fig.~\ref{spectra_damm_para}), we have verified if the 
assumption is correct: first, we have fit the hyperfine structure 
of the \pDAMM\ line, and found that all detected lines are optically thin. 
Because most of the \oDAMM\ lines detected are either optically thin
or have $\tau \leq 1$, we have decided to compare the integrated areas 
of the two transitions under the channels with signal. 
These are reported in Table~\ref{line_par_op}. As we can see, the ratio 
$\int T_{\rm MB}{\rm d}v [{\it o}]\,/ \int T_{\rm MB}{\rm d}v [{\it p}]$
is consistent with 3 within the errors in most of the sources: the 
mean value is 2.6, with standard deviation 0.6, hence consistent with 
three.

Shah \& Wootten~(\citeyear{sew}) have found similar results in a sample of
protostellar cores, in which they compare the integrated intensity of the 
same two transitions, and derived a mean value of the 
o/p ratio of 3.2 (with a larger standard deviation of $\sim 1.3$).
Comparable values have been also found by Pillai et al.~(\citeyear{pillai07})
in infrared dark clouds and by Tin\'e et al.~(\citeyear{tine00}) in
the two dark molecular clouds L134N and TMC1.
%On the other hand, ratios around $\sim 1$ have been measured in diffuse gas 
%surrounding the high-mass star form region W49N, through 
%Herschel-HIFI observations of {\it ortho-} and {\it para-}ammonia lines 
%(Persson et al.~\citeyear{persson12}), which can be 
%explained through non-standard, revisited gas-phase formation
%of Nitrogen hydrides in dark cloud conditions (Le Gal et 
%al.~\citeyear{legal14}). 
%Due to this, we speculate that the o/p ratio of \DAMM\
%found in our work is a further evidence of emission due to molecules
%produced mostly on grain mantles rather than in the gas, in
%which the expected o/p ratio should be around 1 according to
%the updated chemical models of Le Gal et al.~(\citeyear{legal14}). 

\begin{table}
\begin{center}
\caption[]{Integrated area of the {\it ortho-} and {\it para-} 
lines (Cols.~1 and 2, respectively), and their ratio (Col.~3).
Note that the integrated area of the {\it ortho-} line is equal 
to the integrated area of the best-fit Gaussian (Table~\ref{line_par_damm}) 
within the errors).}
\label{line_par_op}
\begin{tabular}{llll}
\hline \hline
    & $\int T_{\rm MB}{\rm d}v$ [{\it o}] & $\int T_{\rm MB}{\rm d}v$ [{\it p}] & $\frac{\int T_{\rm MB}{\rm d}v [{\it o}]}{\int T_{\rm MB}{\rm d}v [{\it p}]}$  \\
    &  (K \kms )     &  (K \kms )     &  \\
\cline{1-4}
\multicolumn{4}{c}{HMSC}   \\
I00117--MM2 & 0.66(0.03) & 0.16(0.03) & 4.1(0.9) \\
AFGL5142--EC & 2.03(0.04) & 0.69(0.03) & 2.9(0.2) \\
05358--mm3  & 0.83(0.04) & 0.33(0.03) & 2.5(0.3) \\ 
G034--G2(MM2) & 0.48(0.02) & & \\ 
G034--F2(MM7) & 0.55(0.03) & & \\
G034--F1(MM8) & 0.34(0.04) & & \\
G028--C1(MM9) & 0.74(0.02) & 0.28(0.03) & 2.6(0.4) \\
G028--C3(MM11)  & 0.12(0.02) & & \\
I20293--WC   & 2.77(0.03) & 1.15(0.03) & 2.4(0.1) \\ 
I22134--G    & 0.18(0.02)& & \\
I22134--B    & 0.20(0.02) & 0.13(0.02) & 1.5(0.4) \\
\cline{1-4}
\multicolumn{4}{c}{HMPO}   \\
I00117--MM1    & 0.38(0.03) & & \\
I04579--VLA1   & & & \\
AFGL5142--MM   & 2.25(0.03) & 0.75(0.03) & 3.0(0.2) \\
05358--mm1     & 0.51(0.04) & 0.26(0.03) & 2.0(0.4) \\
18089--1732    & 2.33(0.06) & 1.02(0.04)& 2.3(0.2) \\
18517+0437     & 0.50(0.02) & & \\ 
G75--core      & 0.29(0.03) & & \\
I20293--MM1    & 2.44(0.03) & 0.84(0.03) & 2.9(0.2) \\
I21307         & & &  \\
I23385	       & 0.13(0.03) & & \\ 
\cline{1-4}
\multicolumn{4}{c}{UC \HII }   \\
G5.89--0.39   & 0.54(0.03) & & \\
I19035--VLA1  & 0.62(0.03) & & \\ 
19410+2336    & 1.98(0.03) & 0.72(0.03) & 2.8(0.2) \\
ON1           & 1.33(0.03) & 0.45(0.03) & 3.0(0.3)\\
I22134--VLA1  & 0.11(0.02) & & \\  
23033+5951    &  0.88(0.03)& 0.36(0.02) & 2.4(0.2) \\
NGC7538-IRS9  & 0.16(0.03) & & \\ 
\hline
\end{tabular}
\end{center}
\end{table}

%For the \pDAMM\ line, this
%has been derived following the same approach 
%adopted for the \oDAMM\ line, but since the fit to the
%hyperfine structure always gave $\tau \sim 0.1$, for all sources we
%have used the equation valid for optically thin conditions,
%i.e. Eq.~(A4) of Caselli et al.~(\citeyear{casellib}). 

\subsection{On the \AMM\ and \DAMM\ excitation temperatures}
\label{sect_tex}

\begin{figure}
 \begin{center}
 \includegraphics[angle=-90, width=9cm]{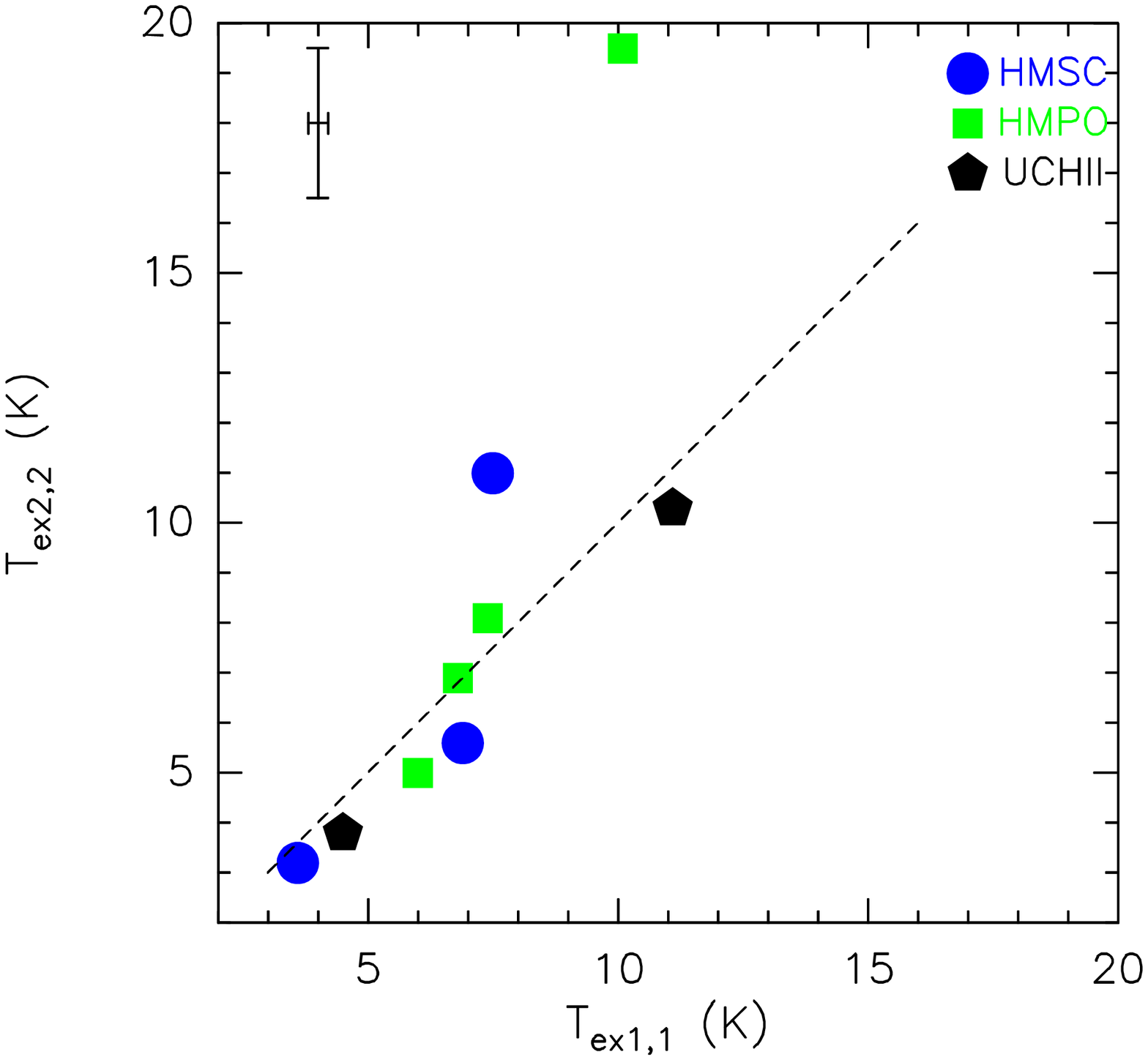}
 \includegraphics[angle=-90, width=9cm]{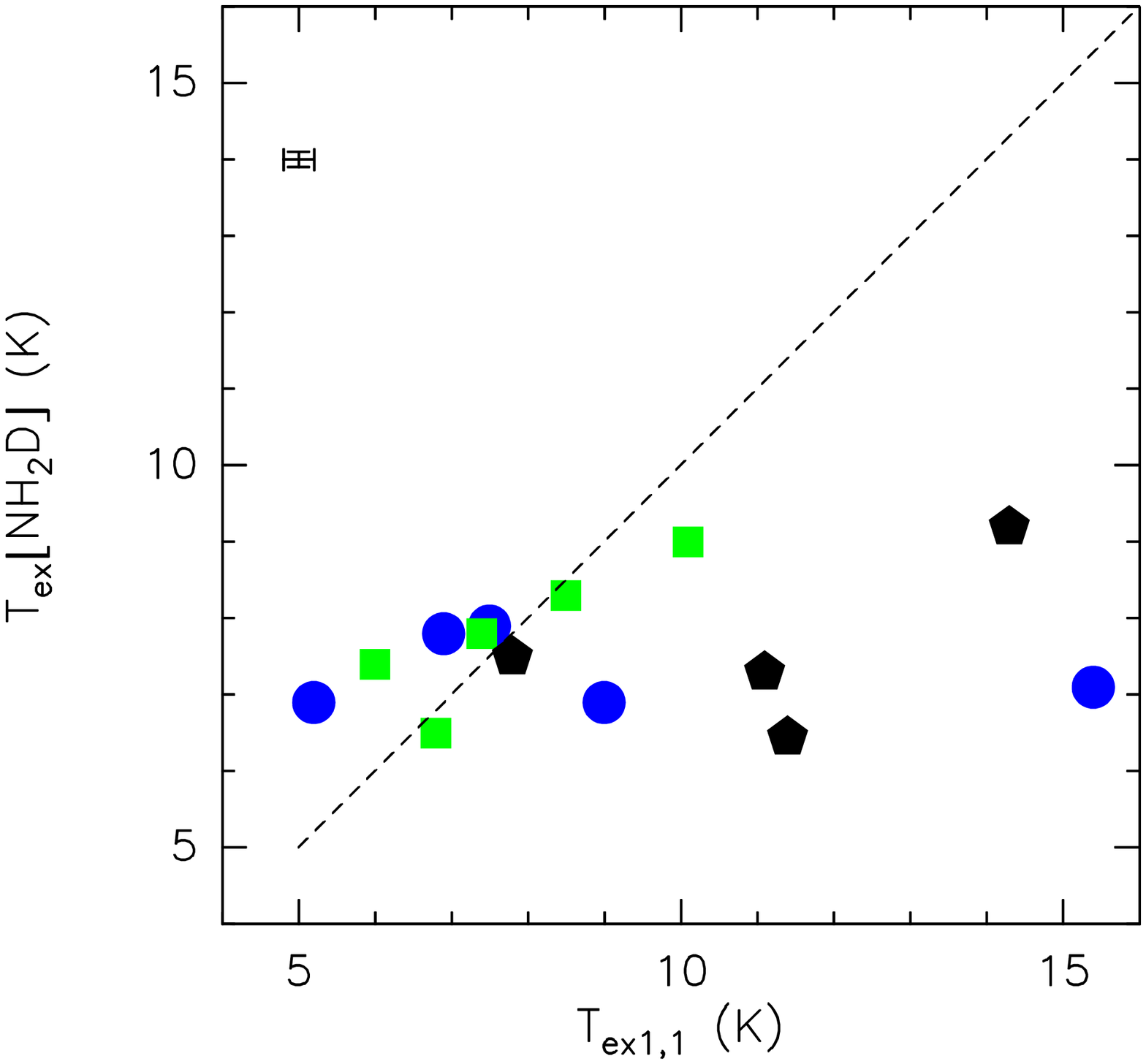}
 \caption[]
 {\label{tex_comp} Comparison between the excitation
 temperatures of \AMM (1,1), \Tex$_{1,1}$, and both
 \Tex$_{2,2}$ (upper panel) and \Tex\ of the \oDAMM\ line
 (lower panel). In both panels, blue circles correspond to HMSCs, 
 green squares show HMPOs, and black pentagons correspond to 
 UC \HII\ regions, and the dashed line indicates $y=x$. Typical
 error bars are indicated in the top-left corner of each panel.
 }
 \end{center}
\end{figure}

The excitation temperatures of the three lines examined in
the previous sections have very similar
mean values: 7.8, 8.2 and 7.5~K for \AMM (1,1), \AMM (2,2) and
\oDAMM , respectively.  The \AMM (1,1) and (2,2) lines are also
well correlated (see upper panel of Fig.~\ref{tex_comp}), while
the excitation temperatures of the \oDAMM\ line and that of 
\AMM (1,1) are not correlated due to the different
dispersion around the average values: in fact, \Tex\ of \AMM (1,1) spans
a range from $\sim 3.5$ K to 15 K, while \Tex\ of the \oDAMM\ lines is distributed tightly
around the average value. 
This may indicate either that \oDAMM\ is in sub-thermal conditions, 
as it was suggested by the asymmetric pattern of the hyperfine structure 
observed in some spectra (see Sect.~\ref{res_line_par}), or
to the fact that we are neglecting the correction for beam dilution. 
%In all cases, the values obtained are smaller
%than the kinetic temperatures in the regions, which are above $\sim 15$~K
%in all sources (see Table~A3 in paperI). This might be due either to the
%fact that we are neglecting the correction for beam dilution, or to
%sub-thermal excitation conditions. 
The former hypothesis seems plausible for the \oDAMM\ line, which 
has a high critical density ($\sim 10^{6}$ \cmc ). 
%but not for \AMM , for which the critical density is $\sim 10^{3}$ \cmc . 
About the possible different beam dilution: as stated in Sect.~\ref{sect_damm}, 
in the few cores in which both \AMM\ (1,1) and \oDAMM ($1_{-1,1}-1_{0,1}$) 
have been mapped at high angular resolution, the emissions have comparable 
extension, despite the different critical densities. Therefore, sub-thermal
conditions of the \oDAMM\ lines seem the most likely explanation to the
different excitation temperatures.

\subsection{Deuterated fraction of \AMM }
\label{sect_dfrac}

\begin{figure}
 \begin{center}
 \includegraphics[angle=-90, width=9cm]{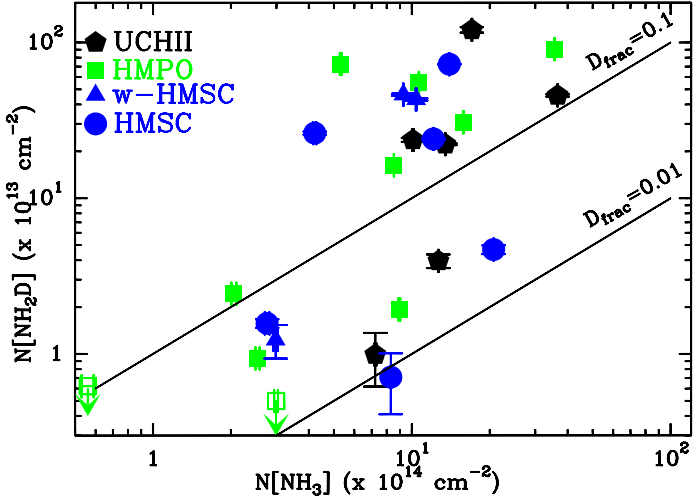}
 \caption[]
 {\label{nnh2d_nnh3} Comparison between the total column 
 density of \DAMM , $N$(\DAMM ), and \AMM , $N$(\AMM ).
Blue symbols correspond to HMSCs (triangles: ÒwarmÓ cores, 
see Sect. 2); green squares show HMPOs (open squares are upper 
limits); black pentagons correspond to UC \HII\ regions. In most
cases, the errorbars are barely visible because comparable
to (or smaller than) the size of the symbol. The straight lines represent 
the loci of \Dfrac (\AMM ) = 0.01 and 0.1.}
 \end{center}
\end{figure}

By dividing $N$(\DAMM ) for $N$(\AMM ) we have computed \Dfrac (\AMM ).
The three parameters are given in Table~\ref{tab_coldens_amm}.
The average values of \Dfrac (\AMM ) for HMSCs, HMPOs ad UC \HII s, 
are 0.26 (0.23 if one excludes the "warm" HMSCs, see Sect.~\ref{obs}), 
0.34 and 0.21 respectively. 
These values are consistent with those obtained
by Pillai et al.~(\citeyear{pillai07}) in a sample of infrared-dark clouds,
for which, however, the evolutionary stage of the embedded sources
was not determined.
The mean \Dfrac (\AMM ) is thus maximum at the HMPO stage, although
the large disperion of the data does not allow to find a statistical difference
between the three groups. This is apparent in Fig.~\ref{nnh2d_nnh3}, 
where we compare the total column densities of \DAMM\ and \AMM : 
the plot shows that the three groups are not clearly separated.
Kolmogorov-Smirnov tests applied to the data confirm that the
difference is not statistically significant. 
If one compares Fig.~\ref{nnh2d_nnh3} with the same plot
made in paperI for \H ,{\it we note clearly that, unlike \Dfrac (\H ),
\Dfrac (\AMM ) does not decrease with core evolution}.
Thus, it is not a tracer of pre--protostellar or young protostellar objects,
because it keeps above 0.1 even in the evolved stage of UC \HII\ region.
%, as previously claimed by several studies ().
Moreover, because for both \H\ and \AMM\ the deuteration in the gas-phase
is linked to H$_2$D$^+$, our results would confirm that the 
formation of \DAMM\ is largely influenced by surface chemistry.

\begin{figure}
 \begin{center}
{\includegraphics[angle=-90, width=9cm]{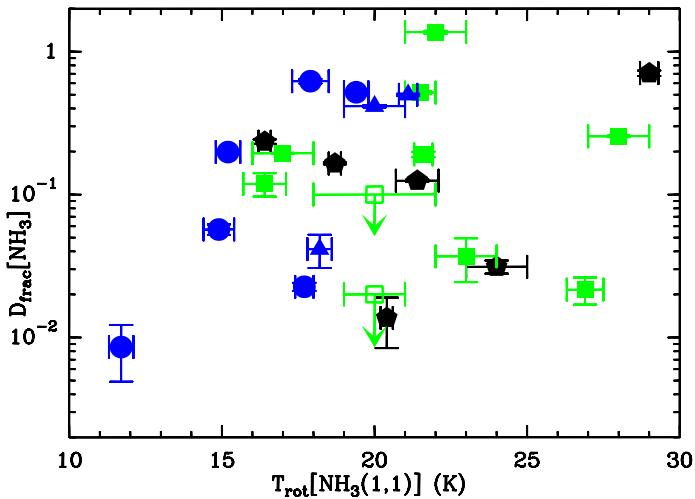}}
 \caption[] {\label{Dfrac_amm_trot} \Dfrac (\AMM ) against the ammonia
 rotation temperature.
 The symbols have the same meaning as in Fig.~\ref{nnh2d_nnh3}.
 No clear (anti-)correlation is found between the two parameters.
 In some cases, the errorbars are barely visible because comparable 
 to (or smaller than) the size of the symbol.}
 \end{center}
\end{figure}

\begin{figure}
 \begin{center}
{\includegraphics[angle=-90, width=9cm]{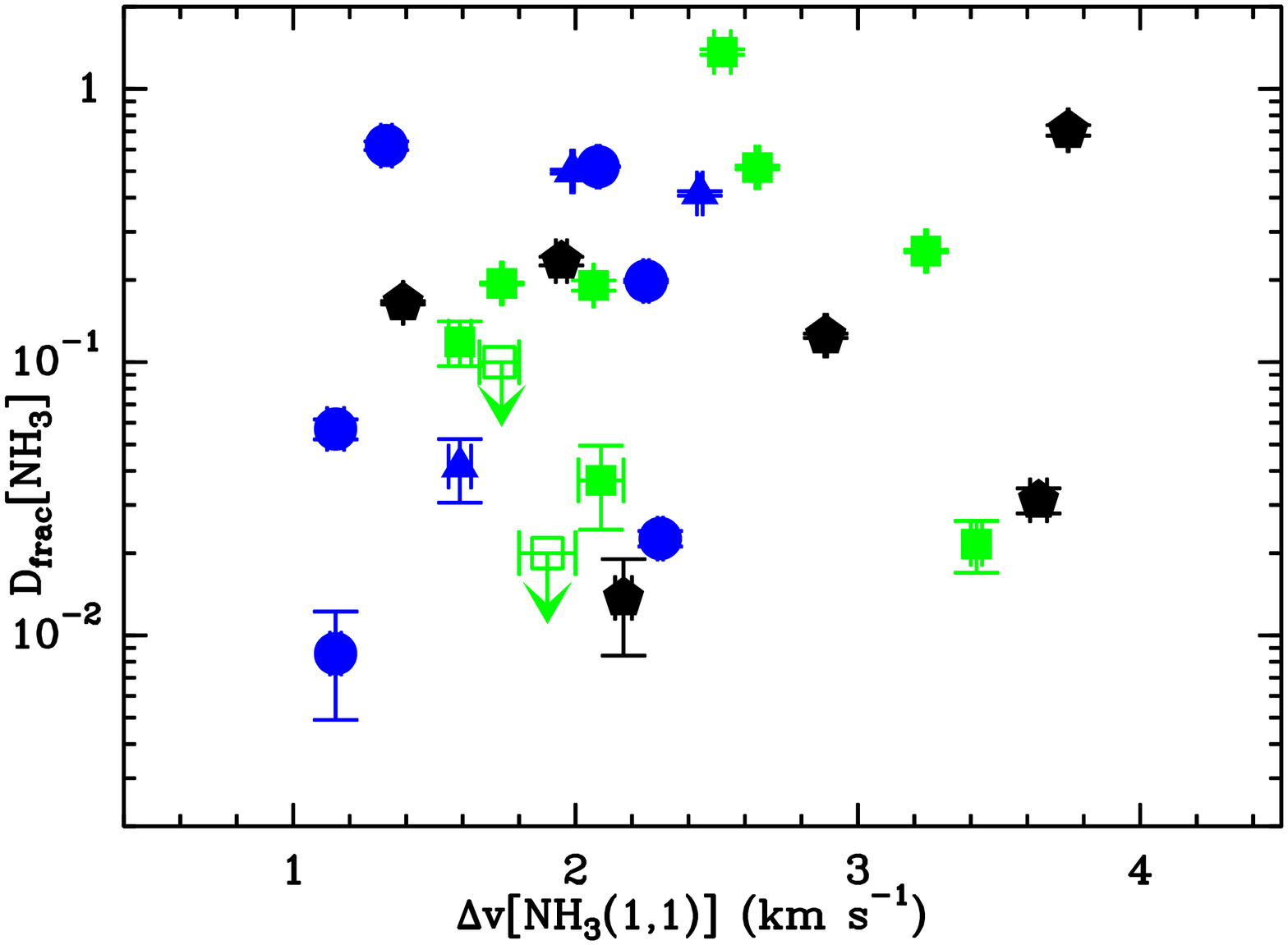}}
 \caption[] {\label{Dfrac_amm_dvnh3} \Dfrac (\AMM ) against the
 \AMM (1,1) line width.
 The symbols have the same meaning as in Fig.~\ref{nnh2d_nnh3}.
 No clear (anti-)correlation is found between the two parameters,
 like in Fig.~\ref{Dfrac_amm_trot}. The errorbars are barely visible 
 because comparable to (or smaller than) the size of the symbol.}
 \end{center}
\end{figure}

Furthermore, \Dfrac (\AMM ) does not show any clear anti-correlation 
with typical indicators of evolution. This is suggested by Figs.~\ref{Dfrac_amm_trot} and 
\ref{Dfrac_amm_dvnh3}, where we plot \Dfrac (\AMM ) against the gas temperature and 
the line widths of the (1,1) transition, both known to 
increase with time (e.g., S\'anchez-Monge et al.~\citeyear{sanchez13}): 
by applying statistical tests, we even find 
that \Dfrac (\AMM ) could be slightly correlated with both the ammonia rotation 
temperature (Pearson's linear correlation coefficient $\rho \sim 0.2$) 
and line widths (Pearson's linear correlation coefficient $\rho \sim 0.4$). 
On the contrary, \Dfrac (\H ) is anti-correlated with both parameters, as
shown in Fig.~2 of paperI. We stress, however, that the $\it p-$value (measure
of the probability of chance correlation) is 0.12 for \Dfrac (\AMM ) vs \Trot , 
and 0.22 for \Dfrac (\AMM ) vs $\Delta v$(1,1). 
Therefore, because typically the significance level under
which one can reject the null hypothesis is $\it p \sim 0.1$, both correlations are
very weak from a statistical point of view.
Nevertheless, the relevant result provided by Figs.~\ref{Dfrac_amm_trot} and 
\ref{Dfrac_amm_dvnh3} is the absence of anti-correlation, contrary to what
found for \Dfrac (\H ).

\begin{table}
\begin{center}
\caption[] {Rotation temperatures, total column densities of 
\AMM\ and \DAMM , and ammonia deuterated fraction
derived as explained in Sect.~\ref{res_amm}.}
\label{tab_coldens_amm}
\scriptsize
\begin{tabular}{lllll}
\hline
source & $T_{\rm rot}$ & $N_{\rm NH_3}$ & $N_{\rm NH_2D}$ & \Dfrac (\AMM )  \\
       &    (K)        &  ($\times 10^{14}$\cmq ) & ($\times 10^{14}$\cmq )  &  \\
\cline{1-5}
\multicolumn{5}{c}{HMSCs}   \\
I00117--MM2 & 17.9(0.6) & 4.22(0.06)  &  2.62(0.05)  &  0.62(0.02)  \\ 
AFGL5142-EC & 20(1)     & 10.39(0.03) &  4.31(0.07)  &  0.41(0.01)  \\
05358--mm3  & 21.1(0.3) & 9.27(0.02)  &  4.62(0.07)  &  0.498(0.008)  \\
G034--G2    & 15.2(0.4) & 12.12(0.04) &  2.40(0.02)  &  0.198(0.002)  \\
G034--F2    & --\tablefootmark{a}   & --\tablefootmark{a}    & 2.1(0.02)  & -- \\
G034--F1    & --\tablefootmark{a}   & --\tablefootmark{a}    & 0.12(0.02) & -- \\
G028--C1    & 17.7(0.3) & 20.70(0.02) &  0.47(0.03)  &  0.023(0.002)  \\
G028--C3       & 11.7(0.4) & 8.29(0.04)  &  0.07(0.02)  &  0.009(0.004)  \\
I20293--WC  & 19.4(0.4) & 13.96(0.02) &  0.78(0.02) &    0.519(0.002) \\  
I22134--G   & 18.2(0.4) & 2.98(0.06)  &  0.12(0.03)  &  0.04(0.01)  \\
I22134--B   & 14.9(0.5) & 2.76(0.06)  &  0.16(0.01)  &  0.057(0.005) \\
\cline{1-5}
\multicolumn{5}{c}{HMPOs}   \\
I00117--MM1  & 16.4(0.7) & 2.05(0.05)  &  0.24(0.04) &  0.12(0.02)  \\
I04579--VLA1 & 20(2)     & 0.56(0.03)  &  $\leq 0.06$ & $\leq 0.1$  \\
AFGL5142--MM & 21.5(0.5) & 10.65(0.02) &  5.50(0.08) &  0.516(0.008)  \\
05358--mm1   & 21.6(0.3) & 8.53(0.03)  &  1.63(0.07) &  0.191(0.008)  \\
18089--1732  & 28(1)     & 35.55(0.02) &  9.1(0.1)   &  0.255(0.003)  \\ 
18517+0437   & 22(1)     & 5.32(0.05)  &  7.3(0.1)   &   1.37(0.03)  \\
G75--core    & 26.9(0.6) & 8.94(0.07)  &  0.19(0.04) &  0.022(0.005)  \\ 
I20293--MM1  & 17(1)     & 15.82(0.03) &  3.07(0.02) &  0.194(0.002)  \\
I21307       & 20(1)     & 2.99(0.05)  &  $\leq 0.05$ &  $\leq 0.02$\\
I23385       & 23(1)     & 2.53(0.05)  &   0.09(0.03) &  0.04(0.01) \\
\cline{1-5}
\multicolumn{5}{c}{UC \HII s}   \\ 
G5.89--0.39   & 29.0(0.3) &  17.06(0.04)  & 12.0(0.5)  &    0.71(0.03) \\
I19035--VLA1  & 24(1)     &  12.67(0.05)  & 0.40(0.04)  &  0.031(0.003) \\
19410+2336    & 18.7(0.2) &  13.47(0.05)  & 2.2(0.03)   &    0.165(0.002)  \\
ON1           & 21.4(0.7) &  36.58(0.03)  & 4.57(0.08)   &   0.125(0.002)  \\
I22134--VLA1  & --\tablefootmark{a}     & --\tablefootmark{a}    & 0.08(0.02)  & -- \\
23033+5951    & 16.4(0.2) &  10.10(0.07)  & 2.37(0.07)   &  0.234(0.009)  \\
NGC7538--IRS9 & 20.4(0.2) &  7.23(0.06)   & 0.10(0.04)   &  0.014(0.005)  \\
\hline
\end{tabular}
\end{center}
\tablefoot{
\tablefoottext{a}{Not observed.}
}
\end{table}
\normalsize

\subsection{Deuterated fraction of methanol}
\label{sect_dfrac_2}

In the six objects detected in \DMETH , we have computed \Dfrac (\METH )=$N$(\DMETH )/$N$(\METH ). 
The results are listed in the last column of Table~\ref{tab_coldens_meth}.
The average \Dfrac (\METH ) in the three HMPOs detected is $\sim 0.04$
if one uses $N$(\DMETH ) derived from rotation diagrams, $\sim 0.01$ if
we use the simplified approach from one line only (see Sect.~\ref{sect_meth_par}). 
In the two "warm" HMSCs is $\sim 0.0025$. 
G034--G2 is the unique quiescent HMSC detected in 
\DMETH , and in this core \Dfrac (\METH ) is $\sim 0.015$.
For the cores undetected in \DMETH , the large majority of the targets
observed, we have estimated upper limits of \Dfrac (\METH )
in this way: we have computed the $3\sigma$ level in the spectrum
of the ($2_{0,2}-1_{0,1}$)e0 line, which is the transition
with the smallest energy of the upper level ($E_u\sim 6.5$ K) at 3~mm,
and estimated the upper limit to the integrated area from the relation 
$\int{T_{\rm MB}{\rm d}v}=3\sigma\frac{\Delta V}{2 \sqrt{\ln 2/\pi}}$,
valid for a Gaussian line having peak temperature $= 3\sigma$.
We have assumed $\Delta V$=1 \kms , which is the average value of the
detected \DMETH\ lines both in the HMSCs and
the HMPOs (see Table~\ref{tab_fit3mm}); then, the upper limit
on the \DMETH\ column density has been computed using the 
same equations as for the detected sources.
We have followed the same method to compute the upper limits
on the \METHI\ lines, using this time the ($2_{0,2}-1_{0,1}++$) line.

The case of the HMSC G034--G2 is quite peculiar, because it is
the only quiescent starless core detected in \DMETH\ (in one line only), 
and its \Dfrac (\METH )
exceeds 0.01, while the upper limits found in the other quiescent HMSCs
are lower. Its detection is thus quite doubtful. We have checked for possible 
contamination from other species by running synthetic spectra 
(see Sect.~\ref{sect_meth_ide}) of molecules that possess transitions 
at a similar frequency, and concluded that indeed contributions from lines of 
CH$_3$OOH, and HCCCH$_2$OH are possible.
Therefore, this detection remains doubtful.

Although the number of detections is low, and the results are affected by
the faintness of the \DMETH\ lines,
these findings suggest that high values of \Dfrac (\METH  ) tend to be associated
with "warm" HMSCs and HMPOs rather than with cores very young
(quiescent HMSCs) or evolved (UC \HII s), although the remarkable value
derived in G034--G2 (if confirmed) suggests that the story could not be so simple. 
Parise et al.~(\citeyear{parise06}) measured values of \Dfrac (\METH ) 
higher than ours by at least an order of magnitude in a sample of low-mass
protostellar cores. Nevertheless, due to the smaller linear resolution
of their observations (most of their cores are in Perseus, at a distance
of $\sim 200$~pc), their measurements should be less affected than ours by 
non-deuterated gas along the line of sight. Moreover,
our \Dfrac (\METH ) are consistent with the upper
limits found by Loinard et al.~(\citeyear{loinard03}) in high-mass
protostellar objects (where, however, they observed D$_2$CO and
derived [D$_2$CO] / [H$_2$CO] $<$ 0.5\%), 
as well as with observations of
deuterated methanol in the intermediate-mass protostar
NGC7129-FIRS2 (Fuente et al.~\citeyear{fuente14}) and in Orion BN/KL
(Peng et al.~\citeyear{peng12}).

\subsection{Deuteration and core evolution: the role of surface chemistry}
\label{deuteration}

\begin{figure}
 \begin{center}
{\includegraphics[angle=0, width=9cm]{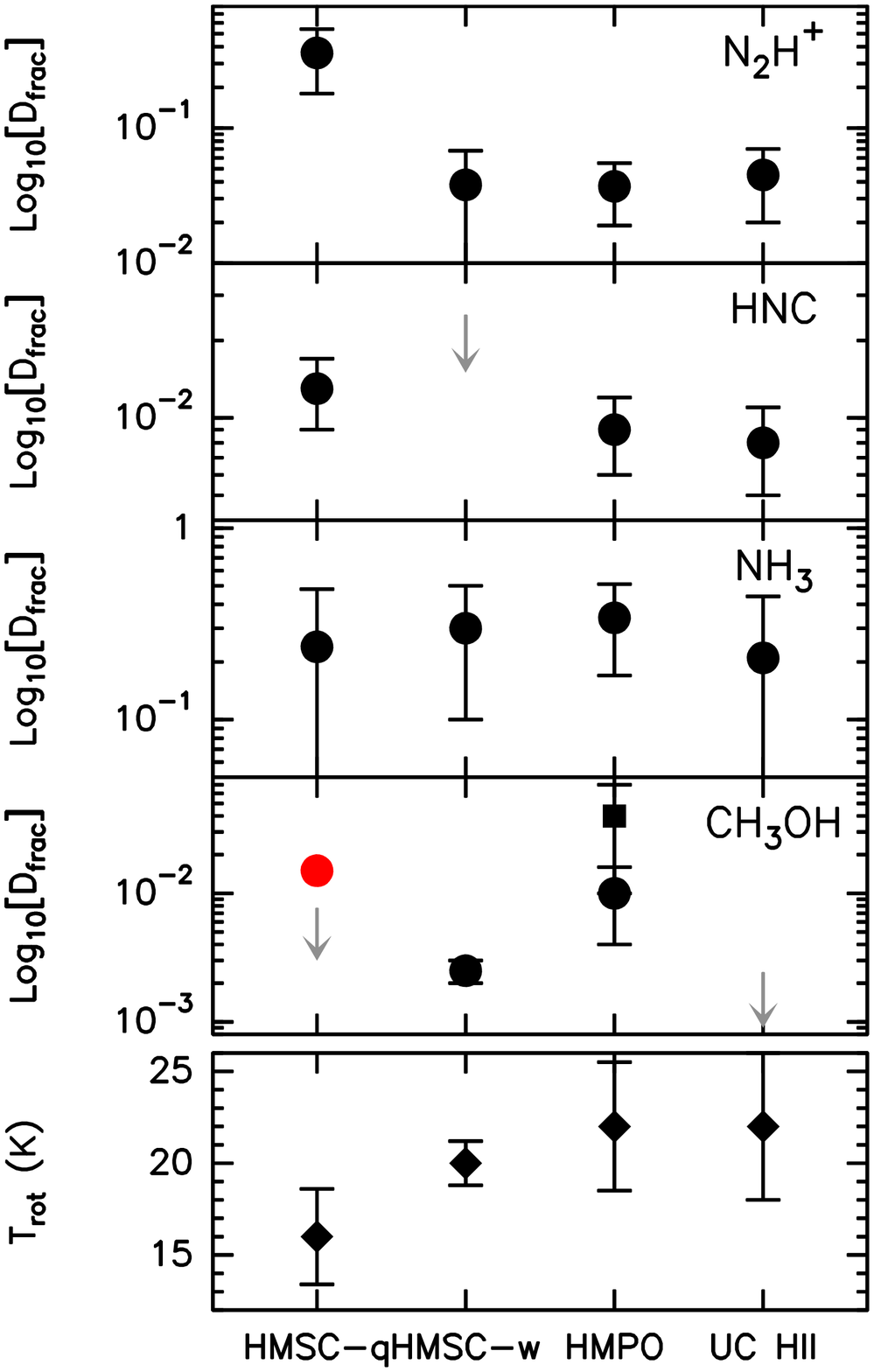}}
 \caption[] {\label{Alvaro} {\it Panels one to four, from top:} comparison between 
 the mean deuterated fractions (black dots) of \H\ (first panel, paperI), HNC (second
 panel, paperII)
\AMM\ and \METH\ (third and fourth panels, this work). 
The mean values have been computed for each evolutionary group. 
Quiescent HMSCs (HMSC-q) and warm HMSCs (HMSC-w) have been 
treated separately. The errorbars indicate the standard deviation. The 
grey arrows represent mean upper limits for those evolutionary groups
in which no sources have been detected. The red dot in the fourth
panel represents the doubtful \DMETH\ detection in G034-G2 
(see Sect.~\ref{sect_dfrac_2}), while the square indicates the mean
\Dfrac\ of HMPOs when $N$(\DMETH ) is derived from rotation diagrams
for AFGL5142--MM and 18089--732.

{\it Bottom panel:} mean rotation temperatures (filled diamonds) derived 
from ammonia in the four groups (see Table~\ref{tab_coldens_amm}).}
 \end{center}
\end{figure}

In Fig.~\ref{Alvaro} we report the mean values of \Dfrac\ obtained in HMSCs,
HMPOs and UC \HII s for the four molecular species investigated so far towards 
our source sample: \H\ (paperI), HNC (paperII), \AMM\ and \METH\ (this work). 
We show separately the values derived for quiescent HMSCs and "warm"
HMSCs to underline the effect of nearby star formation. We also include the 
mean values (with standard deviation) of the ammonia rotation temperatures
derived in this work to highlight possible (anti-)correlation betweem \Dfrac\ and
gas temperature. Inspection of Fig.~\ref{Alvaro} leads to these immediate results: 
(i) only \Dfrac (\H ) shows a net decrease from the HMSC stage to the 
HMPO stage, associated to a temperature enhancement; (ii) 
\Dfrac (\AMM ) remains nearly constant in all stages; (iii) 
\Dfrac (\METH ) is maximum in the HMPO stage, although this result
must be interpreted carefully due to the low number of detections and
the caveats on the methods to derive \Dfrac\ (see Sects.~\ref{sect_meth_par});
(iv) the behaviour of \Dfrac(HNC) is something in between
that of \Dfrac(\H ) and that of \Dfrac(\AMM ), because its maximum value is found
in the HMSC phase, like \Dfrac (\H ), but the statistically significant decrease when
going to the HMPO stage is not seen. In paperII we have already
discussed this difference, and attributed it to a slower
process of destruction of DNC into the warm gas with respect to \D .

As stated in Sect.~\ref{Introduction}, \Dfrac (\H ) and \Dfrac (\METH ) 
should represent the two "extreme" situations
under which deuteration can occur: in the gas only and on
grain mantles only, respectively. In the classical framework, both ammonia and 
methanol (and their deuterated isotopologues) are produced efficiently 
on grain mantles during the pre--stellar phase through hydrogenation 
of N and CO, respectively. Specifically, hydrogenation of CO forms 
sequentially formaldehyde first and then methanol:
thus, as time proceeds, the formation of methanol and their deuterated 
isotopologues is boosted, until the energy released by the nascent protostellar object 
in the form of radiation increases the temperature of its environment, causing the 
evaporation of the grain mantles and the release of these molecules into the gas. 
As the temperature increases and the protostar evolves towards the UC \HII\ 
region phase, the deuterated species are expected to be gradually destroyed 
due to the higher efficiency of the backward endothermic reactions 
(see Caselli \& Ceccarelli~\citeyear{cec} for a review). 
The trends shown in Fig.~\ref{Alvaro} are consistent with this classic framework, and 
show clearly that high deuterated fractions of ammonia cannot
be used as evolutionary indicator of a high-mass star forming core.
On the other hand, \Dfrac (\METH ) may be potentially a tracer of the very early stages
of the protostellar evolution, at which the evaporation/sputtering of the grain 
mantles is most efficient. Our results, however, suffers from a too low statistics,
and needs to be reinforced by other observations of deuterated methanol 
at higher sensitivity. 

Chemical models of low-mass star-forming cores predict how the abundance 
of several deuterated species varies during the evolution, including the 
amount formed on ices during the early cold phase (e.g.~Taquet et al.~\citeyear{taquet12}, 
Aikawa et al.~\citeyear{aikawa12}). Aikawa et al.~(\citeyear{aikawa12}) show that
the relative abundance ratios [\DAMM ]/[\AMM] and [CH$_2$DOH]/[CH$_3$OH]
in the ices during the pre--stellar phase are both in between 0.01 and 0.1. 
These values are consistent with \Dfrac(\METH ) measured in this work, 
and confirm that methanol and its deuterated forms are products of the 
evaporation of grain mantles. On the other hand, \Dfrac (\AMM ) measured 
in our work ($\geq 0.1$) is larger than the values predicted on ices by Aikawa 
et al.~(\citeyear{aikawa12}), suggesting that the emission we see must include 
a contribution from material formed through gas-phase reactions. 
Awad et al.~(\citeyear{awad}) modeled the deuterium chemistry of star-forming
cores using both gas-phase and grain-surface reactions, but focus on the 
protostellar phase, when the evaporation of the icy mantles of dust grains is 
maximum. The model that best reproduces a HMPO predicts 
\Dfrac (\AMM )$\sim 10^{-3}-10^{-2}$ and \Dfrac (\METH )$\leq 4 \times 10^{-3}$,
both smaller than our observed values. However, the abundance of 
deuterated species strongly depends on the density of the gas:
lower-density cores have lower abundances of deuterated species, due
to a smaller degree of CO depletion. Therefore, larger core densities could
be able to reproduce the larger deuterated fractions that we measure.

In any case,
the huge dispersion of the data do not allow us to derive firmer quantitative 
conclusions, and push us to interpret any comparison with chemical
models with caution. 
Moreover, the chemical models of Taquet et al.~(\citeyear{taquet12})
and Aikawa et al.~(\citeyear{aikawa12}) neglect the spin
states of the deuterated species, which can significantly
influence the deuterium fractionation depending on the ortho-to-para H$_2$
ratio (Flower et al.~\citeyear{flower06}).
Nevertheless, the clear different trend between \Dfrac (\AMM ) and \Dfrac (\H ) 
indicates undoubtedly that gas-phase chemistry cannot play a dominant role
in the production of \DAMM . 

%Finally, the results of this work and of paperI supports the idea that 
%{\it deuteration evolves similarly in star-forming cores from the low- to the 
%high-mass regime}.

\begin{table*}
\begin{center}
\caption[] {Rotation temperatures and total column densities for
\METH , \METHI\ and \DMETH\ derived from rotation diagrams,
unless when specified differently.
For \DMETH , in the sources where only one line has
been detected, we have computed the total column density 
from Eq. (A4) of Caselli et al.~(\citeyear{casellib}), assuming the 
temperatures obtained from \METH .}
\label{tab_coldens_meth}
\normalsize
\begin{tabular}{cccccccc}
\hline \hline
source & \multicolumn{2}{c}{CH$_3$OH} & \multicolumn{2}{c}{$^{13}$CH$_3$OH} & \multicolumn{2}{c}{CH$_2$DOH} & \\
             &  $T_{\rm rot}$ & $N$ &  $T_{\rm rot}$ & $N$ & $T_{\rm rot}$ & $N$ & \Dfrac (\METH )  \\
            &  K      &  ($\times 10^{14}$)\cmq\    &    K    &    ($\times 10^{13}$)\cmq\    &   K   &   ($\times 10^{14}$)\cmq\  &  \\
\cline{1-8}
\multicolumn{8}{c}{HMSC} \\
I00117--MM2   & 19.0 & 1.80 & & $\leq 0.3$ & & $\leq 0.013$ & $\leq 0.007$  \\
AFGL5142--EC  & 41.5 & 61.5 & 14.1 & 5.27 & --\tablefootmark{a}  & 0.11\tablefootmark{a} & 0.002(0.001) \\
05358--mm3  & 26.1 & 24.9 & 5.1 & 1.53 & & 0.08 & 0.003(0.001) \\ 
G034--G2(MM2)  & 6.0 & 1.75 & --\tablefootmark{a} & 0.09\tablefootmark{a} & & 0.03 & 0.015(0.07) \\
G034--F2(MM7)  & 5.7 &  0.95 & & $\leq 0.15$ & & $\leq 0.007$ & $\leq 0.007$ \\
G034--F1(MM8)  & 17.5 & 2.24 & & $\leq 0.3$ & & $\leq 0.01$ & $\leq 0.006$ \\
G028--C1(MM9)  & 14.2 & 2.69 & 6.8 & 0.71 & & $\leq 0.01$ & $\leq 0.004$ \\
%G028b(HMSC) & 18:42:43.98  & $-$04:01:54.4 & $+78.3$ \\
I20293--WC   & 24.4 & 3.44 & & $\leq 0.5$ & & $\leq 0.02$ & $\leq 0.005$ \\
I22134--G   & 18.1 & 2.87 & & $\leq 0.3$ & & $\leq 0.01$ & $\leq 0.004$ \\
I22134--B  &  7.8 & 0.35 & & $\leq 0.2$ & & $\leq 0.007$ & $\leq 0.02$ \\
% 22134+5834-3 (HMSC) & 22:15:06.77   &  +58:48:49.30 & \\
\cline{1-8}
\multicolumn{8}{c}{HMPO}   \\
I00117--MM1  & 27.7 & 1.22 & & $\leq 0.6$ & & $\leq 0.02$ & $\leq 0.02$ \\
%I04579--VLA1  & 112.6 & 262 \\
AFGL5142--MM  & 112.6 & 262.7 & 7.9 & 5.81 & 10.4 & 19.0\tablefootmark{b};2.1\tablefootmark{c} & 0.07(0.03)\tablefootmark{b}; 0.008(0.004)\tablefootmark{c} \\
05358--mm1  & 84.0 & 125.1 & --\tablefootmark{a} & 6.1\tablefootmark{a} & & $\leq 0.13$ & $\leq 0.001$ \\
18089--1732  & 158.6 & 318.1 & 153.0 & 64.2 & 56 & 14.0\tablefootmark{b};4.0\tablefootmark{c} & 0.04(0.02)\tablefootmark{b}; 0.01(0.01)\tablefootmark{c}  \\
18517+0437  & 137.6 & 209.2 & 44.5 & 25.6 & & $\leq 0.2$ & $\leq 0.001$ \\
G75--core  & 108. 5 & 150.6 & --\tablefootmark{a} & 4.2\tablefootmark{a} & & 0.55 & 0.005(0.003) \\ 
I20293--MM1  & 35.1 & 27.5 & --\tablefootmark{a} & 0.9\tablefootmark{a} & & $\leq 0.04$ & $\leq 0.001$ \\
I21307  & 29.4 & 6.54 &  & $\leq 0.5$ & & $\leq 0.03$ & $\leq 0.004$ \\ 
%I22198-VLA2\tablefootmark{a}  & 22:21:26.7 &	+63:51:38 & $-11.2$  & {\bf 0.76} & {\bf $10^{2.6}$} & (14) \\
I23385  & 25.3 & 18.0 & --\tablefootmark{a} & 0.3\tablefootmark{a} & & $\leq 0.02$ & $\leq 0.01$ \\
\cline{1-8}
\multicolumn{8}{c}{UC \HII }   \\
%G75.78+0.74 (HC HII) & 20:21:44.01 &	+37:26:37.6 & \\
%05137(UC)\tablefootmark{c}  &  05:17:13.3 &  +39:22:14 & $-25.4$  & \\
G5.89--0.39  & 64.1 & 128.1 & 37.9 & 14.0 & & $\leq 0.14$ & $\leq 0.001$ \\
I19035--VLA1  & 30.7 & 16.4 & 28.6 & 4.80 & & $\leq 0.03$ & $\leq 0.002$ \\
19410+2336  & 31.1 & 20.2 & 20.8 & 5.79 & & $\leq 0.03$ & $\leq 0.001$ \\
ON1  & 31.3 & 32.4 & 25.5 & 8.98 & & $\leq 0.02$ & $\leq 0.0007$ \\
I22134--VLA1  &  19.4 & 1.64 & & $\leq 0.3$ & & $\leq 0.02$ & $\leq 0.009$ \\
23033+5951  & 24.2 & 12.0 & 37.4 & 8.33 & & $\leq 0.02$ & $\leq 0.002$ \\
NGC7538--IRS9  & 28.7 & 17.6 & --\tablefootmark{a} & 0.5\tablefootmark{a} & & $\leq 0.02$ & $\leq 0.001$ \\
\hline
\end{tabular}
\end{center}
\tablefoot{
\tablefoottext{a}{Only lines with very close upper energies are detected, and the
rotation diagram provides a meaningless negative \Trot . Therefore, the column density has been
derived from the transition $(2_{0,2}-1_{0,1})$++ assuming LTE conditions and \Trot\
from methanol;}
\tablefoottext{b}{derived from rotation diagrams;}
\tablefoottext{c}{derived from the transition ($5_{2,3}-4_{1,4}$)e1 assuming LTE conditions
and \Trot\ from methanol.}
}
\end{table*}

\section{Summary and conclusions}
\label{summary}

The deuterated fraction of species that can be formed on dust grains
(in part, like \AMM , or uniquely, like \METH ) has been investigated
towards a sample of dense cores harbouring different evolutionary stages
of the high-mass star formation process. As expected, the deuterated fraction
of these species and those of molecules totally or predominantly
formed in the gas, like \H\ and HNC, evolve differently with time and
with temperature: 
\Dfrac (\AMM ) does not show statistically significant changes with evolution, 
unlike \Dfrac (\H ) and \Dfrac (HNC), which decrease (especially \Dfrac (\H ))
when temperature increases. Few lines of \DMETH\ and \METHD\ 
are clearly detected, and only towards protostellar cores or externally heated 
starless cores. Only one line of \DMETH\ could have been detected in
a quiescent starless core, but the detection is doubtful. 
No lines of deuterated methanol species are detected in 
UC \HII\ regions. This work clearly supports the scenario in which the contribution 
of surface chemistry to the formation of deuterated forms of ammonia is relevant, 
and hence that \Dfrac (\H ) remains the best indicator of massive 
starless cores. High values of \Dfrac (\METH ) 
seem suitable to trace the earliest protostellar phases, 
at which the evaporation/sputtering of the grain mantles is most efficient,
but this result needs to be supported by further, higher sensitivity observations.
The data presented in this work represent an excellent starting
point for higher angular resolution studies to address further questions. 
In particular: if the various deuterated molecules are formed with different
mechanisms, do we expect a different distribution of the emission too? 

{\it Acknowledgments.}  FF and AP are grateful to the IRAM-30m staff
for their help in the observations at the IRAM-30m telescope. 
GB is grateful to Amanda Kepley for her help during the GBT
observations, and to Jeff Magnum for providing the procedures to
convert the GBT spectra from GBTIDL to CLASS format.
AP acknowledges the financial support from UNAM, and CONACyT, M\'exico.
GB is supported by the Spanish MICINN grant AYA2008--06189--C03--01
(co-funded with FEDER funds) and
by the Italian Space Agency (ASI) fellowship under contract 
number I/005/07/01.
PC acknowledges the financial support of the European Research Council
(ERC; project PALs 320620).
AS-M is supported by the Deutsche Forschungsgemeinschaft (DFG) through 
the collaborative research grant SFB 956 "Conditions and Impact of Star Formation", project area A6.
The research leading to these results has received funding from the European 
Commission Seventh Framework Programme (FP/2007-2013) under grant 
agreement N¡ 283393 (RadioNet3).

%Many thanks to the anonymous referee for his/her useful comments
%and suggestions.

{}

%%%%%%%%%%%%%%%%%%%%%%%%%%%%%%%%%%%%%%%%

\normalsize

\clearpage

\renewcommand{\thefigure}{A-\arabic{figure}}
\setcounter{figure}{0}
\section*{Appendix A: \AMM\ and \DAMM\ spectra}
%\section{Tables}
\label{appa}

\clearpage 

\begin{figure*}
 \begin{center}
 \includegraphics[angle=-90, width=18cm]{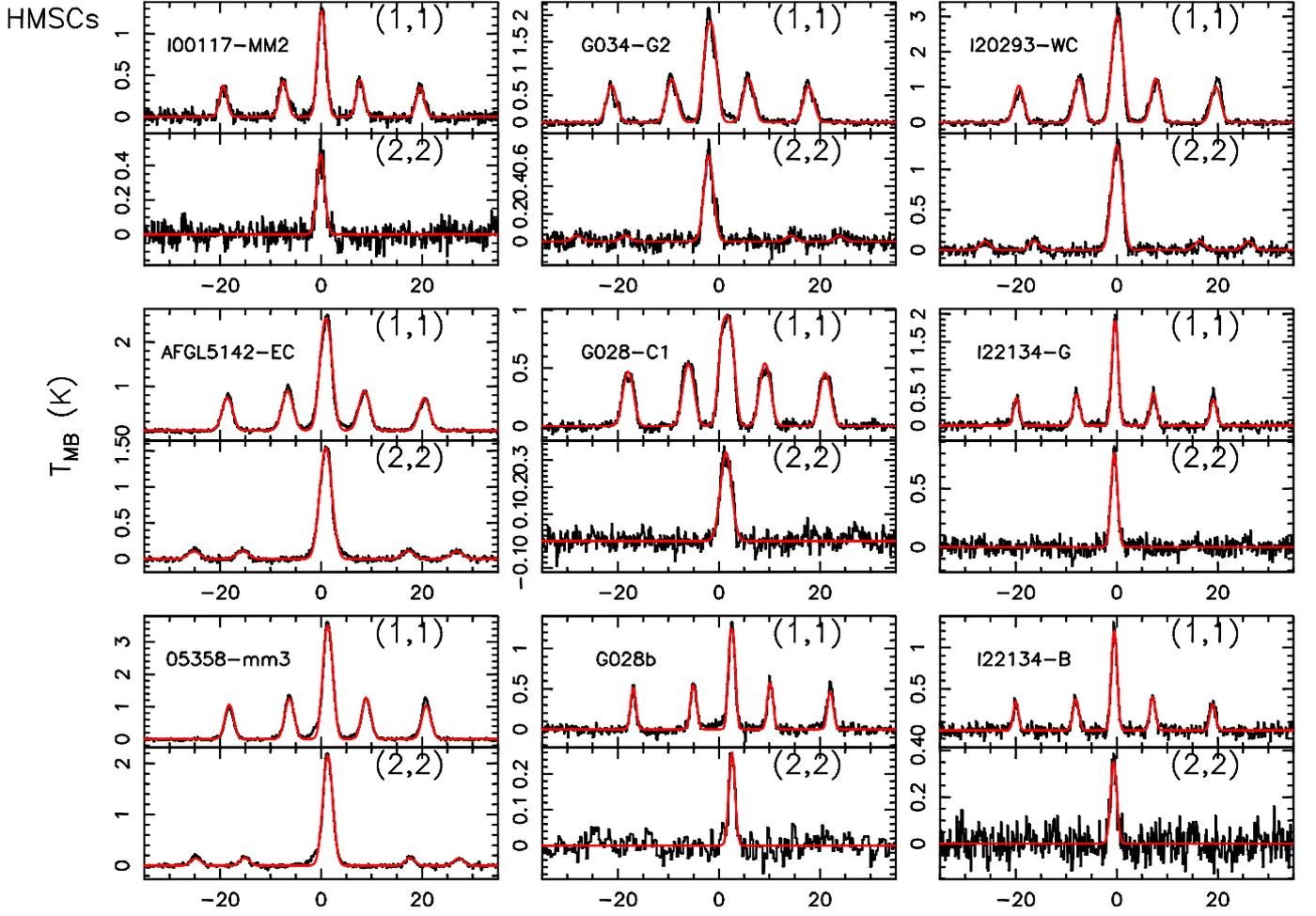}
 \caption[]
 {\label{spectra_nh3_HMSC} GBT spectra of \AMM (1,1) and (2,2) obtained 
 towards the sources classified as HMSCs. For each spectrum, the x-axis
 represents a velocity interval of $\pm 35$ \kms\ from the systemic velocity listed in 
 Table~\ref{tab_sources}. 
 The y-axis shows the intensity scale in main beam brightness temperature units.
 In each spectrum, the red curve indicates the best fit either obtained
 by fitting the hyperfine structure, when possible, or with a single Gaussian
 (see Sect.~\ref{res_amm}).
 }
 \end{center}
\end{figure*}

\begin{figure*}
 \begin{center}
 \includegraphics[angle=-90, width=18cm]{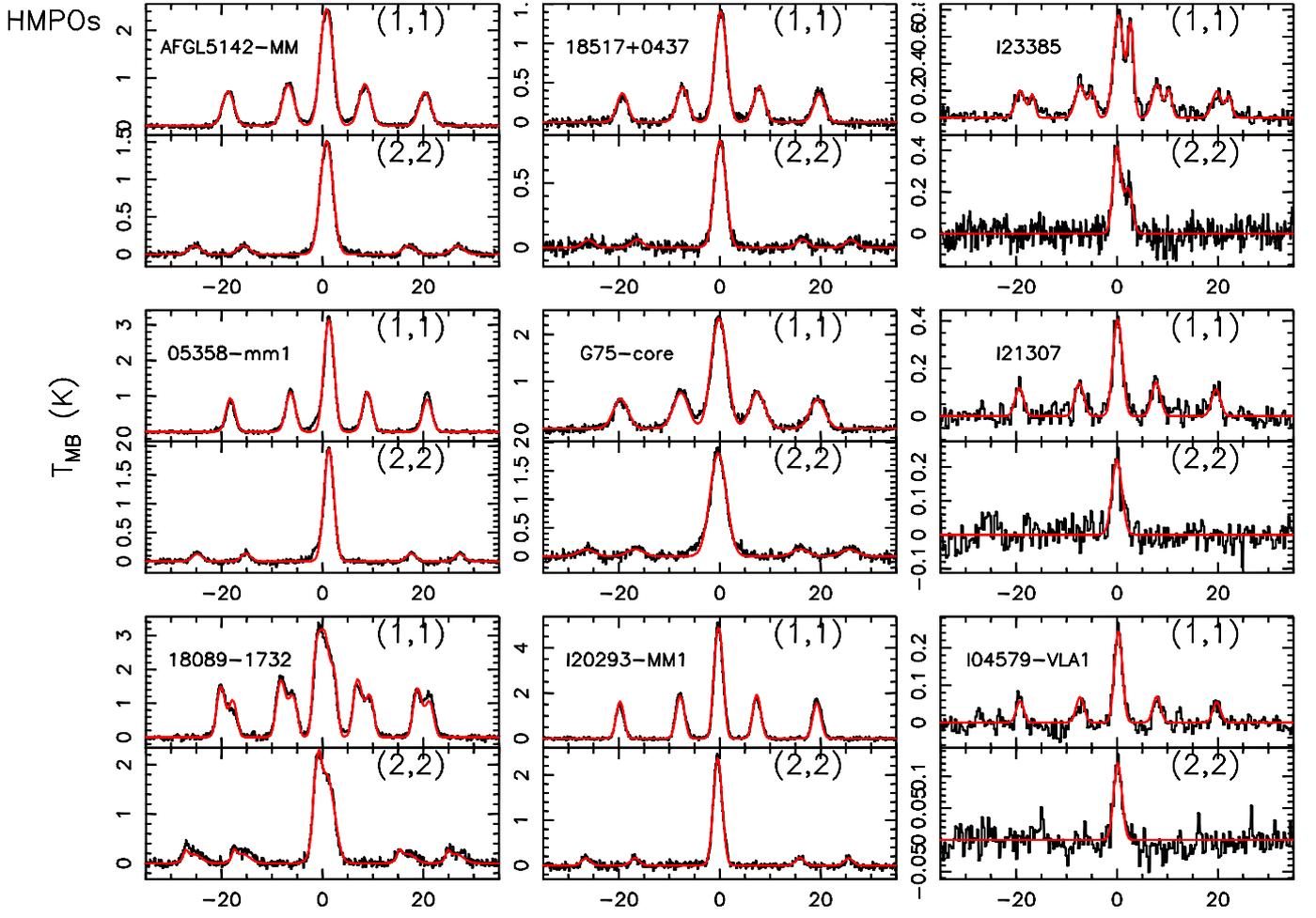}
 \caption[]
 {\label{spectra_nh3_HMPO} Same as Fig.~\ref{spectra_nh3_HMSC}
 for the sources classified as HMPOs. Note that for the spectra of
 I23385 and 18089--1732, a fit with two velocity components has been 
 performed.
 }
 \end{center}
\end{figure*}

\begin{figure*}
 \begin{center}
 \includegraphics[angle=-90, width=18cm]{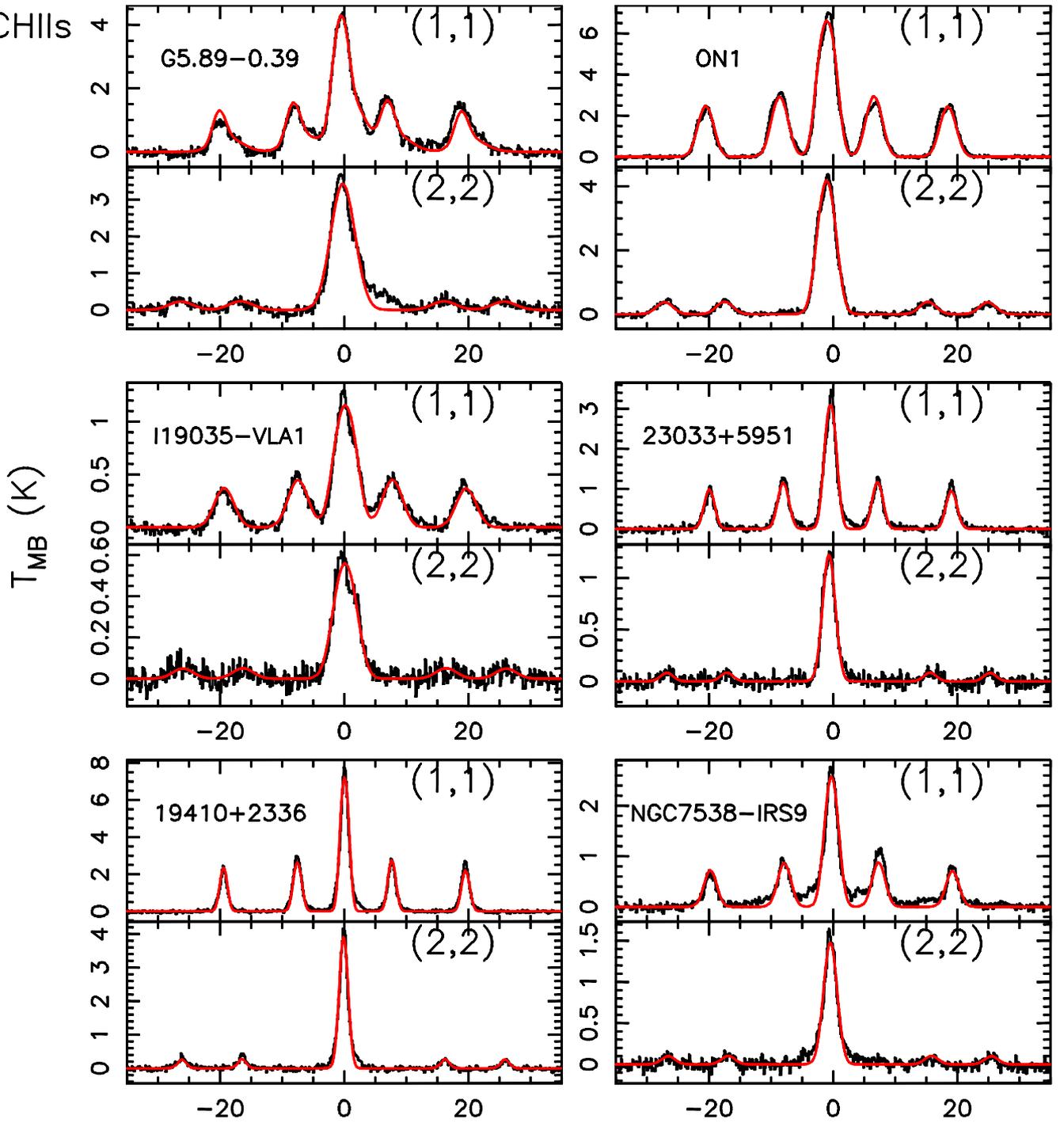}
 \caption[]
 {\label{spectra_nh3_UCHII} Same as Fig.~\ref{spectra_nh3_HMSC}
 for the sources classified as UC \HII s. }
 \end{center}
\end{figure*}

\begin{figure*}
 \begin{center}
 \includegraphics[angle=0, width=16.5cm]{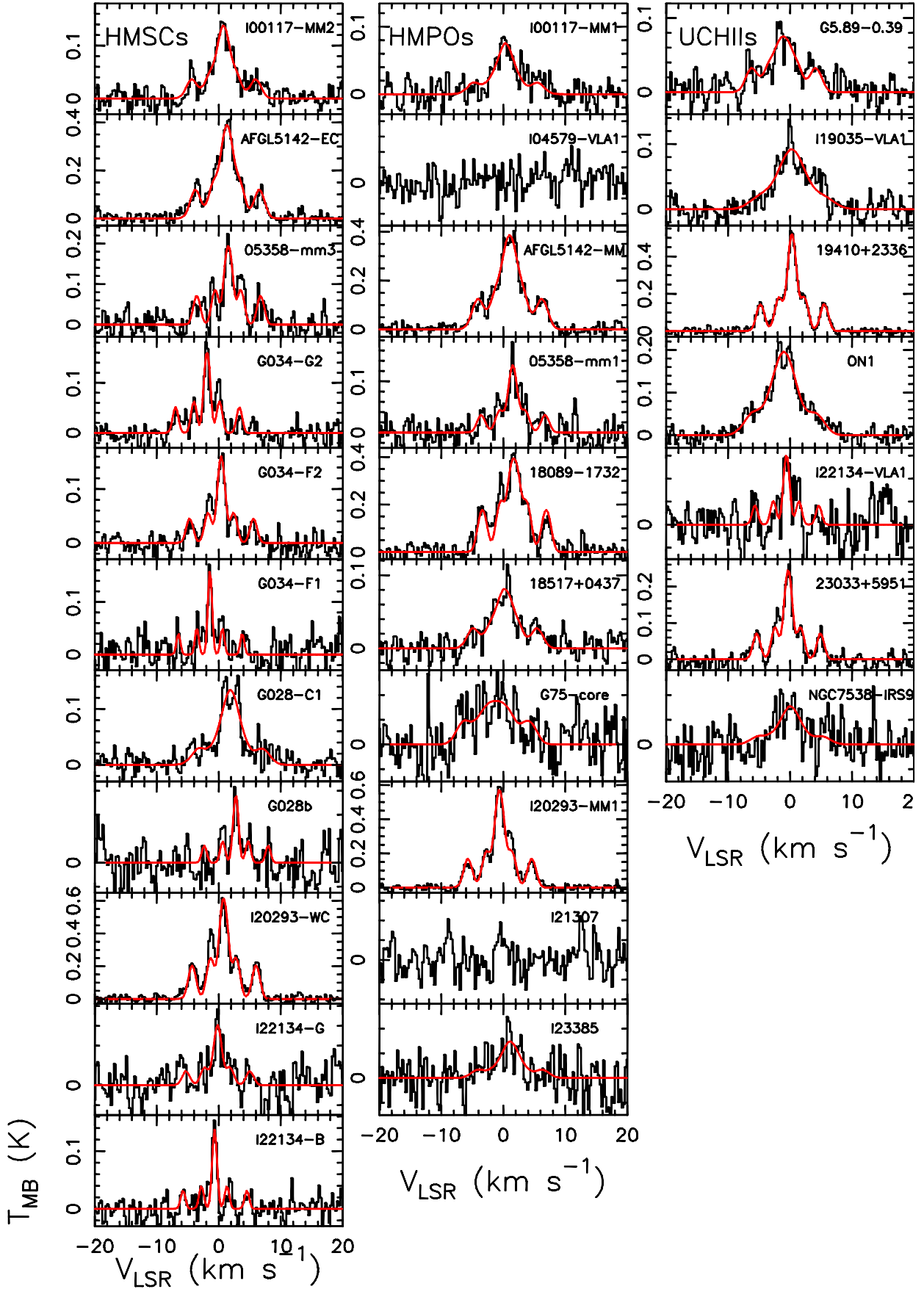}
 \caption[]
 {\label{spectra_damm} IRAM-30m spectra of \oDAMM ($1_{1,1}-1_{0,1}$) 
 obtained towards all sources observed. We show the HMSCs in the left column,
 the HMPOs in the central column, and the UC \HII s in the right column, from
 top to bottom in the same order as they appear in Table~\ref{tab_sources}. 
 For each spectrum, the x-axis represents a velocity interval of $\pm 20$ \kms\ 
 from the systemic velocity listed in Table~\ref{tab_sources}. 
 The y-axis shows the intensity scale in main beam brightness temperature units.
 In each spectrum, the red curve indicates the best fit either obtained
 by fitting the hyperfine structure, when possible, or with a single Gaussian
 (see Sect.~\ref{res_amm}).
 }
 \end{center}
\end{figure*}

\begin{figure*}
 \begin{center}
 \includegraphics[angle=0, width=16.5cm]{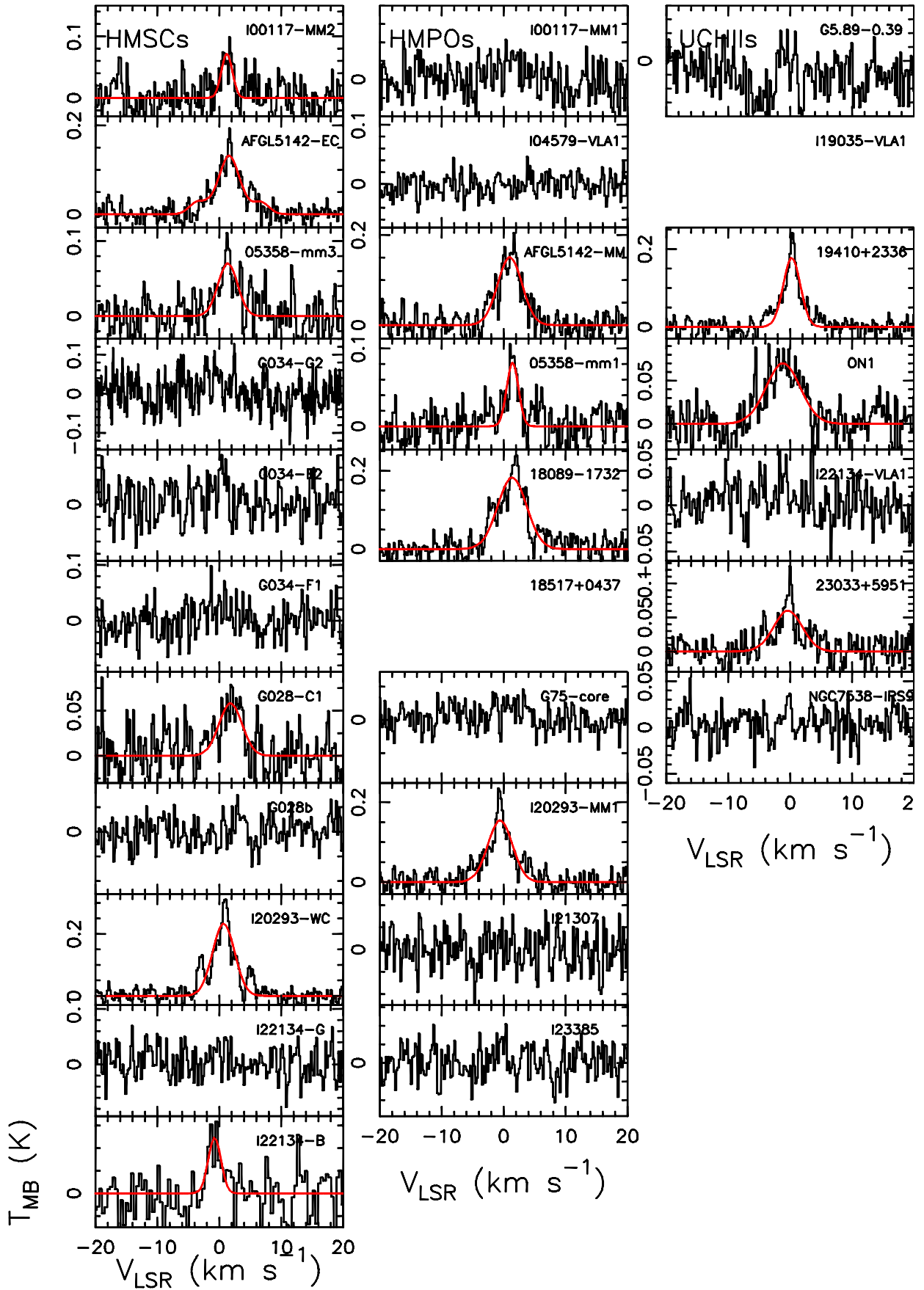}
 \caption[]
 {\label{spectra_damm_para} Same as Fig.~\ref{spectra_damm} for \pDAMM ($1_{1,1}-1_{0,1}$).
Note that two sources (18517+0437 and I19035--VLA1) have not been observed.
In each spectrum, the red curve indicates the best fit (see Sect.~\ref{res_amm}).
 }
 \end{center}
\end{figure*}

%%%%%%%%%%%%%%%%%%%%%%%%%%%%%%%%%%%%%%%%

\normalsize

\clearpage

\renewcommand{\thetable}{B-\arabic{table}}
\setcounter{table}{0}
\section*{Appendix B: Tables}
%\section{Tables}
\label{appb}

\clearpage 

\addtocounter{table}{1}

%\begin{table*}[t!]
\longtab{1}{
\begin{longtable}{lllll}
\caption{\label{tab_fit3mm} Transitions of \METH , \METHI , \DMETH\ and \METHD\ detected at 3~mm, and
line parameters derived from Gaussian fits: line integrated intensity ($\int T_{\rm MB}{\rm d}v$),
full width at half maximum (\deltav ) and main beam temperature at line peak ($T_{\rm pk}$).} \\
%\label{tab_fit3mm}
%\begin{tabular}{ccccc}
\hline \hline
  freq &   transition    &   $\int T_{\rm MB}{\rm d}v$ &   \deltav\ &   $T_{\rm pk}$  \\
  MHz &                     &  K km s$^{-1}$                    &  km s$^{-1}$    & K \\  
\hline
\endfirsthead
\caption{continued.} \\
\hline \hline
  freq &   transition    &   $\int T_{\rm MB}{\rm d}v$ &   \deltav\ &   $T_{\rm pk}$  \\
  MHz &                     &  K km s$^{-1}$                    &  km s$^{-1}$    & K \\  
 \hline
 \endhead
\hline
%\endfoot
\cline{1-5}
\multicolumn{5}{c}{HMSCs}   \\
\cline{1-5}
\multicolumn{5}{c}{I00117--MM2} \\
\cline{1-5}
%89407.91 & CH$_2$DOH 2(0,2)--1(0,1)e0 & 1.40504E-02(0.083) &  0.610(22.055) &  2.16557E-02 {\bf tentative}   \\ %! tentative			
%90262.31 & $^{13}$CH$_3$OH 11(--1,11)--10(--2,9)  & 7.71660E-03(0.003) &  0.391(2.978) &  1.85546E-02  \\ %!tentative!might be 15N2H+(1-0)!!!!
94405.16 & $^{13}$CH$_3$OH 2(--1,2)--1(--1,1) &  0.014(0.004) &  1.1(0.4) &  0.012  \\
95169.46 & CH$_3$OH 8(0,8)--7(1,7)++ &     0.031(0.006) &  2.4(0.5) &  0.012 \\
96739.36 & CH$_3$OH 2(--1,2)--1(--1,1)  &    0.35(0.02) &   0.7(0.2) &  0.45131 \\
96741.38 & CH$_3$OH 2(0,2)--1(0,1)++ &    0.48(0.02) &   0.7(0.2) &  0.62447  \\
96744.55 & CH$_3$OH 2(0,2)--1(0,1) &	     0.06(0.02) & 0.6(0.2) &  0.09  \\
96755.51 & CH$_3$OH 2(1,1)--1(1,0)  &      0.02(0.01) & 0.7(0.3) &  0.02 \\
\cline{1-5}
\multicolumn{5}{c}{AFGL5142--EC} \\
\cline{1-5}
%89407.91 & CH$_2$DOH 2(0,2)--1(0,1)e0 & 1.72122E-02(0.097) &   1.382(454.387) &  1.16984E-02 \\ %!tentative
89505.78 & CH$_3$OH 8(--4,5)--9(--3,7) &  0.07(0.02) &   1.9(0.8) & 0.03 \\
%90262.31 & $^{13}$CH$_3$OH 11(--1,11)--10(--2,9) & 1.52106E-02(0.006) &  0.991(1.221) &  1.44147E-02 \\ %!might be 15N2H+(1-0)!!!!
94405.16 & $^{13}$CH$_3$OH 2(--1,2)--1(--1,1) &   0.08(0.01) &  1.5(0.2) &  0.05 \\
94407.13 & $^{13}$CH$_3$OH 2(0,2)--1(0,1)++ &  0.06(0.02) &  1.1(0.3) &  0.06 \\
94411.02 & $^{13}$CH$_3$OH 2(0,2)--1(0,1) &   0.04(0.03) &   4.1(0.9) &  0.01 \\
94420.45 & $^{13}$CH$_3$OH 2(1,1)--1(1,0) &   0.014(0.007) &   1.7(0.6) &  0.01 \\
94541.76 & CH$_3$OH 8(3,5)--9(2,7)      &   0.083(0.009) &    1.7(0.2) &  0.05 \\
95169.46 & CH$_3$OH 8(0,8)--7(1,7)++ &  1.937(0.003) &    1.1(0.4) &   1.7 \\
95914.31 & CH$_3$OH 2(1,2)--1(1,1)++ &  0.6(0.2) &    1.4(0.5) &  0.4 \\
96739.36 & CH$_3$OH 2(--1,2)--1(--1,1) &  4.3(0.1) &     1.6(0.4) &   2.6 \\
96741.38 & CH$_3$OH 2(0,2)--1(0,1)++ &  4.4(0.1) &      2.0(0.4) &   3.4 \\
96744.55 & CH$_3$OH 2(0,2)--1(0,1)   & 1.7(0.1) &     1.9(0.4) &  0.8 \\
96755.51 & CH$_3$OH 2(1,1)--1(1,0)   &	 0.6(0.1) &    1.413(0.4) &  0.4 \\
\cline{1-5}
\multicolumn{5}{c}{05358--mm3} \\
\cline{1-5}
89407.91 & CH$_2$DOH 2(0,2)--1(0,1)e0 &  0.020(0.006) &  1.2(0.5) &  0.02 \\
%90262.31 & $^{13}$CH$_3$OH 11(--1,11)--10(--2,9) & 5.05686E-02(0.005) &   1.112(0.118) &  3.8166E-02 \\ %!might be 15N2H+(1-0)!!!!
91586.97 & CH$_2$DOH 4(1,3)--4(0,4) &   0.03(0.007) &   1.7(0.6) &  0.02 \\
%92075.51 & CH3OD 2(1,1)--1(1,0)A-- &  1.90353E-02(0.021) &   1.345(1.762) &  1.32964E-02   \\ %! tentative
94405.16 & $^{13}$CH$_3$OH 2(--1,2)--1(--1,1) & 0.040(0.007) &   1.0(0.2) &  0.03  \\
94407.13 & $^{13}$CH$_3$OH 2(0,2)--1(0,1)++ & 0.101(0.009) &   2.0(0.2) &  0.05  \\
94411.02 & $^{13}$CH$_3$OH 2(0,2)--1(0,1)  &  0.008(0.003) &   0.4(0.2) &  0.02 \\
94541.76 & CH$_3$OH 8(3,5)--9(2,7)    &   0.021(0.006) &   1.4(0.5) &  0.02 \\
95169.46 & CH$_3$OH 8(0,8)--7(1,7)++ &  1.400(0.005) &    0.839(0.004) &  1.6 \\
95914.31 & CH$_3$OH 2(1,2)--1(1,1)++ &  0.33(0.04) &    1.2(0.2) &  0.25 \\
96739.36 & CH$_3$OH 2(--1,2)--1(--1,1) &   2.4(0.1) &    1.4(0.4) &   1.7 \\
96741.38 & CH$_3$OH 2(0,2)--1(0,1)++ &  4.0(0.1) &    1.6(0.4) &   2.35 \\
96744.55 & CH$_3$OH 2(0,2)--1(0,1)  &	 0.9(0.1) &    1.5(0.4) &  0.55 \\
96755.51 & CH$_3$OH 2(1,1)--1(1,0)  &	 0.4(0.1) &    1.6(0.4) &  0.22 \\
\cline{1-5}
\multicolumn{5}{c}{G034--G2} \\
\cline{1-5}
89407.91 & CH$_2$DOH 2(0,2)--1(0,1)e0 & 0.02(0.01) &  0.8(0.2) &  0.03 \\
94405.16 & $^{13}$CH$_3$OH 2(--1,2)--1(--1,1) & 0.008(0.004) &  1.0(0.4) &  0.007 \\
94407.13 & $^{13}$CH$_3$OH 2(0,2)--1(0,1)++ & 0.010(0.003) &  0.5(0.15) &  0.018 \\
96739.36 & CH$_3$OH 2(--1,2)--1(--1,1)  &   0.764(0.004) &    1.138(0.008) &   0.63 \\
96741.38 & CH$_3$OH 2(0,2)--1(0,1)++ &   0.909(0.004) &    1.009(0.005) &   0.84 \\
96744.55 & CH$_3$OH 2(0,2)--1(0,1)	&  0.072(0.002) &   0.91(0.05) &  0.07 \\
\cline{1-5}
\multicolumn{5}{c}{G034--F2} \\
\cline{1-5}
96739.36 & CH$_3$OH 2(--1,2)--1(--1,1) &   0.37(0.02) &  0.8(0.4) &  0.46 \\
96741.38 & CH$_3$OH 2(0,2)--1(0,1)++ &   0.46(0.02) &  0.8(0.4) &  0.58 \\
96744.55 & CH$_3$OH 2(0,2)--1(0,1) &    0.05(0.02) &  1.4(0.4) &  0.034 \\
\cline{1-5}
\multicolumn{5}{c}{G034--F1} \\
\cline{1-5}
95169.46 & CH$_3$OH 8(0,8)--7(1,7)++ &  0.043(0.005) &  1.3(0.2) &  0.03 \\
95914.31 & CH$_3$OH 2(1,2)--1(1,1)++ &  0.035(0.009) &  1.8(0.6) &  0.02 \\
96739.36 & CH$_3$OH 2(--1,2)--1(--1,1) &   0.65(0.025) &   1.3(0.4) &  0.46 \\
96741.38 & CH$_3$OH 2(0,2)--1(0,1)++ &  0.68(0.025) &   1.2(0.4) &  0.55 \\
96744.55 & CH$_3$OH 2(0,2)--1(0,1) &	 0.11(0.025) &   1.7(0.4) &  0.06 \\
96755.51 & CH$_3$OH 2(1,1)--1(1,0) &	 0.02(0.1) &   1.7(0.9) &  0.015 \\
\cline{1-5}
\multicolumn{5}{c}{G028--C1} \\
\cline{1-5}
94405.16 & $^{13}$CH$_3$OH 2(--1,2)--1(--1,1) &   0.013(0.003) &  0.4(0.2) &  0.03 \\
94407.13 & $^{13}$CH$_3$OH 2(0,2)--1(0,1)++ &   0.039(0.005) &  0.9(0.1) &  0.04 \\
95169.46 & CH$_3$OH 8(0,8)--7(1,7)++ &     0.033(0.015) &   1.8(0.6) &  0.017 \\
96739.36 & CH$_3$OH 2(--1,2)--1(--1,1) &    0.94(0.04) &  1.1(0.4) &  0.84 \\
96741.38 & CH$_3$OH 2(0,2)--1(0,1)++ &   1.12(0.04) &  1.0(0.4) &   1.03  \\
96744.55 & CH$_3$OH 2(0,2)--1(0,1) &	  0.32(0.04) &  2.9(0.4) &  0.1 \\
96755.51 & CH$_3$OH 2(1,1)--1(1,0)  &    0.02(0.008) &   0.9(0.4) &  0.02 \\
\cline{1-5}
\multicolumn{5}{c}{I20293--WC} \\
\cline{1-5}
95169.46 & CH$_3$OH 8(0,8)--7(1,7)++ &  0.205(0.005) &    0.94(0.03) &  0.2 \\
95914.31 & CH$_3$OH 2(1,2)--1(1,1)++ &  0.06(0.02) &  2.3(0.9) &  0.023 \\
96739.36 & CH$_3$OH 2(--1,2)--1(--1,1) &   0.84(0.03) &    1.6(0.4) &  0.48   \\   % !blue wing!
96741.38 & CH$_3$OH 2(0,2)--1(0,1)++ &   0.73(0.03) &    1.0(0.4) &  0.66 \\
96744.55 & CH$_3$OH 2(0,2)--1(0,1)    &    0.14(0.03) &    1.4(0.4) &  0.09 \\
96755.51 & CH$_3$OH 2(1,1)--1(1,0)  &    0.05(0.03) &  1.6(0.4) &  0.03 \\ %!blue wing!
\cline{1-5}
\multicolumn{5}{c}{I22134--G} \\
\cline{1-5}
95169.46 & CH$_3$OH 8(0,8)--7(1,7)++ &  0.030(0.005) &  1.1(0.2) &  0.026 \\
96739.36 & CH$_3$OH 2(--1,2)--1(--1,1) &   0.28(0.02) &   0.8(0.4) &  0.32 \\
96741.38 & CH$_3$OH 2(0,2)--1(0,1)++ &  0.39(0.02) &   0.7(0.4) &  0.49 \\
96744.55 & CH$_3$OH 2(0,2)--1(0,1)  &	  0.06(0.02) &  0.7(0.4) &  0.08 \\
96755.51 & CH$_3$OH 2(1,1)--1(1,0) &    0.02(0.02) &  0.9(0.4) &  0.02 \\
\cline{1-5}
\multicolumn{5}{c}{I22134--B} \\
\cline{1-5}
96739.36 & CH$_3$OH 2(--1,2)--1(--1,1) &   0.113(0.002) &  0.60(0.02) &  0.18 \\
96741.38 & CH$_3$OH 2(0,2)--1(0,1)++ &  0.157(0.002) &  0.60(0.01) &  0.25  \\
96744.55 & CH$_3$OH 2(0,2)--1(0,1)" &	  0.018(0.002) &  0.57(0.07) &  0.03 \\
\cline{1-5}
\multicolumn{5}{c}{HMPOs}   \\
\cline{1-5}
\multicolumn{5}{c}{I00117--MM1}   \\
\cline{1-5}
95169.46 & CH$_3$OH 8(0,8)--7(1,7)++  &  0.024(0.005) &  1.2(0.3) &  0.02 \\	
%95914.31 & CH$_3$OH 2(1,2)--1(1,1)++ &  0.22939(0.013)  & 0.99(0.251)  &  
96739.36 & CH$_3$OH 2(--1,2)--1(--1,1)  &  0.23(0.01) &  0.8(0.4) &  0.27 \\
96741.38 & CH$_3$OH 2(0,2)--1(0,1)++ &  0.35(0.01) &  0.8(0.4) &  0.39 \\
96744.55 & CH$_3$OH 2(0,2)--1(0,1)  &	  0.04(0.01) &  0.7(0.4) &  0.052 \\
96755.51 & CH$_3$OH 2(1,1)--1(1,0)  &    0.01(0.01) &  0.8(0.9) &  0.015 \\				
\cline{1-5}
\multicolumn{5}{c}{AFGL5142--MM} \\
\cline{1-5}
89505.78 & CH$_3$OH 8(--4,5)--9(--3,7) &  0.08(0.05) &  2.0(0.9) &  0.04 \\
%90262.31 & $^{13}$CH$_3$OH 11(--1,11)--10(--2,9) & 2.40801E-02(0.005) & 0.986(0.232) &  2.29323E-02 \\ %!might be 15N2H+(1-0)!!!!
91586.97 & CH$_2$DOH 4(1,3)--4(0,4)  &   0.021(0.007) &  1.4(0.4) &  0.014 \\
94405.16 & $^{13}$CH$_3$OH 2(--1,2)--1(--1,1)  &   0.064(0.009) & 1.3(0.2) &  0.046 \\
94407.13 & $^{13}$CH$_3$OH 2(0,2)--1(0,1)++ &   0.09(0.01) & 1.3(0.2) &  0.06 \\
94411.02 & $^{13}$CH$_3$OH 2(0,2)--1(0,1)  &   0.015(0.007) &   1.0(0.6) &  0.014 \\
94420.45 & $^{13}$CH$_3$OH 2(1,1)--1(1,0)  &   0.006(0.007) &   0.7(0.6) &  0.008 \\
94541.76 & CH$_3$OH 8(3,5)--9(2,7)          &    0.10(0.01) &   1.6(0.2) &  0.055 \\
%94814.99 & CH$_3$OH 19(7,13)--20(6,14)++ &  1.12637E-02(0.004) &  0.491(0.184) &  2.15415E-02 \\
95169.46 & CH$_3$OH 8(0,8)--7(1,7)++   &   2.403(0.008) &  0.935(0.004) &   2.4 \\
95914.31 & CH$_3$OH 2(1,2)--1(1,1)++   &   0.71(0.06) &  1.45(0.15) &  0.46 \\
96739.36 & CH$_3$OH 2(--1,2)--1(--1,1)  &   4.0(0.15) &  1.5(0.4) &   2.45 \\
96741.38 & CH$_3$OH 2(0,2)--1(0,1)++  &   4.2(0.15) &  1.3(0.4) &   3.1 \\
96744.55 & CH$_3$OH 2(0,2)--1(0,1)    &   1.6(0.15) &  1.7(0.4) &  0.87 \\
96755.51 & CH$_3$OH 2(1,1)--1(1,0)   &   0.7(0.15) &  1.6(0.4) &  0.4 \\
\cline{1-5}
\multicolumn{5}{c}{05358--mm1}   \\
\cline{1-5}
89505.78 & CH$_3$OH 8(--4,5)--9(--3,7) &  2.66615E-02(0.317) &  1.751(125.117) &  1.43062E-02 \\
%90262.31 & $^{13}$CH$_3$OH 11(--1,11)--10(--2,9) &  4.92390E-02(0.007) &  0.923(0.153) &  5.01396E-02 \\  %!might be 15N2H+(1-0)!!!!
94405.16 & $^{13}$CH$_3$OH 2(--1,2)--1(--1,1) &   5.01113E-02(0.008) &   1.129(0.184) &  4.17009E-02 \\
94407.13 & $^{13}$CH$_3$OH 2(0,2)--1(0,1)++ &   5.25924E-02(0.010) &   1.632(0.425) &  3.02682E-02 \\
95914.31 & CH$_3$OH 2(1,2)--1(1,1)++ &   0.28432(0.035) &  1.289(0.199) &  0.20717 \\
96739.36 & CH$_3$OH 2(--1,2)--1(--1,1)  &    2.2854(0.107) &  1.366(0.391) &   1.5720 \\
96741.38 & CH$_3$OH 2(0,2)--1(0,1)++ &   3.7474(0.107) &  1.615(0.391) &   2.1795 \\
96744.55 & CH$_3$OH 2(0,2)--1(0,1)  &	  0.91252(0.107) &  1.713(0.391) &  0.50046 \\
96755.51 & CH$_3$OH 2(1,1)--1(1,0)  &  0.33703(0.107) &  1.605(0.391) &  0.19726 \\
95169.46 & CH$_3$OH 8(0,8)--7(1,7)++  &   1.4000(0.006) &  0.663(0.003) &   1.9824 \\
\cline{1-5}
\multicolumn{5}{c}{18089--1732}   \\
\cline{1-5}
89275.41 & CH$_2$DOH 2(0,2)--1(0,1)e1 &  0.01(0.01) &   0.7(0.3) &  0.014 \\
89505.78 & CH$_3$OH 8(--4,5)--9(--3,7)   &   0.18(0.09) &   1.4(0.8) &  0.12 \\
%90262.31 & $^{13}$CH$_3$OH 11(--1,11)--10(--2,9)  &   6.14246E-02(0.010) &   2.460(0.478) &  2.34582E-02 \\  %!might be 15N2H+(1-0)!!!!
90384.31 & $^{13}$CH$_3$OH 13(1,13)--12(2,10)  &    0.027(0.007) &   1.4(0.4) &  0.018 \\
92588.70 & $^{13}$CH$_3$OH 7(2,6)--8(1,7)--- &   0.030(0.007) &   0.8(0.3) &  0.034 \\
93619.46 & $^{13}$CH$_3$OH 2(1,2)--1(1,1)++ &   0.04(0.01) &   1.4(0.4) &  0.03 \\
94405.16 & $^{13}$CH$_3$OH 2(--1,2)--1(--1,1)  &   0.095(0.008) &    1.8(0.2) &  0.05 \\
94407.13 & $^{13}$CH$_3$OH 2(0,2)--1(0,1)++ &    0.084(0.007) &    1.1(0.1) &  0.07 \\
94411.02 & $^{13}$CH$_3$OH 2(0,2)--1(0,1)  &      0.072(0.005) &    1.5(0.1) &  0.04 \\
94420.45 & $^{13}$CH$_3$OH 2(1,1)--1(1,0)  &     0.048(0.005) &    1.3(0.2) &  0.03 \\
94541.76 & CH$_3$OH 8(3,5)--9(2,7)             &     0.230(0.006)  &    1.61(0.05) &  0.13 \\
94814.99 & CH$_3$OH 19(7,13)--20(6,14)++ &   0.050(0.006) &   1.4(0.2) &  0.033 \\
%94912.69 & $^{13}$CH$_3$OH 15(1,14)--15(1,15)--+ &  5.69567E-02(0.019) &  2.560(0.391) &  2.08990E-02 \\
95169.46 & CH$_3$OH 8(0,8)--7(1,7)++"       & 1.27(0.02) &  1.2(0.4) &   1.015 \\
95208.66 & $^{13}$CH$_3$OH 2(1,1)--1(1,0)-- &    0.06(0.02) &  1.8(0.4) &  0.029 \\
95273.44 & $^{13}$CH$_3$OH 6(--2,5)--7(--1,7)  &    0.05(0.02) &  1.7(0.4) &  0.027 \\
95914.31 & CH$_3$OH 2(1,2)--1(1,1)++ &       0.60(0.09) &   1.4(0.3) &  0.40 \\
96739.36 & CH$_3$OH 2(--1,2)--1(--1,1)  &     1.58(0.07) &    1.3(0.4) &   1.14 \\
96741.38 & CH$_3$OH 2(0,2)--1(0,1)++ &     1.95(0.07) &    1.2(0.4) &   1.5 \\
96744.55 & CH$_3$OH 2(0,2)--1(0,1)  &      0.87(0.07) &    1.4(0.4) &  0.58 \\
96755.51 & CH$_3$OH 2(1,1)--1(1,0)  &      0.57(0.07) &    1.5(0.4) &  0.36 \\ 
\cline{1-5}
\multicolumn{5}{c}{18517+0437}   \\
\cline{1-5}
89505.78 & CH$_3$OH 8(--4,5)--9(--3,7)           &   0.07(0.2) &  1.5(0.4) &  0.04 \\
%90262.31 & $^{13}$CH$_3$OH 11(--1,11)--10(--2,9)  &  1.90106E-02(0.004) &  0.811(0.175) &  2.20270E-02 \\  %!might be 15N2H+(1-0)!!!!
94405.16 & $^{13}$CH$_3$OH 2(--1,2)--1(--1,1) &     0.055(0.004) &  1.4(0.4) &  0.04 \\
94407.13 & $^{13}$CH$_3$OH 2(0,2)--1(0,1)++ &    0.066(0.004) &   1.1(0.4) &  0.06 \\
94411.02 & $^{13}$CH$_3$OH 2(0,2)--1(0,1)  &      0.017(0.004) &   1.0(0.4) &  0.015 \\
94541.76 & CH$_3$OH 8(3,5)--9(2,7)  &	      0.138(0.004) &   2.0(0.4) &  0.06 \\
%94814.99 & CH$_3$OH 19(7,13)--20(6,14)++ &  3.48396E-02(0.007) &  2.666(0.519) &  1.22759E-02 \\
95169.46 & CH$_3$OH 8(0,8)--7(1,7)++ &   0.648(0.003) &  0.821(0.006) &  0.74 \\
95914.31 & CH$_3$OH 2(1,2)--1(1,1)++ &   0.44(0.04) &  1.3(0.1) &  0.32 \\
96739.36 & CH$_3$OH 2(--1,2)--1(--1,1) &    1.93(0.09) &   1.2(0.4) &   1.5 \\
96741.38 & CH$_3$OH 2(0,2)--1(0,1)++ &   2.70(0.09) &   1.2(0.4) &   2.12 \\
96744.55 & CH$_3$OH 2(0,2)--1(0,1) &   0.85(0.09) &   1.2(0.4) &  0.64 \\
96755.51 & CH$_3$OH 2(1,1)--1(1,0) &   0.46(0.09) &   1.4(0.4) &  0.31 \\
\cline{1-5}
\multicolumn{5}{c}{G75--HCHII}  \\
\cline{1-5}
89505.78 & CH$_3$OH 8(--4,5)--9(--3,7)  &   0.1(0.1) &  0.9(0.8) &  0.06 \\
94405.16 & $^{13}$CH$_3$OH 2(--1,2)--1(--1,1)  &   0.033(0.009) &  1.6(0.5) &  0.02 \\
94407.13 & $^{13}$CH$_3$OH 2(0,2)--1(0,1)++ &    0.025(0.008) &  1.0(0.3) &  0.024 \\
94541.76 & CH$_3$OH 8(3,5)--9(2,7)           &       0.074(0.005) &  1.4(0.1) &  0.05 \\
%94912.69 & $^{13}$CH$_3$OH 15(1,14)--15(1,15)--+ &   3.60183E-02(0.005) &   1.276(0.194) &  2.65083E-02 \\  %!could be C2H3CN (vynil cyanide)
95169.46 & CH$_3$OH 8(0,8)--7(1,7)++  &	 0.934(0.006) &   1.98(0.02) &  0.44 \\
95914.31 & CH$_3$OH 2(1,2)--1(1,1)++ &     0.37(0.06) &   1.4(0.25) &  0.26 \\
96739.36 & CH$_3$OH 2(--1,2)--1(--1,1)  &    1.13(0.06) &   1.5(0.4) &  0.70 \\
96741.38 & CH$_3$OH 2(0,2)--1(0,1)++ &    2.00(0.06) &   1.7(0.4) &   1.07 \\
96744.55 & CH$_3$OH 2(0,2)--1(0,1)  &     0.81(0.06) &   1.8(0.4) &  0.42 \\
96755.51 & CH$_3$OH 2(1,1)--1(1,0)  &     0.34(0.06) &   1.3(0.4) &  0.24 \\
\cline{1-5}
\multicolumn{5}{c}{I20293--MM1}  \\
\cline{1-5}
%90262.31 & $^{13}$CH$_3$OH 11(--1,11)--10(--2,9)  &  2.56986E-02(0.005) &  0.710(0.155) &  3.40198E-02 \\  %!might be 15N2H+(1-0)!!!!
94405.16 & $^{13}$CH$_3$OH 2(--1,2)--1(--1,1)  &    0.033(0.008) &  1.6(0.5) &  0.02 \\
94407.13 & $^{13}$CH$_3$OH 2(0,2)--1(0,1)++ &    0.024(0.007) &  0.7(0.2) &  0.03 \\
94541.76 & CH$_3$OH 8(3,5)--9(2,7)    &     0.04(0.01) &  5(1) &  0.007 \\
95169.46 & CH$_3$OH 8(0,8)--7(1,7)++ &   1.387(0.007) &  1.126(0.007) &   1.16 \\
95914.31 & CH$_3$OH 2(1,2)--1(1,1)++ &   0.15(0.03) &  1.5(0.4) &  0.09 \\
96739.36 & CH$_3$OH 2(--1,2)--1(--1,1)  &   1.28(0.06) &  1.2(0.4) &   1.01 \\
96741.38 & CH$_3$OH 2(0,2)--1(0,1)++ &   1.81(0.06) &  1.2(0.4) &   1.4 \\
96744.55 & CH$_3$OH 2(0,2)--1(0,1)  &	  0.81(0.06) &  2.9(0.4) &  0.26 \\
96755.51 & CH$_3$OH 2(1,1)--1(1,0) &	  0.12(0.06) &  1.1(0.4) &  0.1 \\
\cline{1-5}
\multicolumn{5}{c}{I21307}   \\
\cline{1-5}
%90743.56 & CH3OD 2(1,1)--1(1,0)E &   1.12503E-02(0.004) &  0.621(0.248) &  1.70129E-02 {\bf tentative} \\ %! tentative
95169.46 & CH$_3$OH 8(0,8)--7(1,7)++ &   0.201(0.004) &  0.64(0.02) &  0.3 \\
95914.31 & CH$_3$OH 2(1,2)--1(1,1)++ &   0.03(0.01) &  1.2(0.6) &  0.02 \\
96739.36 & CH$_3$OH 2(--1,2)--1(--1,1) &    0.220(0.004) &  0.84(0.02) &  0.25 \\
96741.38 & CH$_3$OH 2(0,2)--1(0,1)++ &   0.321(0.004) &  0.86(0.02) &  0.35  \\
96744.55 & CH$_3$OH 2(0,2)--1(0,1)  &   0.080(0.005) &  0.97(0.06) &  0.08 \\
96755.51 & CH$_3$OH 2(1,1)--1(1,0)  &   0.026(0.004) &  0.9(0.2) &  0.03 \\
\cline{1-5}
\multicolumn{5}{c}{I23385}    \\
\cline{1-5}
94405.16 & $^{13}$CH$_3$OH 2(--1,2)--1(--1,1)  &   0.015(0.004) &  0.9(0.2) &  0.016 \\
94407.13 & $^{13}$CH$_3$OH 2(0,2)--1(0,1)++ &   0.013(0.004) &  0.9(0.3) &  0.013 \\
95169.46 & CH$_3$OH 8(0,8)--7(1,7)++ &    0.464(0.005) &   1.15(0.02) &  0.38 \\
95914.31 & CH$_3$OH 2(1,2)--1(1,1)++ &    0.11(0.02) &   1.25(0.3) &  0.08 \\
96739.36 & CH$_3$OH 2(--1,2)--1(--1,1) &    0.77(0.04) &  1.3(0.4) &  0.57 \\
96741.38 & CH$_3$OH 2(0,2)--1(0,1)++ &    1.13(0.04) &  1.3(0.4) &  0.8 \\
96744.55 & CH$_3$OH 2(0,2)--1(0,1)     &    0.33(0.04) &  1.8(0.4) &  0.2 \\
96755.51 & CH$_3$OH 2(1,1)--1(1,0)    &    0.08(0.04) &  0.75(0.4) &  0.1 \\
\cline{1-5}
\multicolumn{5}{c}{UC \HII s}   \\
\cline{1-5}
\multicolumn{5}{c}{G5.89--0.39}  \\   
\cline{1-5}
%89251.16 & CH$_2$DOH 2(0,2)--1(0,1)o1 &    4.51870E-02(0.008) &   1.012(0.204) &  4.19489E-02 \\
94405.16 & $^{13}$CH$_3$OH 2(--1,2)--1(--1,1)  &   0.050(0.004) &   1.8(0.4) &  0.025 \\
94407.13 & $^{13}$CH$_3$OH 2(0,2)--1(0,1)++ &   0.100(0.004) &   2.0(0.4) &  0.04 \\
94411.02 & $^{13}$CH$_3$OH 2(0,2)--1(0,1)  &     0.03(0.004) &   1.0(0.4)  &  0.03 \\
94420.45 & $^{13}$CH$_3$OH 2(1,1)--1(1,0)  &     0.011(0.004) &   1.3(0.4)  &  0.01 \\
94541.76 & CH$_3$OH 8(3,5)--9(2,7)        &       0.052(0.004) &   1.4(0.4)   &  0.03 \\
95169.46 & CH$_3$OH 8(0,8)--7(1,7)++   &    2.726(0.007) &  1.638(0.007) &   1.56 \\
95914.31 & CH$_3$OH 2(1,2)--1(1,1)++   &    0.97(0.05) &    1.8(0.1) &  0.49 \\
96739.36 & CH$_3$OH 2(--1,2)--1(--1,1)  &    3.8(0.1) &    2.5(0.4) &   1.43 \\
96741.38 & CH$_3$OH 2(0,2)--1(0,1)++ &    2.4(0.1) &    1.4(0.4) &   1.66 \\
96744.55 & CH$_3$OH 2(0,2)--1(0,1)     &      1.6(0.1) &    1.8(0.4) &  0.83 \\
96755.51 & CH$_3$OH 2(1,1)--1(1,0)     &      0.9(0.1) &    1.9(0.4) &  0.47 \\
\cline{1-5}
\multicolumn{5}{c}{I19035--VLA1}    \\
\cline{1-5}
%89251.16 & CH$_2$DOH 2(0,2)--1(0,1)o1 &   4.93293E-02(0.311) &    3.158(30.107) &  1.46734E-02 \\
94407.13 & $^{13}$CH$_3$OH 2(0,2)--1(0,1)++ &  0.032(0.005) &    1.2(0.2) &  0.024 \\
94411.02 & $^{13}$CH$_3$OH 2(0,2)--1(0,1)  &    0.008(0.003) &    0.4(0.3) &  0.02  \\ %!tentative
94420.45 & $^{13}$CH$_3$OH 2(1,1)--1(1,0)  &    0.014(0.006) &    2.0(0.8) &  0.007  \\
94541.76 & CH$_3$OH 8(3,5)--9(2,7)        &      0.033(0.009) &    4(1) &  0.008  \\
95169.46 & CH$_3$OH 8(0,8)--7(1,7)++ &    0.147(0.007) &     1.9(0.1) &  0.07 \\
95914.31 & CH$_3$OH 2(1,2)--1(1,1)++ &    0.12(0.02) &     1.7(0.4) &  0.065 \\
96739.36 & CH$_3$OH 2(--1,2)--1(--1,1)  &    0.98(0.045) &     1.5(0.4) &  0.62 \\
96741.38 & CH$_3$OH 2(0,2)--1(0,1)++ &     1.70(0.045) &     1.8(0.4) &  0.89 \\
96744.55 & CH$_3$OH 2(0,2)--1(0,1)  &      0.52(0.045) &     2.1(0.4) &  0.23 \\
96755.51 & CH$_3$OH 2(1,1)--1(1,0)  &      0.17(0.045) &     1.9(0.4) &  0.08 \\
\cline{1-5}
\multicolumn{5}{c}{19410+2336}   \\
\cline{1-5}
%89275.41 & CH$_2$DOH 2(0,2)--1(0,1)e1 &   5.93523E-03(0.003) &   0.391(0.887) &  1.42747E-02  \\ %!tentative
94405.16 & $^{13}$CH$_3$OH 2(--1,2)--1(--1,1) &  0.026(0.004) &   0.9(0.2) &  0.03 \\
94407.13 & $^{13}$CH$_3$OH 2(0,2)--1(0,1)++ &  0.044(0.004) &   1.0(0.1) &  0.04 \\
94541.76 & CH$_3$OH 8(3,5)--9(2,7)         &      0.028(0.005) &   1.6(0.4) &  0.016 \\
95169.46 & CH$_3$OH 8(0,8)--7(1,7)++ &    1.014(0.004) &    0.641(0.003) &   1.49 \\
95914.31 & CH$_3$OH 2(1,2)--1(1,1)++ &    0.24(0.05) &   1.0(0.3) &  0.22 \\
96739.36 & CH$_3$OH 2(--1,2)--1(--1,1)  &     1.56(0.08) &   1.0(0.4) &   1.51 \\
96741.38 & CH$_3$OH 2(0,2)--1(0,1)++ &     2.32(0.08) &   1.0(0.4) &   2.16 \\
96744.55 & CH$_3$OH 2(0,2)--1(0,1)  &      0.61(0.08) &   1.0(0.4) &  0.57 \\
96755.51 & CH$_3$OH 2(1,1)--1(1,0)  &      0.21(0.08) &   0.8(0.4) &  0.27 \\
\cline{1-5}
\multicolumn{5}{c}{ON1}   \\
\cline{1-5}
%89505.78 & CH$_3$OH 8(--4,5)--9(--3,7)  &    5.16884E-02(0.336) &  1.114(22.951) & 4.35915E-02 \\
94405.16 & $^{13}$CH$_3$OH 2(--1,2)--1(--1,1)  &  0.048(0.009) &   1.5(0.2) &  0.03 \\
94407.13 & $^{13}$CH$_3$OH 2(0,2)--1(0,1)++ &  0.11(0.01) &    1.9(0.2) &  0.053 \\
94411.02 & $^{13}$CH$_3$OH 2(0,2)--1(0,1)  &    0.010(0.004) &   0.9(0.5) &  0.011 \\
94420.45 & $^{13}$CH$_3$OH 2(1,1)--1(1,0)  &    0.014(0.005) &   2.3(0.8) &  0.006 \\
94541.76 & CH$_3$OH 8(3,5)--9(2,7)              &      0.065(0.004) &   1.3(0.1) &  0.047 \\
95169.46 & CH$_3$OH 8(0,8)--7(1,7)++       &    0.730(0.007) &    0.97(0.01) &  0.70 \\
%95208.66 & $^{13}$CH$_3$OH 2(1,1)--1(1,0)--- &  1.81454E-02(0.040) &    2.779(7.952) &  6.13443E-03 \\
95914.31 & CH$_3$OH 2(1,2)--1(1,1)++ &    0.39(0.04) &    1.8(0.2) &  0.21 \\
96739.36 & CH$_3$OH 2(--1,2)--1(--1,1)  &     2.1(0.1) &    1.3(0.4) &   1.48 \\
96741.38 & CH$_3$OH 2(0,2)--1(0,1)++ &     3.9(0.1) &    1.8(0.4) &   2.01 \\
96744.55 & CH$_3$OH 2(0,2)--1(0,1)  &       1.3(0.1) &    2.2(0.4) &  0.55 \\
96755.51 & CH$_3$OH 2(1,1)--1(1,0)  &      0.5(0.1) &    1.9(0.4) &  0.23 \\
\cline{1-5}
\multicolumn{5}{c}{I22134--VLA1}   \\
\cline{1-5}
%89251.16 & CH$_2$DOH 2(0,2)--1(0,1)o1 &   1.40158E-02(0.007) &    1.663(0.998) &  7.91811E-03  \\ %!tentative
95169.46 & CH$_3$OH 8(0,8)--7(1,7)++ &    0.022(0.003) &    0.7(0.1) &  0.03 \\
95914.31 & CH$_3$OH 2(1,2)--1(1,1)++ &    0.02(0.02) &    1.4(0.9) &  0.01  \\ %! tentative
96739.36 & CH$_3$OH 2(--1,2)--1(--1,1)  &    0.23(0.01) &    0.9(0.4) &  0.25 \\
96741.38 & CH$_3$OH 2(0,2)--1(0,1)++ &    0.30(0.01) &    0.744(0.4) &  0.38  \\
96744.55 & CH$_3$OH 2(0,2)--1(0,1)  &      0.06(0.01) &    0.7(0.4) &  0.085 \\
96755.51 & CH$_3$OH 2(1,1)--1(1,0)  &      0.01(0.01) &    0.5(0.4) &  0.026 \\
\cline{1-5}
\multicolumn{5}{c}{23033+5951}  \\
\cline{1-5}
%90262.31 & $^{13}$CH$_3$OH 11(--1,11)--10(--2,9)  &   1.52855E-02(0.005) &   0.860(0.343) &  1.67043E-02  \\ %!might be 15N2H+(1-0)!!!!
%90703.78 & CH3OD 2(--1,2)--1(--1,1)E &     7.48618E-03(0.003) &   0.391(7.150) &  1.80053E-02 \\
94405.16 & $^{13}$CH$_3$OH 2(--1,2)--1(--1,1)  &    0.022(0.005) &   1.2(0.3) &  0.017 \\
94407.13 & $^{13}$CH$_3$OH 2(0,2)--1(0,1)++ &    0.034(0.004) &    1.2(0.3) &  0.026 \\
95169.46 & CH$_3$OH 8(0,8)--7(1,7)++ &      0.634(0.003) &    0.647(0.004) &  0.92 \\
95914.31 & CH$_3$OH 2(1,2)--1(1,1)++ &      0.09(0.03) &    1.1(0.5) &  0.07 \\
96739.36 & CH$_3$OH 2(--1,2)--1(--1,1)  &      1.59(0.07) &    1.1(0.4) &   1.33 \\
96741.38 & CH$_3$OH 2(0,2)--1(0,1)++ &      1.94(0.07) &    1.0(0.4) &   1.75 \\
96744.55 & CH$_3$OH 2(0,2)--1(0,1)  &        0.32(0.07) &    1.0(0.4) &  0.31 \\
96755.51 & CH$_3$OH 2(1,1)--1(1,0)  &        0.1(0.2)    &    1.1(0.9) &  0.09 \\
\cline{1-5}
\multicolumn{5}{c}{NGC7538--IRS9}   \\
\cline{1-5}
94405.16 & $^{13}$CH$_3$OH 2(--1,2)--1(--1,1)  &   0.020(0.004) &   1.1(0.3) &  0.02 \\
94407.13 & $^{13}$CH$_3$OH 2(0,2)--1(0,1)++ &   0.019(0.004) &   0.7(0.2) &  0.03 \\
95169.46 & CH$_3$OH 8(0,8)--7(1,7)++ &     0.904(0.003) &    0.737(0.002) &   1.15 \\
95914.31 & CH$_3$OH 2(1,2)--1(1,1)++ &     0.15(0.03) &    1.1(0.2) &  0.13 \\
96739.36 & CH$_3$OH 2(--1,2)--1(--1,1)  &     0.95(0.05) &    1.2(0.4) &  0.77 \\
96741.38 & CH$_3$OH 2(0,2)--1(0,1)++ &      1.67(0.05) &    1.5(0.4) &   1.06 \\
96744.55 & CH$_3$OH 2(0,2)--1(0,1)  &       0.46(0.05) &    1.6(0.4) &  0.28 \\
96755.51 & CH$_3$OH 2(1,1)--1(1,0)  &       0.17(0.05) &    1.4(0.4) &  0.11 \\
% \hline
 \end{longtable}
 %$^{\bf a}$ might be a transition of CH$_3$OOH. \\ 
 }
 
\addtocounter{table}{2}

%\begin{table*}[t!]
\longtab{2}{
\begin{longtable}{lllll}
\caption{\label{tab_fit1mm} Same as Table~\ref{tab_fit3mm} for the transitions detected at 1~mm.} \\
%\label{tab_fit1mm}
%\begin{tabular}{ccccc}
\hline \hline
  freq &   transition    &   $\int T_{\rm MB}{\rm d}v$ &   \deltav\ &   $T_{\rm pk}$  \\
  MHz &            &  K km s$^{-1}$                    &  km s$^{-1}$    & K \\  
\hline
\endfirsthead
\caption{continued.} \\
\hline \hline
  freq &   transition    &   $\int T_{\rm MB}{\rm d}v$ &   \deltav\ &   $T_{\rm pk}$  \\
  MHz &        &  K km s$^{-1}$                    &  km s$^{-1}$    & K \\  
 \hline
 \endhead
\hline
\cline{1-5}
\multicolumn{5}{c}{HMSCs}   \\
\cline{1-5}
\multicolumn{5}{c}{I00117--MM2} \\
\cline{1-5}
218440.05 & CH$_3$OH 4(2,2)--3(1,2) &    0.29(0.03) &  1.9(0.2) & 0.14	\\
\cline{1-5}
\multicolumn{5}{c}{AFGL5142--EC} \\
\cline{1-5}
216945.6 & CH$_3$OH 5(1,4)--4(2,2) &     1.9(0.3) &  3.3(0.6) & 0.53 \\
%217886.39 & CH$_3$OH 20(1,19)--20(0,20) &   0.39660(0.047) &  4.510(0.502) & 8.26184E-02 \\
218440.05 & CH$_3$OH 4(2,2)--3(1,2) &      10.1(0.4) &  2.9(0.1) &  3.23 \\
%219983.99 & CH$_3$OH 25(3,22)--24(4,20) &  6.77714E-02(0.420) &	 2.674 (261.299) & 2.38080E-02  \\%! tentative
%219993.94 & CH$_3$OH 23(5,19-22(6,17) &   0.11638(0.998) &   3.894(143.464) & 2.80787E-02   \\%!tentative
220078.5 & CH$_3$OH 8(0,8)--7(1,6)    &      2(1) &   3.7(0.9) & 0.59 \\
223071.3 & CH$_2$DOH 5(2,3)--4(1,4)e1 &    0.02(0.01) &   0.4(0.8) & 0.06 \\
223107.3 & CH$_2$DOH 5(0,5)--4(0,4)o1 &    0.03(0.02) &   0.6(0.3) & 0.04 \\
223153.7 & CH$_2$DOH 5(3,2)--4(3,1)o1 &    0.06(0.03) &   2.1(0.7) & 0.03 \\
%223308.57 & CH$_3$OD 5(1,5)--4(1,4)A++ &   0.10358(0.028) &   2.946(0.981) & 3.30312E-02 \\
223315.4 & CH$_2$DOH 5(2,3)--4(2,2)e1 &    0.05(0.02) &   2.0(0.7) & 0.024$^{\bf a}$ \\
223422.3 & CH$_2$DOH 5(2,4)--4(2,3)e0 &    0.09(0.02) &   2.6(0.6) & 0.032$^{\bf a}$ \\
\cline{1-5}
\multicolumn{5}{c}{05358--mm3} \\
\cline{1-5}
216945.60 & CH$_3$OH 5(1,4)--4(2,2) &      0.67(0.06) &   2.2(0.3) & 0.29 \\
%217886.39 & CH$_3$OH 20(1,19)--20(0,20) &  3.29780E-02(0.021) &   0.605(0.465) & 5.12137E-02 \\
218440.05 & CH$_3$OH 4(2,2)--3(1,2) &      5.1(0.2) &   1.96(0.08) &  2.46 \\
220078.49 & CH$_3$OH 8(0,8)--7(1,6) &      0.8(0.4) &   2(1)        & 0.32 \\
\cline{1-5}
\multicolumn{5}{c}{G034--G2} \\
\cline{1-5}
218440.05 & CH$_3$OH 4(2,2)--3(1,2) &   0.11(0.02) &   2.4(0.6) & 0.045 \\   
\cline{1-5}
\multicolumn{5}{c}{G034--F2} \\
\cline{1-5}
 -- &  -- & -- &  -- &  -- \\ 
\cline{1-5}
\multicolumn{5}{c}{G034--F1} \\
\cline{1-5}
218440.05 & CH$_3$OH 4(2,2)--3(1,2) &  0.10(0.02) &   2.100(0.001) & 0.043 \\
\cline{1-5}
\multicolumn{5}{c}{G028--C1} \\
\cline{1-5}
218440.05 & CH$_3$OH 4(2,2)--3(1,2) &    0.09(0.014) &   1.188(0.001) & 0.074 \\
%223422.3 & CH$_2$DOH 5(2,4)--4(2,3)e0 &  1.97493E-02(0.010) &   0.339(0.000) & 5.47947E-02 \\ %!tentative @ 2.5 sigma
\cline{1-5}
\multicolumn{5}{c}{I20293--WC} \\
\cline{1-5}
218440.05 & CH$_3$OH 4(2,2)--3(1,2) &    0.05(0.015) &   2.0(0.7) & 0.024 \\
%223153.7 & CH2DOH 5(3,2)--4(3,1)o1 &  3.11180E-02(0.008) &   0.849(0.210) & 3.44519E-02 \\
\cline{1-5}
\multicolumn{5}{c}{I22134--G} \\
\cline{1-5}
218440.05 & CH$_3$OH 4(2,2)--3(1,2) &   0.23(0.02) &   1.1(0.1) & 0.20 \\
\cline{1-5}
\multicolumn{5}{c}{I22134--B} \\
\cline{1-5}
218440.05 & CH$_3$OH 4(2,2)--3(1,2) &    0.06(0.02) &   2.1(0.7) & 0.03 \\
\cline{1-5}
\multicolumn{5}{c}{HMPOs} \\
\cline{1-5}
\multicolumn{5}{c}{I00117--MM1} \\
\cline{1-5}
218440.05 & CH$_3$OH 4(2,2)--3(1,2) &   0.19(0.02) &   1.9(0.3) & 0.10 \\
\cline{1-5}
\multicolumn{5}{c}{AFGL5142--MM}  \\
\cline{1-5}
216945.60 & CH$_3$OH 5(1,4)--4(2,2)     &     1.9(0.2) &   3.2(0.5) & 0.55  \\
217886.39 & CH$_3$OH 20(1,19)--20(0,20) &   0.59(0.06) &   5.8(0.7) & 0.09 \\
218440.05 & CH$_3$OH 4(2,2)--3(1,2)     &     9.9(0.4) &   2.7(0.1) &  3.46 \\
219983.99 & CH$_3$OH 25(3,22)--24(4,20) &  0.10(0.09) &   2.7(0.6) & 0.037 \\
219993.94 & CH$_3$OH 23(5,19)--22(6,17) &   0.04(0.03) &  1.7(0.5) & 0.02 \\
220078.49 & CH$_3$OH 8(0,8)--7(1,6)     &      2(1) &   3(1) & 0.6 \\
223071.3 & CH$_2$DOH 5(2,3)--4(1,4)e1 &    0.06(0.02) &  1.6(0.7) & 0.034$^{\bf a}$ \\ %! tentative at 2 sigma 
223308.57 & CH$_3$OD 5(1,5)--4(1,4)A++ &   0.08(0.02) &  1.4(0.4) & 0.05$^{\bf a}$ \\ %! at 3 sigma
\cline{1-5}
\multicolumn{5}{c}{05358--mm1}  \\
\cline{1-5}
216945.60 & CH$_3$OH 5(1,4)--4(2,2) &       0.92(0.07) &  3.3(0.4) & 0.26 \\
217886.39 & CH$_3$OH 20(1,19)--20(0,20) &   0.25(0.08) &  8(2) & 0.03 \\
218440.05 & CH$_3$OH 4(2,2)--3(1,2) &       4.6(0.2) &   2.5(0.2) &  1.75 \\
220078.49 & CH$_3$OH 8(0,8)--7(1,6) &       1.0(0.4) &   3(1) & 0.33 \\
\cline{1-5}
\multicolumn{5}{c}{18089--1732} \\
\cline{1-5}
%216857.86 & CH$_3$OH 6(3,3)--7(1,6) &         6.07221E-02(0.128) &   2.308(6.146) & 2.47132E-02 \\
216945.60 & CH$_3$OH 5(1,4)--4(2,2) &  	 3.3(0.2) &    3.5(0.2) & 0.89 \\
%217044.62 & $^{13}$CH$_3$OH 14(1,13)--13(2,12)--- &  0.44045(0.168) &   3.061(1.461) & 0.13519 \\  %!could be blended with 13CN(2-1) @ 217045.6GHz
217399.54 & $^{13}$CH$_3$OH 10(2,8)--9(3,7)A++ &    0.96(0.08) &   4.4(0.4) & 0.20 \\
%217525.07 & CH$_3$OH 16(1,15)--15(3,12)"	& 9.73294E-02(0.061) &  2.916(1.664) & 3.13541E-02 \\
217886.39 & CH$_3$OH 20(1,19)--20(0,20)	& 1.39(0.08) &    3.8(0.3) & 0.34 \\
218440.05 & CH$_3$OH 4(2,2)--3(1,2)  &       8.6(0.3) &   3.3(0.1) &  2.44 \\
219983.99 & CH$_3$OH 25(3,22)--24(4,20) &   0.4(0.6) &   3.6(0.9) & 0.09 \\
219993.94 & CH$_3$OH 23(5,19-22(6,17) &    0.4(0.6) &	   4.0(0.9) & 0.09 \\
220078.49 & CH$_3$OH 8(0,8)--7(1,6) &       3.7(0.7) &	3.7(0.8) & 0.93 \\
223071.3 & CH$_2$DOH 5(2,3)--4(1,4)e1 &    0.08(0.06) &   2.4(0.3) & 0.03 \\
223107.3 & CH$_2$DOH 5(0,5)--4(0,4)o1 &    0.08(0.06) &   1.3(0.5) & 0.057 \\
223153.7 & CH$_2$DOH 5(3,2)--4(3,1)o1 &    0.07(0.06) &   1.0(0.5) & 0.06 \\
222468.34 & $^{13}$CH$_3$OH 21(1,20)--21(0,21) &  0.06(0.1) &   1.7(0.5) & 0.03 \\
223308.57 & CH$_3$OD 5(1,5)--4(1,4)A++ &     0.21(0.06) &   2.6(0.7) & 0.08$^{\bf b}$ \\
223315.4 & CH$_2$DOH 5(2,3)--4(2,2)e1 &      0.06(0.04) &   1.2(0.7) & 0.05	  \\
\cline{1-5}
\multicolumn{5}{c}{18517+0437}  \\
\cline{1-5}
216945.60 & CH$_3$OH 5(1,4)--4(2,2) &      2.3(0.2) &   3.6(0.3) & 0.59 \\
%217044.62 & $^{13}$CH$_3$OH 14(1,13)--13(2,12)--- &  0.10651(0.148) &   5.809(8.749) & 1.72239E-02 \\ %!could be blended with 13CN(2-1) @ 217045.6GHz
%217399.54 & $^{13}$CH$_3$OH 10(2,8)--9(3,7)A++ &   0.22557(0.072)	5.363(2.126) & 3.95137E-02 \\
217886.39 & CH$_3$OH 20(1,19)--20(0,20) &     0.97(0.07) &	4.6(0.4) & 0.20 \\
218440.05 & CH$_3$OH 4(2,2)--3(1,2) &        5.8(0.3) &  2.7(0.2) &  2.0 \\
219983.99 & CH$_3$OH 25(3,22)--24(4,20) &    0.14(0.09)	& 3.8(0.8) & 0.034 \\
219993.94 & CH$_3$OH 23(5,19)--22(6,17) &    0.13(0.09)	& 3.3(0.5) & 0.038 \\
220078.49 & CH$_3$OH 8(0,8)--7(1,6)        &        2.1(0.7)	& 3.9(0.9) & 0.52\\
221285.24 & $^{13}$CH$_3$OH 8(--1,8)--7(0,7) &    0.39(0.03) &   3.8(0.3) & 0.095 \\
\cline{1-5}
\multicolumn{5}{c}{G75--HCHII} \\
\cline{1-5}
216945.60 & CH$_3$OH 5(1,4)--4(2,2) &     2.3(0.2) &   3.1(0.2) & 0.69 \\
217399.54 & $^{13}$CH$_3$OH 10(2,8)--9(3,7)A++ & 0.57(0.06) &	  2.7(0.4) & 0.20 \\
217886.39 & CH$_3$OH 20(1,19)--20(0,20) &   0.45(0.06) &	3.0(0.5) & 0.14 \\
218440.05 & CH$_3$OH 4(2,2)--3(1,2)   &      6.4(0.3) &   3.1(0.2) &  1.92 \\
%219983.99 & CH$_3$OH 25(3,22)--24(4,20) &  0.04(0.2) &   1.0(0.6) & 2.77006E-02 \\
220078.49 & CH$_3$OH 8(0,8)--7(1,6) &      2.2(0.8) &   3.1(0.9) & 0.66 \\
223107.3 & CH$_2$DOH 5(0,5)--4(0,4)o1 &    0.07(0.03) &   2.2(0.9) & 0.03 \\
223422.3 & CH$_2$DOH 5(2,4)--4(2,3)e0 &    0.04(0.02) &   0.8(0.3)  & 0.05 \\
\cline{1-5}
\multicolumn{5}{c}{I20293--MM1}  \\
\cline{1-5}
216945.60 & CH$_3$OH 5(1,4)--4(2,2) &       0.18(0.06) &   2.9(0.9) & 0.06 \\
218440.05 & CH$_3$OH 4(2,2)--3(1,2) &       0.22(0.03) &   2.7(0.4) & 0.075 \\
220078.49 & CH$_3$OH 8(0,8)--7(1,6) &      0.4(0.2) &  5.4(0.9) & 0.06 \\
\cline{1-5}
\multicolumn{5}{c}{I21307}  \\
\cline{1-5}
216945.60 & CH$_3$OH 5(1,4)--4(2,2) &      0.10(0.02) &   1.9(0.4) & 0.05 \\
218440.05 & CH$_3$OH 4(2,2)--3(1,2) &      0.99(0.05) &   2.0(0.1) & 0.47 \\
220078.49 & CH$_3$OH 8(0,8)--7(1,6) &      0.1(0.2) &   2.5(0.9) & 0.04 \\
\cline{1-5}
\multicolumn{5}{c}{I23385}  \\
\cline{1-5}
216945.60 & CH$_3$OH 5(1,4)--4(2,2) &      0.22(0.04) &   3.0(0.8) & 0.07 \\
218440.05 & CH$_3$OH 4(2,2)--3(1,2) &      1.9(0.07) &   2.7(0.1) & 0.66 \\
220078.49 & CH$_3$OH 8(0,8)--7(1,6) &      0.3(0.2) &   3.2(0.9) & 0.08 \\
\cline{1-5}
\multicolumn{5}{c}{UC \HII s} \\
\cline{1-5}
\multicolumn{5}{c}{G5.89--0.39}  \\
\cline{1-5}
216945.60 & CH$_3$OH 5(1,4)--4(2,2) &      4(1) &   5(1) & 0.82 \\
%217399.54 & $^{13}$CH$_3$OH 10(2,8)--9(3,7)A++ &   2.2822(0.257) &   3.582(0.501) & 0.59847 \\
217886.39 & CH$_3$OH 20(1,19)--20(0,20) &   0.08(0.08) &   3.7(0.6) & 0.02 \\
218440.05 & CH$_3$OH 4(2,2)--3(1,2) &       20(2) &   4.8(0.7) &  3.87 \\
220078.49 & CH$_3$OH 8(0,8)--7(1,6) &      5.3(0.8) &   4.8(0.7) &  1.04 \\
221285.24 & $^{13}$CH$_3$OH 8(--1,8)--7(0,7) &   0.35(0.05) &   4.8(0.8) & 0.07 \\
%223308.57 & CH$_3$OD 5(1,5)--4(1,4)A++ &   0.25515(0.156) &   5.004(5.152) & 4.79012E-02  \\
\cline{1-5}
\multicolumn{5}{c}{I19035--VLA1}  \\
\cline{1-5}
216945.60 & CH$_3$OH 5(1,4)--4(2,2) &      0.50(0.05) &   4.2(0.5) & 0.11  \\
218440.05 & CH$_3$OH 4(2,2)--3(1,2) &      2.01(0.07) &   3.6(0.15) & 0.52 \\   
220078.49 & CH$_3$OH 8(0,8)--7(1,6) &      0.6(0.3) &   5.0(0.9) & 0.12 \\    
\cline{1-5}
\multicolumn{5}{c}{19410+2336}  \\
\cline{1-5}
216945.60 & CH$_3$OH 5(1,4)--4(2,2) &      0.55(0.04) &   3.3(0.3) & 0.16 \\
218440.05 & CH$_3$OH 4(2,2)--3(1,2) &      0.38(0.02) &   2.1(0.2) & 0.17 \\
220078.49 & CH$_3$OH 8(0,8)--7(1,6) &      0.6(0.4) &   3.8(0.9) & 0.14 \\
\cline{1-5}
\multicolumn{5}{c}{ON1}  \\
\cline{1-5}
216945.60 & CH$_3$OH 5(1,4)--4(2,2) &    1.0(0.1) &   3.6(0.5) & 0.25 \\
218440.05 & CH$_3$OH 4(2,2)--3(1,2) &    0.44(0.03) &   3.1(0.2) & 0.14 \\
220078.49 & CH$_3$OH 8(0,8)--7(1,6) &    1.1(0.6) &   4.2(0.9) & 0.25 \\
221285.24 & $^{13}$CH$_3$OH 8(--1,8)--7(0,7) &  0.12(0.02) &   3.4(0.6) & 0.033 \\
%223071.3 & CH2DOH 5(2,3)--4(1,4)e1 &  7.89176E-02(0.023)	 & 4.884(1.398) & 1.51793E-02 \\
%223107.3 & CH2DOH 5(0,5)--4(0,4)o1 &  7.35382E-02(0.016) &	 2.095(0.577) & 3.29686E-02 \\
%223153.7 & CH2DOH 5(3,2)--4(3,1)o1 &  4.31117E-02(0.011)	 & 1.199(0.329) & 3.37873E-02 \\
\cline{1-5}
\multicolumn{5}{c}{I22134--VLA1}  \\
\cline{1-5}
218440.05 & CH$_3$OH 4(2,2)--3(1,2) &   0.23(0.03) &   1.4(0.2) & 0.16 \\
220078.49 & CH$_3$OH 8(0,8)--7(1,6) &   0.03(0.05) &   1.1(0.7) & 0.03 \\
\cline{1-5}
\multicolumn{5}{c}{23033--UCHII}  \\
\cline{1-5}
216945.60 & CH$_3$OH 5(1,4)--4(2,2) &   0.21(0.04) &   2.9(0.7) & 0.07 \\
218440.05 & CH$_3$OH 4(2,2)--3(1,2) &   1.49(0.06) &   2.3(0.1) & 0.6  \\
220078.49 & CH$_3$OH 8(0,8)--7(1,6) &   0.3(0.3) &   4.7(0.9) & 0.06 \\
\cline{1-5}
\multicolumn{5}{c}{NGC7538--IRS9}  \\
\cline{1-5}
216945.60 & CH$_3$OH 5(1,4)--4(2,2) &   0.45(0.04) &   3.1(0.3) & 0.14 \\
218440.05 & CH$_3$OH 4(2,2)--3(1,2) &   2.42(0.09) &   2.4(0.1) & 0.94 \\
220078.49 & CH$_3$OH 8(0,8)--7(1,6) &   0.5(0.3) &   3.3(0.9) & 0.15 \\
\hline
$^{\bf a}$ tentative detection in between 2 and 3$\sigma$ rms; \\
$^{\bf b}$ partially blended with (CH$_2$OH)$_2$ (ethylene-glycol). \\
 \end{longtable}
 }

\normalsize

\clearpage

\end{document}